\documentclass[12pt]{article}

 \usepackage{hyperref}
\hypersetup {
pdfpagemode = {UseNone},
pdftitle = {CollapseInM},
pdfauthor = {Christian Beck},
pdflang = {en-GB}
}

\usepackage[dvips,letterpaper,margin=0.75in,bottom=0.5in]{geometry}
\makeatother % End of region containing @ commands
\renewcommand{\v}[1]{\ensuremath{\boldsymbol{#1}}} % for vectors
\newcommand{\gv}[1]{\ensuremath{\mbox{\boldmath$ #1 $}}} 
\newcommand{\abs}[1]{\left| #1 \right|} % for absolute value
\newcommand{\avg}[1]{\left< #1 \right>} % for average
\renewcommand{\d}[2]{\frac{d #1}{d #2}} % for derivatives
\newcommand{\pd}[2]{\frac{\partial #1}{\partial #2}} 
\newcommand{\grad}[1]{\gv{\nabla} #1} % for gradient
\renewcommand{\div}[1]{\gv{\nabla} \cdot #1} % for divergence
\let\baraccent=\= % rename builtin command \= to \baraccent
\renewcommand{\=}[1]{\stackrel{#1}{=}} % for putting numbers above =

\usepackage{dsfont,a4wide,amsthm,amscd,mathrsfs,graphicx,bbm}

\usepackage{savesym}
\usepackage{amsmath}
\savesymbol{iint}
\usepackage{txfonts}
\restoresymbol{TXF}{iint}

\newenvironment{Acknowledgement}{%
   \abstract}{%
  \endabstract
}

\usepackage[format=plain,labelsep=newline,font=sl,textfont=small,labelfont=bf,justification=centerlast]{caption}
\newcommand{\RM}[1]{\MakeUppercase{\romannumeral #1}}

\usepackage{amsfonts}
\usepackage{amssymb}

\usepackage{fancyhdr}
\fancyhfoffset{0.0pt}

\begin{document}
%\vspace*{1.0cm}
\cfoot{}

\renewcommand\thefootnote{\fnsymbol{footnote}}

%\noindent
\newcommand{\LMUTitle}[9]{
  \thispagestyle{empty}
  \vspace*{.5cm}
  {\parindent0cm
   \rule{\linewidth}{.3ex}}
  \begin{center}
  	\vspace*{.5cm}
   	\sffamily\bfseries\LARGE #1\\
%	\vspace*{\stretch{0.5}}
%	\large #9\\
   	\vspace*{.5cm}
   	\sffamily\bfseries\large
   	#2
   	\vspace*{.5cm}
  \end{center} 
  \rule{\linewidth}{.3ex}
  \vspace*{1cm}
\begin{abstract}
\noindent
The implications of the relativistic space-time structure for a physical description by quantum mechanical wave-functions are investigated. On the basis of a detailed analysis of Bell's concept of local causality, which is violated in quantum theory, we argue that this is a subtle, as well as an important effort. A central requirement appearing in relativistic quantum mechanics, namely local commutativity, is analyzed in detail and possible justifications are given and discussed. The complexity of the implications of wave function reduction in connection with Minkowski space-time are illustrated by a quantum mechanical measurement procedure which was proposed by Aharonov and Albert. This procedure and its relativistic implications are explicitly analyzed and discussed in terms of state evolution. This analysis shows that the usual notion of state evolution fails in relativistic quantum theory. Two possible solutions of this problem are given. In particular, it is shown that also a theory with a distinguished foliation of space-time into space-like leafs -- accounting for nonlocality -- makes the right predictions for the Aharonov-Albert experiment. We will repeatedly encounter that an analysis of the wave-function alone does not suffice to answer the question of relativistic compatibility of the theory, but that the actual events in space time, which are predicted and described by the theory, are crucial. Relativistic versions of quantum mechanical theories which precisely describe actual processes in space-time are briefly described and discussed in the appendix.          
\end{abstract}}

\LMUTitle
      {\huge{\textbf{Wavefunctions and Minkowski Space-Time}} \\ \large{On the Reconciliation of Quantum Theory with Special Relativity}}		%Titel - #1
      {Christian Beck\footnote{beck@math.lmu.de}}                   % Vor- und Nachname des Autors - #2
      {M\"unchen}                   	% Geburtsort des Autors - #3
      {Fakult\"at f\"ur Physik}		    % Name der Fakultaet - #4
      {M\"unchen 2010}                 	% Ort und Jahr der Erstellung - #5
      {29. Juli 2010}                   % Tag der Abgabe - #6
      {Prof. Dr. D. D\"urr}            	% Name des Erstgutachters - #7
      {Prof. Dr. D. L\"ust}              	% Name des Zweitgutachters - #8

\renewcommand{\thefootnote}{\arabic{footnote}}

\newpage

\renewcommand{\baselinestretch}{1.1}

\cfoot{}

\tableofcontents

\newpage

\lhead{Clarification \& Notation}

\section*{Clarification \& Notation}

\begin{center}
\textbf{Space-Time} 
\end{center}

Minkowski space-time, denoted by $\mathscr{M}$, is $\mathbb{R}^4$ endowed with the Minkowski metric $g^{\mu\nu}$, where our choice of signature is $(1,-1,-1,-1)$. Pairs of events $x,y\in\mathscr{M}$ are said to be space-like separated if $(x-y)_{\mu}(x-y)^{\mu}<0$ and time-like separated if $(x-y)_{\mu}(x-y)^{\mu}>0$. Two non intersecting space-time regions $\mathcal{A}$ and $\mathcal{B}$ are said to be space-like (time-like) separated if each space-time point $x\in\mathcal{A}$ is space-like (time-like) separated with respect to each space-time point in $y\in\mathcal{B}$. The absolute (causal) future of some space-time point $x$, denoted by $\mathscr{F}(x)$, is the set of all points $y\in\mathscr{M}$ for which $(x-y)_{\mu}(x-y)^{\mu}>0$ and $x^0<y^0$ (since the time-order of time-like separated events is Lorentz invariant this set is also invariant under Lorentz transformations). $\mathscr{F}(x)$ is the interior of the forward light-cone of event $x$. In contrast the absolute (causal) past of $x$, denoted by $\mathscr{P}(x)$, is the set of points $y\in\mathscr{M}$ for which also $(x-y)_{\mu}(x-y)^{\mu}>0$ but $x^0>y^0$. This (also Lorentz invariant) set is given by the interior of the backward light-cone of event $x$.   

\begin{center}
\textbf{Operators \& Values}
\end{center}

To avoid confusion within the calculations I will distinguish between some \textsl{physical quantity} $\mathscr{A}$, its corresponding selfadjoint \textsl{operator} $\hat{A}$ (acting on Hilbertspace) and its \textsl{actual (``measured``) value} $\tilde{\alpha}$, where necessary. This is done by clapping  operators a hat on and indicating actual values appropriately (where necessary). Physical quantities and their generic (variable) values are not distinguished in such a way, the meaning of the symbols should be clear from the context in this case.

\begin{center}
\textbf{Wave Equations}
\end{center}

All wave-functions appearing within this work are assumed to be solutions of Lorentz invariant equations, like the Dirac equation (nonetheless I will sometimes refer to the corresponding unitary evolution, e.g.\! generated by the Dirac-Hamiltonian, as Schr\"odinger-evolution). Negative-energy-states will not be considered here. In addition we assume that all relevant length-scales (e.g.\! the scales of interaction of particles with some device) are much larger than the Compton-wavelength of the involved particles, such that we can neglect particle creation and annihilation effects. The Planck constant $\hbar$ and the velocity of light $c$ are set equal to one.

\begin{center}
\textbf{Spin}
\end{center}

Considerations will be often illustrated by considering spin-$\frac{1}{2}$-particles. Since the only aim is to shed some light on delicate conceptual implications of wave-function collapse and quantum nonlocality in a relativistic framework I will omit to deal with transformation properties of the corresponding wave-functions under Lorentz transformations. Ghirardi argued \cite{ghirardilessons, ghirardinonlocal} that this can be justified by exchanging the spin with some physical quantity which has an analog underlying mathematical description, but which is a scalar under Lorentz transformations, like the isospin. Nevertheless I will use the terminology of \textsl{spin} and describe corresponding measurement situations in which devices like Stern-Gerlach magnets are involved. 

The mathematical description in brief is the following: Consider some particle with a degree of freedom which is described by the elements of a two-dimensional Hilbertspace $\mathcal{H}_S$. Further this degree of freedom is associated with three non-commuting self-adjoint operators acting on $\mathcal{H}_S$ (and describing the measurement statistics of various associated experiments) which we shall denote by $\hat{\sigma}_x$, $\hat{\sigma}_y$ and $\hat{\sigma}_z$. These operators obey the algebra $\hat{\sigma}_{x_i}\hat{\sigma}_{x_j}=\delta_{ij}\mathds{1}_{\mathcal{H}_S}+i\varepsilon_{ijk}\hat{\sigma}_{x_k}$ (i.e.\! they are (essentially) the infinitesimal generators of group $SU(2)$) and we shall call them \textsl{spin-operators}. Each operator has eigenvalues $\pm1$. Let us denote the eigenstates of $\hat{\sigma}_z$ (which will be most used in calculations) in the following way
\begin{equation}
 \hat{\sigma}_z \mid \uparrow \rangle = + \mid \uparrow \rangle \quad \mbox{ and } \quad \hat{\sigma}_z \mid \downarrow \rangle = - \mid \downarrow \rangle \mbox{ ,}  
\end{equation}
(i.e.\! in spectral representation we have $\hat{\sigma}_z = \mid \uparrow \rangle \langle \uparrow \mid - \mid \downarrow \rangle \langle \downarrow \mid$). The eigenstates of the remaining two operators are denoted analogously, only with indicated $x$ and $y$, respectively, i.e.\!
\begin{equation}
\begin{gathered}
 \hat{\sigma}_x \mid \uparrow \rangle_x = + \mid \uparrow \rangle_x \quad \mbox{ and } \quad \hat{\sigma}_x \mid \downarrow \rangle_x = - \mid \downarrow \rangle_x\\
 \hat{\sigma}_y \mid \uparrow \rangle_y = + \mid \uparrow \rangle_y \quad \mbox{ and } \quad \hat{\sigma}_y \mid \downarrow \rangle_y = - \mid \downarrow \rangle_y 
\end{gathered}
\end{equation}
From this, appropriate basis transformations can be calculated: Denote $\hat{\boldsymbol{\sigma}}:= (\hat{\sigma}_x,\hat{\sigma}_y,\hat{\sigma}_z)$ and consider the unit-vector $\boldsymbol{u}\in\mathbb{R}^3$ which is characterized by the polar angles $\theta$ and $\varphi$. Then the eigenvectors of the operator $\hat{\sigma}_{\boldsymbol{u}}:=\boldsymbol{u}\cdot \hat{\boldsymbol{\sigma}}$ expanded in the eigenbasis  of $\hat{\sigma}_z$ read
\begin{equation}
\begin{gathered} \label{sigmatrans}
 \mid \uparrow \rangle_{\boldsymbol{u}} = \cos{\frac{\theta}{2}} e^{-i\frac{\varphi}{2}}\mid \uparrow \rangle + \sin{\frac{\theta}{2}} e^{i\frac{\varphi}{2}}\mid \downarrow \rangle \quad \mbox{ with } \quad \hat{\sigma}_{\boldsymbol{u}} \mid \uparrow \rangle_{\boldsymbol{u}} = +\mid \uparrow \rangle_{\boldsymbol{u}} \\
\mid \downarrow \rangle_{\boldsymbol{u}} =-\sin{\frac{\theta}{2}} e^{-i\frac{\varphi}{2}}\mid \uparrow \rangle + \cos{\frac{\theta}{2}} e^{i\frac{\varphi}{2}}\mid \downarrow \rangle \quad \mbox{ with } \quad \hat{\sigma}_{\boldsymbol{u}} \mid \downarrow \rangle_{\boldsymbol{u}} = -\mid \downarrow \rangle_{\boldsymbol{u}}
\end{gathered}
\end{equation}

The complete Hilbertspace is the tensor-product (the product-space) of $\mathcal{H}_S$ with the Hilbert-space related to the other degrees of freedom (in particular the spatial ones) of the particle under consideration. But we will ignore other degrees of freedom in calculations dealing with spin.

Most of the illustrative examples will concern a two-particle spin-$\frac{1}{2}$-system with the four-dimensional Hilbertspace $\mathcal{H}_S=\mathcal{H}_{S_1}\otimes\mathcal{H}_{S_2}$. If appropriate we will use as the orthonormal basis of $\mathcal{H}_S$ either the common eigenstates of the commuting operators $\hat{\sigma}^{(1)}_z\equiv\hat{\sigma}^{(1)}_z\otimes\mathds{1}_{\mathcal{H}_{S_2}}$ and $\hat{\sigma}^{(2)}_z\equiv\mathds{1}_{\mathcal{H}_{S_1}}\otimes\hat{\sigma}^{(2)}_z$ given by 
\begin{equation}
 \mid \uparrow \uparrow \rangle \quad , \quad \mid \uparrow \downarrow \rangle \quad , \quad \mid \downarrow \uparrow \rangle \quad , \quad \mid \downarrow \downarrow \rangle 
\end{equation}
 (where we denote $\mid a \: b \rangle := \mid a \rangle^{(1)} \otimes \mid b \rangle^{(2)}$) or the common eigenstates of the commuting operators $\hat{\sigma}^{tot}_z=\hat{\sigma}^{(1)}_z+\hat{\sigma}^{(2)}_z$ and $(\hat{\boldsymbol{\sigma}}^{tot})^2=(\hat{\sigma}^{tot}_x)^2+(\hat{\sigma}^{tot}_y)^2+(\hat{\sigma}^{tot}_z)^2$ given by
\begin{equation}
\mid \uparrow \uparrow \rangle \quad , \quad \mid \Psi_+ \rangle := \frac{1}{\sqrt{2}}(\mid \uparrow \downarrow \rangle + \mid \downarrow \uparrow \rangle) \quad , \quad \mid \downarrow \downarrow \rangle \quad \mbox{and} \quad \mid \Psi_- \rangle := \frac{1}{\sqrt{2}}(\mid \uparrow \downarrow \rangle - \mid \downarrow \uparrow \rangle) \mbox{ .}
\end{equation}
The latter state $\mid \Psi_- \rangle$ is the singlet state which obeys $(\hat{\boldsymbol{\sigma}}^{tot})^2 \mid \Psi_- \rangle = 0$. It will be of some importance for our considerations that it is the only element of $\mathcal{H}_{S}$ which is a common eigenstate of the non-commuting operators $\hat{\sigma}^{tot}_x$, $\hat{\sigma}^{tot}_y$ and $\hat{\sigma}^{tot}_z$ (with eigenvalue zero in each case).   

\newpage

\begin{center}
\textbf{EPR}
\end{center}

A crucial gedankenexperiment to illustrate our considerations will be Bohm's version of the EPR-experiment \cite{bohmqt} (EPRB). It is about a system of two spatially separated spin-$\frac{1}{2}$-particles in the singlet state $\mid \Psi_- \rangle$, which -- at some time -- are exposed to inhomogeneous magnetic fields, respectively. The fields are caused by Stern-Gerlach magnets (SGMs). If not stated otherwise the SGMs are both to be thought of as oriented in the z-direction. Consequently, if the system is in an eigenstate of $\hat{\sigma}^{(i)}_z$, particle(i) is deflected upwards by SGM(i) if the corresponding eigenvalue is $+1$ and downwards if the eigenvalue is $-1$. By measuring the position of the respective particle afterwards, the z-component of its spin (or the respective component if the SGM is in a different orientation) is determined. Whenever there is no risk of misunderstanding the subsequent position measurement will not be mentioned explicitly and the interaction of particle(i) with SGM(i) will be called a measurement of the respective component of the spin of that particle. But when it might serve to gain more clarity this measurement will be decomposed into the unitary interaction with the (external) magnetic field of the SGM and a subsequent position measurement (e.g.\! the particle hits a photographic plate).

\begin{center}
 \textbf{Pictures}
\end{center}

Sometimes EPRB-like situations will be illustrated in space-time diagrams. In these pictures the SGMs are only indicated at the time of interaction with the respective particles, the source only at the time of emission, and their world-lines are not sketched. The deflections of the particles at the SGMs are indicated symbolically, so the particles' world-lines after the passage of the SGMs are not to be taken literally. But also the world-lines of the particles prior to the interactions with the SGMs are rather symbolically: For example all inferred findings will be also valid for the relativistic collapse-theory \textsl{rGRWf} \cite{rGRWf} (see chapter \ref{rgrwf} in which it is appropriate to say that (in the vast majority of cases) there are no particles at all in the space-time region laying between the event of particle emission (from some particle source) and the events of interaction with the SGMs.

\newpage
\lhead{\rightmark}

\cfoot{\thepage}
\setcounter{page}{1}

\section{Introduction}

\subsection{Relativistic Quantum Theory: The Meaning of ``Relativistic''}

Relativity principle, invariance of physical laws under Lorentz-/Poincare-transformations, causality, covariance of physical description, micro-causality, macro-causality, impossibility of faster than light signaling,  locality, impossibility of superluminal matter-/energy-transport, no preferred Lorentz-frame of reference, local causality, physical description without resorting to concepts related to simultaneity, no extra structures on Minkowski space-time apart from the Lorentz-metric... \paragraph*{}

The requirements a certain theory has to fulfill in order to deserve the label ``relativistic theory''\footnote{Within this work I will confine myself to the subject of special relativity and that is the sense in which I use the term ``relativistic''. But a huge part of the reasoning within these lines has also implications for analogous issues, if it is applied to general relativistic space-time (with appropriate substitutions of respective terms and concepts).} are formulated in a variety of different expressions, some of which are tautologically equivalent, some of which manifest their logical connections or equivalence after some deductive reasoning and some of which need disambiguation. Indeed, all these notions of ``relativistic'' turn out to be (more or less) equivalent if we take Einsteins conception of causal structures connecting different parts of the physical universe as a basis, i.e.\! that all causal connections between spatially separated objects of physical description must be conveyed by something which propagates in space (a continuous chain of cause and effect), like it is without question the case for ordinary physical interactions\footnote{This assertion is a little bit sloppy. Indeed one can only deduce the existence of an invariant (not necessarily maximal) velocity from the relativity principle (see e.g.\! \cite{Sexl}) and for the actual equivalence of the above ``relativistic requirements``, Einsteins causal conception must be supplemented, e.g.\! by the requirement that causes must precede their effects in directed causal processes (with appropriate definition of a directed causal process -- I will come to this later). For a detailed discussion on e.g.\! superluminal signals within a relativistic theory see \cite{Maudlin}.}. 

\begin{center}
 \textbf{Quantum Theory}
\end{center}

But if we are concerned with quantum theory and take the consequences of physical description by a wave-function, living not on physical but on configurational space, seriously -- and Bell's theorem suggests that we have to -- these things become extraordinarily subtle. Then it is not true anymore that the above requirements of relativity are all equivalent and it is possible, and indeed necessary, to construct theories which conform to some of them but not to others. Therefore, one could say that there are different ``degrees of compatibility with relativity''. And if one follows the discussions and investigations about that issue it becomes indeed a bit unclear what the true notion of ``relativistic'' is, to be set in stone.      
   
%\newpage

\subsection{Relativistic Quantum Theory: The Meaning of Events}

In reasoning about quantum mechanics within a special relativistic realm it becomes also very striking that one has to specify what the events in space-time actually are, which are predicted and described by the theory: We will encounter that different answers to this question will yield very different relativistic properties of the theory, especially very different ``degrees of compatibility'' of quantum theory with relativistic space-time structure. 

There is an ongoing arduous debate in quantum mechanics about the nature and conceptual status of these events. Apart from speculative sophistications about the nature (or non-nature) of reality within such discussions, we should at least regard the positions of pointers (and the readouts on computer screens etc.) in laboratories as somehow objectively occurring ``events in space-time`` predicted by quantum theory: If the theory shall provide predictions for possible human perceptions (e.g.\! in the laboratory) there must be objects in the theory which correspond to objects of possible sensual perceptions which are confined to space-time. John Bell \cite{Bell} called these objects of the theory \textit{local beables} and various authors introduced the concept of \textit{primitive ontology} \cite{commstruc, quantequi, BMoperators, unrompics, sm} to comprise these elements of the theory. 

\begin{center}
 \textbf{Primitive Ontology}
\end{center}

The epithet ''primitive`` expresses the correspondence of these objects of the theory to the ''primitive objects'' which fill the world of our perceptions: the stuff pointers, tables, chairs and trees are made of, i.e.\!\! matter. It is controversial, for example, if it is adequate to give the wave-function ontological status in quantum theory but (without entering this discussion here) there might be arguments which are applicable to justify such an interpretation of the wave-function. However, the pure wave-function as a function on configuration space cannot immediately be regarded as a kind of ``field'' on physical space and therefore (although it might be part of the ontology of the theory) it cannot be part of the primitive ontology. And further -- as I will argue in detail -- within a relativistic theory one cannot create the primitive ontology out of the wave-function in a naive way (i.e.\! without an substantial amount of additional structure) without creating inconsistency.

Thus, for our interest here this is the crucial starting point: In order to gain predictions for events occurring in physical space out of an abstract mathematical description by a function $\Psi(x_1,...,x_N,t)$ living on configurational space\footnote{More appropriate: a function $\Psi(x_1,t_1,...,x_N,t_N)$ living on configuational space-time $\mathscr{M}^N$ within every precise relativistic many-particle-theory with a wave-function.} $\mathbb{R}^{3N}$, a mechanism must be specified which maps this level of mathematical description onto the level of physical predictions for events occurring in $\mathbb{R}^3$. 

\begin{center}
 \textbf{Measurement Problem}
\end{center}

In textbooks quantum mechanics such a mechanism is usually given by some \textit{primitive ontology of measurement events} which enters into the basic formulation of the theory (forcing -- in addition to the linear, deterministic unitary Schr\"odinger evolution -- as a second dynamical principle the wave-function to undergo a nonlinear, stochastic dynamics) together with \textit{Born's probability rule} as the probabilistic prediction for these primitive measurement events: the physical prediction given by the theory. It is well known and extensively discussed in the corresponding literature that this ontology of measurement events is far too vague as to serve as a basic ingredient of the theory's formulation without creating inconsistencies: The concept breaks down as soon as we require the particles, the measurement device consists of, to be guided by the laws of quantum theory, too\footnote{There is a widespread belief that a dynamical mechanism already contained in standard quantum mechanics -- namely decoherence -- is sufficient to solve this puzzle. I do not doubt that decoherence is essential to explain why quantum interference effects do not show up for macroscopic objects. But this is not the problem here. The point is that -- and this is perhaps the right place to use Bohr's and Heisenberg's terminology -- decoherence processes do not yield ``transitions form the potential to the actual'': If there is no fundamental dynamics but the linear Schr\"odinger evolution, superpositions will evolve to superpositions, no matter if we ignore the (dynamically relevant) environment in the appropriate way (i.e.\! by tracing it out). If we admit that Schr\"odinger type dynamics are fundamental, linearity with its grievous consequences is inescapable: We cannot deduce a fundamental nonlinear stochastic dynamics from a fundamental linear deterministic dynamics by simply ignoring a part of the system (in the appropriate way). See also \cite{diss}.}. For further arguments and discussions on that issue I would like to refer the reader to the mentioned literature (especially Schr\"odinger \cite{Schroedinger} as the ingenious starting point and Bell \cite{Bell} as a big highlight of these discussions) and leave it here by citing a phrase (a phrase, it is worth to think about more than a few minutes...) from the first sides of a famous quantum mechanics textbook: 

\begin{quotation}
\textit{``...Thus quantum mechanics occupies a very unusual place among physical theories: it contains classical mechanics as a limiting case, yet at the same time it requires this limiting case for its own formulation...''} \cite{landaulif}.
\end{quotation}

With John Bell I see two considerable types of candidate theories to avoid these conceptual inconsistencies without departing (to much) from the predictions of quantum-mechanics: \paragraph*{}

\textit{i}) Either we stick to the desire to construct the primitive ontology exclusively out of the wave-function. Then we have to give a precise law for its evolution which accounts also for the nonlinear part (in addition to the Schr\"odinger-evolution) without resorting to vague concepts like ``measurement''. And once we have this, we have to give a law for physical events in space-time (the primitive ontology) into which only the wave-function enters as a free variable. \paragraph*{}

\textit{ii}) Or we start with some (reasonable) primitive ontology and find a law, which determines the time evolution of this variable by the wave-function in such a way, that the predictions of quantum theory can be deduced (preferably simple) from its dynamics.
\begin{center}
 \textbf{GRW}
\end{center}

Theories of candidate theory type \textit{i}) are the so called \textit{collapse theories} or \textit{dynamical reduction models}. The first attempt in this direction was presented by Ghirardi, Rimini and Weber \cite{GRW} (\textit{GRW}) in 1986 and such theories have been developed further until now. They incorporate the stochastic reduction process of the wave-function into the basic formulation of the theory in a precise way by supplementing the deterministic linear Schr\"odinger equation by some well defined stochastic nonlinear evolution process of the wave-function. 

For our purpose it is important to mention here, that the GRW-theory was designed first just to give a precise account of wave-function dynamics (including wave-function collapse) without the need of resorting to some vague ontology of measurements, but no attempt was made to give a new precise connection of the wave-function to physical events, i.e.\! to give a primitive ontology beyond Born's rule. But without such a precise connection between the function on configuration space and events occurring in physical space there remains some vagueness in the theory, such that e.g.\! relativistic properties of the theory remain unclear: I will present different primitive ontologies for the non relativistic GRW theory which are empirically equivalent, i.e.\! which yield Born's probability rule, but which have enormously different degrees of compatibility with relativistic space-time structure and thus, as we will see, have very different potential for a relativistic upgrade. 

The first proposal for a primitive ontology for the GRW theory was given by John Bell \cite{BellJumps}, which is called the \textit{flash-ontology} by some authors \cite{commstruc,rGRWf,unrompics, collandrel, pointproc}. By supplementing GRW with the flash ontology Bell set the stage for the first (to my knowledge and in my opinion) thoroughly relativistic quantum theory which reconciles quantum nonlocality with relativistic space-time without the need of intrinsic space-time structures beside the Lorentz-metric and which at the same time shares the full physical relevance of quantum theory\footnote{There are also other proposals which account for nonlocality without violating or supplementing relativistic space-time structure, but they are either rather toy-models (e.g.\! \cite{Dowker}) or have no potential for reasonable physical predictions \cite{oparr} and thus are just artificial models with conceptual but without physical relevance.}. The formulation of this theory on the basis of Bell's proposal was given by Roderich Tumulka in 2006 \cite{rGRWf} and I will illustrate and discuss this model in more detail in the appendix.

\begin{center}
 \textbf{Bohmian Mechanics}
\end{center}

Theories of candidate theory type \textit{ii}) are misleadingly often called ``hidden variable theories''. The most manifest and full-blown (in the non-relativistic case) theory of this type is the Bohmian theory (called \textit{Bohmian mechanics}, \textit{Bohm-deBroglie theory} or \textit{pilot-wave theory}) where the primitive ontology is given by positions of particles. The theory provides a law, the guiding equation, into which the wave-function enters to generate a vector-field on the particles configuration space which provides the particle dynamics. In the non-relativistic case everything is fine: The theory is clear, transparent and, if you like it, also beautiful and it yields the whole machinery of standard quantum-mechanics, the description of measurements, the operator formalism and so on. And it justifies the statistical interpretation, without conceptual ambiguity. 

It is remarkable, that the theory does not need a second wave-function-dynamics besides the Schr\"odinger-evolution. It turns out naturally as a pure fact of simple analysis \cite{quantequi} that as soon as the different branches of a superposition of the wave-function can no longer interfere (e.g.\! by decoherence), the branch which actually guides the configuration will stay the only dynamically relevant branch -- for all practical purposes forever. This branch is called the \textit{effective wave-function} and the process of emergence of an effective wave-function the \textit{effective collapse}, expressing the correspondence to the collapsed wave-function and wave-function collapse of orthodox quantum mechanics, respectively. But there is actually no need for a true wave-function-collapse (and apart from collapse, in a theory about particles which are there, which always have definite positions in space, it is clear right from the start that we do not have to count on ambiguity).        

But when we try to lift the theory with more than one particle to Minkowski space-time, things get a bit delicate: the nonlocality of the non-relativistic guiding law for the particles implies that the velocity of one particle at some instant of time depends in general on the positions of the other particles at that time; and ``distinct spatial positions at the same instant of time`` is not a Lorentz invariant statement. Thus, it is not possible to simply lift the law for the particle dynamics from configuration space $\mathbb{R}^{3N}$ to configuration space-time $\mathscr{M}^N$ without finding a mechanism which connects some space-time point (the position and proper time of the particle whose velocity we intend to calculate) with $N-1$ other space-time points (positions and proper times of the $N-1$ other particles). And currently it seems that the only way such a mechanism could be given  without destroying the predictive power of the theory, is to employ a distinguished structure of Minkowski space-time, namely a foliation of space-time into space-like leafs with absolute physical significance, i.e.\! different foliations yield different particle velocities and thus different physical theories. 

I will analyze such models in more detail in the appendix, but let me anticipate a few remarks: It might sound like a gross violation of relativity to make usage of such an intrinsic foliation of space-time, because it might suggest that a reintroduction of (generalized) absolute time and simultaneity was performed through the back door. But to interpret the foliation as absolute time is more a subjective decision: Introduction of simultaneity in space-time induces a foliation (into space-like simultaneity slices) but the inversion of that argument is not justified in any case. The space-like leafs in this theory, for example, have absolutely nothing to do with -- and can not be used for -- synchronization of clocks and the like \cite{MaudlinTrick}. The physical arena of the theory is still relativistic space-time, which is not touched, and the foliation is only employed to account for the nonlocal correlations of particles. Thus the foliation is phenomenologically only relevant for subtle physical processes which are based on quantum nonlocality (and it is irrelevant as soon as e.g.\! decoherence creates effective product wave-functions) while the whole phenomenology of special relativity survives. Also it turns out that the statistical predictions of quantum theory (i.e.\! Born's rule) are independent of the foliation and that the shape of the leafs cannot be determined by any physical experiment. 

It is also important to note, that the formal covariance of the theory need not to be touched and that the distinguished structure need not to be added as an extra element to the theory: The foliation can be given by a covariant law into which only quantities enter, that are inherent in the theory right from the start. For example, we will see in the appendix that distinguished foliations are already inherent in the covariant wave-function (actually in every theory with a wave-function) such that the dynamics of the particles can be completely determined by the wave-function. No extra structure is to be added to the theory then. 

\begin{center}
 \textbf{Remark}
\end{center}

I would like to mention here that I am convinced that the relevance of this investigation does not depend on the fact, whether the reader shares my strong confidence that the above mentioned two theory types are the only considerable ways to escape the measurement problem. This thesis is concerned with general analysis of wave-functions and state reduction within a relativistic theory. And for a main part there is no reference to the choice, whether orthodox quantum mechanics with a vague ontology of measurements is applied or, say, GRW with a precise ontology and a precise description of wave-function reduction -- although we will repeatedly encounter problems which cannot be solved if the former choice is made. In the appendix I give a brief description and analysis of relativistic GRW-type theories and relativistic Bohmian mechanics, which shall highlight the relativistic relevance of the choice of a primitive ontology, but at the same time shall give insights into the general issue of relativistic quantum theory (independent on the choice of a version).

\subsection{Relativistic Quantum Theory: Quantum Field Theory (QFT)}

One last introductory remark on why these questions and reasoning cannot be found in any QFT-textbook (apart from some few lines on local commutativity discussed in chapter 2.2), although QFT is said to be the relativistic version of quantum theory and enormously successful in making good predictions. The reason is that QFT is a relativistic theory only from a pragmatic point of view. The transition amplitudes, and thereby probability distributions predicted by the theory, are Lorentz invariant, but the underlying states are not. 

The non-Lorentz invariance of state evolution is due to the wave-function reduction-process, which is a little bit hidden in QFT: The square of the Lorentz invariant transition amplitude yields the probability for the collapse of the initial state, time evolved by the S-matrix, onto some given final state, caused by appropriate measurement (e.g.\! particle detectors in a particle accelerator). The apriori non Lorentz invariance of this process and possible ways to make it Lorentz invariant will be a central part of the reasoning within these lines.                 

For the unitary part of the time evolution, formalisms developed in the early years of QED (until the forties of the 19'th century) by Dirac \cite{Dirac}, Tomonaga \cite{Tomonaga}, Schwinger \cite{Schwinger} and others will prove appropriate for our purpose. We will resort to these formalisms in section \ref{functional}.

\newpage

\section{Causality}

The aim of this chapter is twofold:

First it shall contribute to clarify terms and underlying concepts associated with \textsl{''causality''} and \textsl{``locality''} in the physics literature. Such terms are often used in different contexts and with different meanings. In particular I will highlight Bell's very lucid approach.

Second it shall emphasize that we somehow have to take the wave-function and the consequences of physical description by a wave-function seriously, namely nonlocality. This can be seen as preparation for chapter \ref{chap3}, where implications of Minkowski space-time structure for wave-function dynamics and resulting nonlocality are analyzed. \paragraph*{}

\textsl{Remark:}
Local commutativity (see section \ref{loccom}) is sometimes called \textsl{micro-causality}. In connection with dispersion relations, the requirement that an effect must not precede its cause is sometimes called \textsl{macro-causality} (the reaction of the system to some perturbation must not precede the perturbation).

\subsection{Einsteins Principle of Contiguity and Bells Concept of Local Causality \label{Einstein Bell}}

\begin{center}
 \textbf{Einstein}
\end{center}

In his correspondence with Max Born, Einstein wrote: 

\begin{quotation}
\textsl{If one asks what, irrespective of quantum mechanics, is characteristic of the world of ideas of physics, one is first of all struck by the following: the concepts of physics relate to a real outside world, that is, ideas are established relating to things such as bodies, fields, etc., which claim ``real existence`` that is independent of the perceiving subject -- ideas which, on the other hand, have been brought into as secure a relationship as possible with the sense data. It is further characteristic of these physical objects that they are thought of as arranged in a space-time continuum. An essential aspect of this arrangement of things in physics is that they lay claim, at a certain time, to an existence independent of one another, provided these objects ``are situated in different parts of space``.  Unless one makes this kind of assumption about the independence of the existence (the ''being-thus'') of objects which are far apart from one another in space -- which stems in the first place from everyday thinking -- physical thinking in the familiar sense would not be possible. It is also hard to see any way of formulating and testing the laws of physics unless one makes a clear distinction of this kind... The following idea characterizes the relative independence of objects (A and B) far apart in space: external influence on A has no direct influence on B; this is known as the ``principle of contiguity''...  If this axiom were to be completely abolished, the idea of the existence of (quasi-) enclosed systems, and thereby the postulation of laws which can be empirically checked in the accepted sense, would become impossible.} (\cite{Born} pp. 170-71)
\end{quotation}

Einstein expresses here his strong conviction that a cause and its direct effect must always concur spatially. The pictorial translation ``principle of contiguity'' is due to Born, the literally translation of Einsteins expression would be "principle of local action``. \paragraph*{}

Let us analyze the implications for some given physical theory due to one assertion within this quote, which leaves room for interpretation and which is well suited to clarify concepts and to make some disambiguation: ``\textit{external influence\footnote{\label{extinf}External influence can be given by measurements performed on A or by the imposition of ``external fields`` on A, i.e.\! by coupling some degree of freedom of A unitarily to some (variable) parameter in the equations of motion. This parameter in turn is not considered to be determined by some equation but by the choice of the theoretician (or experimenter) and variation of this parameter produces variations of predictions for A (see also the discussion on ''controllability'' in chapter \ref{sectnosig}). Although in a complete physical description there might be no external influence at all, we can justify the consideration of such ``free parameters`` by the requirement that there are physical processes which occur independent of one another but then produce correlations, such that we can arbitrarily choose one such process and then calculate its effects on the other one. This is deeply related to the requirement that \textit{''nature should not be conspiratorial``} in some sense, which is discussed for the case of the EPRB-experiment at the end of this section in more detail.} on A has no direct influence on B}''. This is true and false at the same time in quantum mechanics:\paragraph*{}

\textit{A)} What does it mean to condition on external influence on A in order to calculate predictions for B? Does it mean just to condition on the presence of an external field, for example, or does it mean to condition on the actual influence of the field on A? This makes a big difference in quantum mechanics. If we condition on the fact that one of the two particles of a singlet-pair passed a Stern-Gerlach magnet (SGM) or if we condition on the actual process which occurred to that particle -- e.g.\! the particle passed the SGM and was deflected upwards (e.g.\! was found in the upper region afterwards) -- will yield completely different probability distributions for possible measurements on the other singlet-particle. \paragraph*{}

\textit{B)} What does direct influence on B mean? Does it mean that an experimenter is able to infer information about whether such influence occurred or not by performing experiments on B or does it mean that what might happen with B is somehow directly determined by the influence? This makes also a big difference in quantum theory. If one of the two singlet-particles passes a SGM and hits a detector afterwards in the upper region, the other particle is immediately determined to be deflected downwards by a second SGM (with same orientation as the first one). But according to local commutativity (which will be analyzed in the next section) of the corresponding two spin operators, an experimenter at the second SGM has no possibility to infer information about the question, whether the first particle actually passed the SGM or not; even from the relative frequencies of an ensemble of such processes no such information can be inferred. \paragraph*{}

It is already clear from the above quote (\textsl{``the independence of the existence (the ''being-thus'') of objects``}) that Einstein left no doubt that his idea of independence of spatially separated systems is about what actually happens (''what there is'') in the respective regions. Here are two (of many) more examples: 

\begin{quotation}
\textsl{...what is present in $\mathcal{B}$ should somehow have an existence independent of what is present in $\mathcal{A}$.\footnote{Here the objects A and B are replaced by corresponding spatial regions $\mathcal{A}$ and $\mathcal{B}$.}} \cite{Howard} 
\end{quotation}

or elsewhere

\begin{quotation}
\textsl{The real situation (state) of system $S_2$ is independent of what is done with system $S_1$, which is spatially separated from the former.}\footnote{This Einstein quote serves Bell as a motivation to write down separability condition \ref{sep} in his famous paper ``On the Einstein-Podolsky-rosen paradox'' \cite{Bell}. } \cite{Howard}
\end{quotation}

 The principle which Einstein hold sacred was not something like: \textit{i) ``The pure fact of external influence on A (regardless of its actual effect on A) has no observable influence on B``}, which indeed would be consistent with quantum theory, but rather: \textit{ii) ''What actually happens with A does not change anything actual for B''} which is not in accord with the description given by quantum mechanics. 

Einstein always insisted enthusiastically (see \cite{Howard}), based on his strong confidence that assertion \textit{ii)} should not be touched, that the description given by quantum mechanics cannot be the whole story. The most famous account (although not the one with the most clarity) of this argument is the celebrated EPR-paper \cite{EPR}, where assertion \textit{ii)} is also tacitly assumed:

\begin{quotation}
\textsl{[...] since at the time of measurement the two systems can no longer interact, no real change can take place in the second system in consequence of anything that may be done to the first system. This is, of course, merely a statement of what is meant by the absence of an interaction between the two systems.} \cite{EPR}
\end{quotation}

\vspace{0.3cm}

\begin{center}
\textbf{Bell}
\end{center}

John Bell introduced the notion \textit{``principle of local causality''} (also simply called \textit{locality}) to grasp Einsteins idea. He analyzed this principle in detail over decades and formalized it attentively in order to have a formal criterion whether a given theory is in accord with it or not (with quantum theory in the back of his mind). In his 1990-article \textit{La nouvelle cuisine} \cite{Bell}, Bell presents one of his most transparent formalizations of this principle and I would like to sketch briefly the central line of thought of this nice article \footnote{See also \cite{TravisBJ} and \cite{TravisLC} for a very nice and clear illustration of Bell's argumentation.}.

Bell establishes the principle of local causality by first phrasing it in an intuitive way as a principle of contiguity:

\begin{quotation}
\textsl{The direct causes (and effects) of events are near by, and even the indirect causes (and effects) are no further away than permitted by the velocity of light.} \cite{Bell}
\end{quotation}

\begin{figure}[htbp]
  \centering
    \includegraphics[scale=0.3]{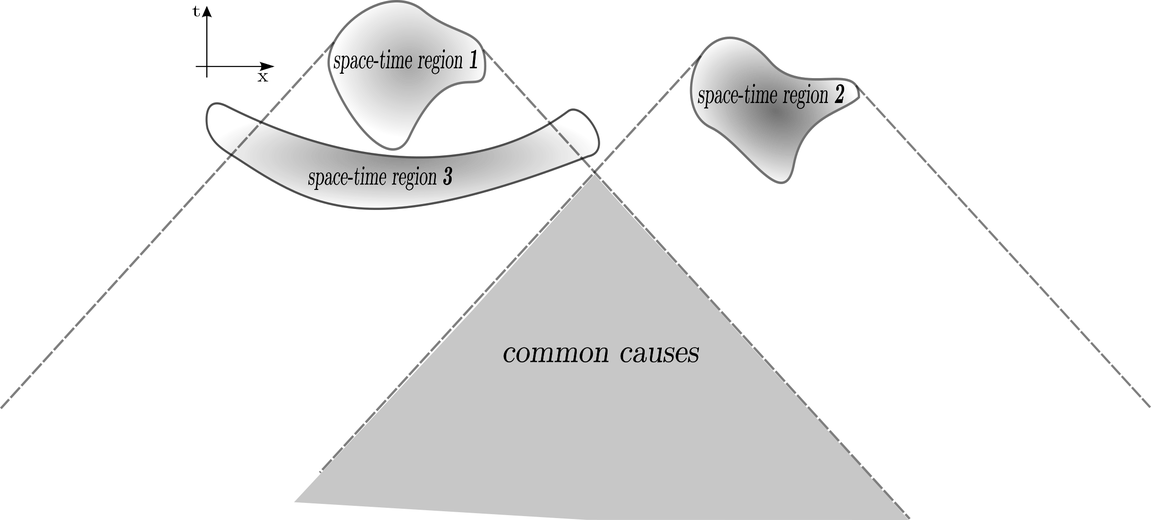}  
  \caption{\textbf{Bell's concept of local causality:} Events in space-time region $1$ are independent of events in some space-like separated space-time region $2$ if conditioned on a sufficient specification of events in space-time region $3$. The latter ''shields of`` completely space-time region $1$ from the intersection of the past light-cones of regions $1$ and $2$, where common causes for events in $1$ and $2$ may lie. Such common causes might be responsible for correlations between events in $1$ and $2$, but in a locally causal theory these correlations vanish by conditioning on a sufficient specification of events in region $3$.}
\label{BellLoc}
\end{figure}

I.e.\! causes of an event are constrained to lie in its past light-cone and effects of an event to lie in its future light-cone. Now comes an ingenious step of rephrasing the content of this assertion not as a criterion for a locally causal ``reality``, but merely as a criterion, which an arbitrary theory has to fulfill in order to be locally causal:

\begin{quotation}
\textsl{A theory is said to be locally causal if the probabilities attached to values of local beables in a space-time region 1 are unaltered by specification of values of local beables in a space-like separated region 2, when what happens in the backward light-cone of 1 is already sufficiently specified, for example by a full specification of local beables in a space-time region 3.} \cite{Bell} (see figure \ref{BellLoc})
\end{quotation}

Some remarks: 

First note that this is a \textit{criterion for some candidate theory to be locally causal} which makes no more reference to rather intuitive concepts like ``cause'' and ''effect''. Every expression within this criterion has a precise counterpart in an arbitrary reasonable physical theory (in a deterministic theory all probabilities are zero or one) where, as described above, the local beables are the elements of the theory which correspond to events occurring in physical space which are predicted and described by the theory, e.g.\! pointers in the laboratory.  

Space-time region 3 -- which ``shields of`` space-time region 1 from the intersection of its past light-cone with the past light-cone of 2 -- is necessary, because events in space-like separated regions can be correlated by time-like causal chains originating from events in the intersection of their backward light-cones: If I know that I have just one glove in my pocket while I left the other one at home -- but I do not know which one -- the probability that the glove at home ($GAH$) is the right one ($R$) will be 
\begin{equation}
 \mathbb{P}(GAH=R)=\frac{1}{2} \mbox{ .}
\end{equation}
On the other hand, if I look into my pocket and find the glove in my pocket ($GIMP$) to be the right one, this probability undergoes an immediate change:
\begin{equation}
 \mathbb{P}(GAH=R \mid GIMP=R)=0 \mbox{ .}
\end{equation}
This is of course not at all a violation of local causality if within my theory the ''state`` (handedness) of the glove at home is not determined by the look into my pocket but only by its past light-cone. Thus if we have a sufficient specification $\lambda$ of local beables in the past light cone of the glove left at home at the time we are interested in its handedness (e.g.\! the shape of the glove at home before that time) the state of the glove in my pocket becomes redundant:
\begin{equation}
 \mathbb{P}(GAH=R \mid GIMP=R,\lambda)=\mathbb{P}(GAH=R \mid \lambda)=0 \mbox{ .}
\end{equation}

Thus -- returning to intuitive language -- Bell's locality-condition is essentially this: \textsl{Correlations between spatially separated events can be explained by common causes} lying in the intersection of their past-light-cones\footnote{This is a little bit sloppy: There is actually a subtle difference between this statement and the concept we developed so far: This statement (which corresponds to Bell's earlier formulations of his concept) means that space-like separated events are independent if it is conditioned on events lying in their common past. In contrast, in the concept we developed so far the independence is due to conditioning on events in space-time region 3. There are exceptional physical situation in which this makes a difference, and it turns out that our dynamical concept of local causality (the one from La Nouvelle Cuisine) is a bit weaker than the assertion that correlations can be explained by common causes. But this subtle distinction is not relevant for all conclusions we shall derive here.}.

\begin{figure}[htbp]
\centering
    \includegraphics[scale=0.16]{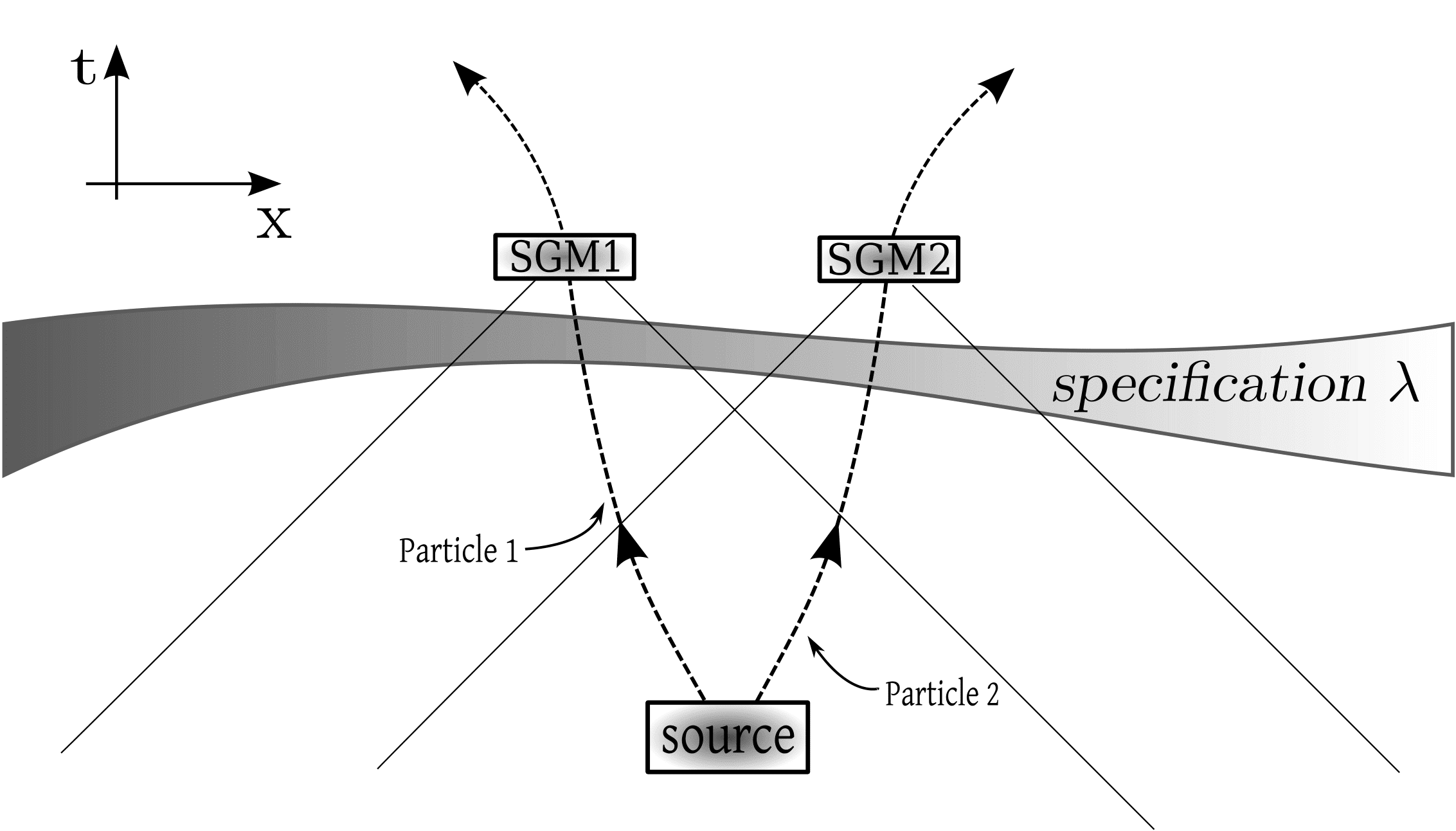}  
  \caption{\textbf{Analysis of the EPRB-experiment from the viewpoint of a locally causal theory:} A pair of singlet particles is emitted from some source, the particles are spatially separated and then some particular component of the one particle spins is measured at the SGMs, respectively. In a locally causal theory a specification $\lambda$ of events in the indicated space-time region can be given, such that the outcome of one of the measurements is independent both, of the outcome as well as of the actual angle of the SGM on the other wing of the experiment.}
  \label{BellEPR}
\end{figure}

Now we are in a position to analyze an actual experiment from the point of view of an arbitrary theory which conforms with the principle of local causality as formalized above. In anticipation of Bell's theorem we shall analyze Bohm's version of the EPR-experiment (EPRB) (figure \ref{BellEPR}): Consider a pair of spin-$\frac{1}{2}$ particles -- Particle(1) and Particle(2) -- prepared in the singlet state
\begin{equation}
\mid \Psi_{-}\rangle:=\frac{1}{\sqrt{2}}(\mid \uparrow \downarrow \rangle - \mid \downarrow \uparrow \rangle) \mbox{ ,}
\end{equation}
which some time after the preparation pass Stern-Gerlach-magnets SGM(1) and SGM(2), respectively, in spatially separated regions. Suppose SGM(1) can be rotated by angles $\vartheta_{(1)}$ and SGM(2) by angles $\varphi_{(2)}$ from some given parallel orientation. Let $\lambda$ be a sufficient specification (in the above sense) of the local beables of the theory in the space-time region indicated in figure \ref{BellEPR} and let us further denote by $\sigma^{(1)},\sigma^{(2)} \in \pm 1$ the respective results of this process corresponding to spin up (+1) or spin down (-1) of the spin-component corresponding to the actual angles, which can be read off e.g.\! from spots on a photographic plate behind the SGMs.

Now, according to the above criterion for every locally causal theory -- given $\lambda$ -- the result on one side of the experimental setup is statistically independent of whatever might happen on the other side, i.e with the input variables defined above we can state that in a locally causal theory 
\begin{equation}\label{loccaus1}
 \mathbb{P}^{\Psi_-}(\sigma^{(1)} \mid \vartheta_{(1)}, \varphi_{(2)}, \sigma^{(2)}, \lambda)=\mathbb{P}^{\Psi_-}(\sigma^{(1)} \mid \vartheta_{(1)}, \lambda)
\end{equation}
\begin{center}
 as well as
\end{center}
\begin{equation}\label{loccaus2}
 \mathbb{P}^{\Psi_-}(\sigma^{(2)} \mid \varphi_{(2)}, \vartheta_{(1)}, \sigma^{(1)}, \lambda)=\mathbb{P}^{\Psi_-}(\sigma^{(2)} \mid \varphi_{(2)}, \lambda)
\end{equation}
must hold\footnote{As two different kinds of quantities enter into conditions \eqref{loccaus1} and \eqref{loccaus2} (namely the angles as ''free'' parameters and the outcomes), they can actually be seen as a conjunction of two different conditions (see Jarrett \cite{Jarrett}), which where called \textit{''parameter-independence''} and \textit{''outcome-independence''} by Abner Shimony \cite{Shimony}. For example the violation of Bell's inequality is due to a violation of parameter-independence in Bohmian mechanics and due to a violation of outcome-independence in GRW; and there is an ongoing debate about the relevance of this decomposition \cite{Maudlin, Ghirardiparam, TravisBJ}. But without question a violation of each condition is a violation of Bell's local causality, so we shall not enter into this discussion here.}. From this we can calculate the famous \textit{separability condition} (with the basic rules of probability calculus), which is thus a mathematical result of any locally causal theory describing the EPRB-experiment:
\begin{equation}
\begin{split} \label{sep}
 \mathbb{P}^{\Psi_-}(\sigma^{(1)}, \sigma^{(2)} \mid \vartheta_{(1)}, \varphi_{(2)}, \lambda) &= \mathbb{P}^{\Psi_-}(\sigma^{(1)} \mid \sigma^{(2)}, \vartheta_{(1)}, \varphi_{(2)}, \lambda) \times \mathbb{P}^{\Psi_-}(\sigma^{(2)} \mid \vartheta_{(1)}, \varphi_{(2)}, \lambda) \\
& = \mathbb{P}^{\Psi_-}(\sigma^{(1)} \mid \vartheta_{(1)}, \lambda) \times \mathbb{P}^{\Psi_-}(\sigma^{(2)} \mid \varphi_{(2)}, \lambda) \mbox{ .}
\end{split}
\end{equation}

Separability condition \eqref{sep} is the starting point for the derivation of Bell's inequality. Many misunderstandings about this condition -- especially concerning the role of the parameter $\lambda$ -- originate from the fact that Bell's own reasoning is not well comprehended. Many authors claim that this condition assumes ''the existence of hidden variables``, ``determinism'', ''realism'', ''counter-factual definiteness,``... which can be refuted one by one by simply following attentively Bell's own argumentation (see e.g.\! \cite{Maudlin, TravisBJ, TravisLC}). 

Misunderstandings might partly have historical reasons: EPR inferred from quantum formalism and tacitly assumed locality that physical description by a wave-function must be completed by additional deterministic variables, restoring locality. The consideration of such variables (together with the opposed intrinsic nonlocality of Bohm's theory) is indeed the motivation of Bell's formalization of the principle of local causality: Can the predictions of quantum mechanics be confirmed by a theory, which contains EPR-like additional local beables $\lambda$ restoring local causality in the sense of its defining principle? And Bell's inequality then shows that this is not the case. But in the formulation of the local causality condition as well as in the separability condition and Bell's inequality deduced from it, there is no reference at all to hidden variables\footnote{David Mermin wrote \textit{``[t]o those for whom nonlocality is anathema, Bell's theorem finally spells the death of the hidden-variables program''} \cite{Mermin}. This quote obviously shows that Mermin did not appreciate the central line of thought of EPR and Bell: \textit{``My own first paper on this subject ... starts with a summary of the EPR argument from locality to deterministic hidden variables. But the commentators have almost universally reported that it begins with deterministic hidden variables.``} \cite{Bell}}. These conditions must hold for any theory, which is able to explain correlations of spatially separated events by common causes (lying in the intersection of their backward-light-cones). And if these expressions turn out to be wrong, no such explanation is possible. $\lambda$ stands for all the local objects provided by the investigated candidate theory corresponding to events in space-time region 3, which matter for calculating the above probabilities. And \textsl{``in ordinary quantum mechanics there just \textbf{\textit{is}} nothing but the wave-function for calculating probabilities. There is then no question of making the result on one side redundant on the other by more fully specifying events in some space-time region 3. We have a violation of local causality''} \cite{Bell} (p.241). 

The remaining step then is to deduce an empirically verifiable or falsifiable prediction from condition \eqref{sep}. This prediction is the celebrated Bell inequality (one of its versions). 

\begin{center}
\textbf{No Conspiracies}
\end{center}

It should be mentioned that besides local causality (separability) a second condition enters into the derivation of the inequality: We have to require that there is a sort of free choice of angles at the SGMs, namely free with respect to $\lambda$.  For example two (deterministic) random number-generators choose one of three possible orientations of the SGMs, respectively, shortly before the particles pass, and the probability distribution of the angles produced by some random number-generator should not be correlated with the distribution of local beables $\lambda$ in space-time region 3 which are relevant for the dynamics of the singlet particles
\begin{equation}\label{nocons}
 \mathbb{P}(\lambda \mid \vartheta_{(1)}, \varphi_{(2)}) = \mathbb{P}(\lambda) \mbox{ .}
\end{equation}
This is indeed a very reasonable assumption, for it simply expresses the requirement that the physics of the macroscopic device (with random number-generator or experimenter included) shall be independent of the physics of the singlet-particles before these two systems got into contact (see also footnote \ref{extinf} on ``external influence'' and the discussion on ``controllability'' in chapter \ref{sectnosig}). Although $\lambda$ may specify local beables in the backward light-cone of the event of adjustment of an angle, we would expect that theoretically possible correlations between local beables influencing the dynamics of the random number generator (choosing the angle) and local beables $\lambda$ influencing the dynamics of the particles, will be for all practical purposes completely suppressed, e.g.\! by interactions of these two systems with the environment. 

This requirement, that parameters of some device can be chosen freely before interaction with some system, is often labeled with the reasonable phrase that \textit{''nature should not be conspiratorial (or super-deterministic)``}. And besides the above thermodynamic justification, there are two more arguments in support of this requirement:
 
First, it is worth noting that the experiments not only show an arbitrary violation of Bell's inequality but coincide impressively good with the predictions of quantum theory which violates local causality \eqref{loccaus1} and \eqref{loccaus2} but conforms with ``no conspiracies''.

And moreover, only the assumption that conspiratorial correlations in the sense of a violation of \eqref{nocons} do not occur in nature, lays the groundwork for reasonable experimental physics:

\begin{quotation}
\textsl{In any scientific experiment in which two or more variables are supposed to be randomly selected, one can always conjecture that some factor in the overlap of the backward light cones has controlled the presumably random choices. But, we maintain, skepticism of this sort will essentially dismiss all results of scientific experimentation. Unless we proceed under the assumption that hidden conspiracies of this sort do not occur, we have abandoned in advance the whole enterprise of discovering the laws of
nature by experimentation.} (Shimony, Clauser and Horn in ``An Exchange on Local Beables'' \cite{exchange})
\end{quotation}

\begin{center}
\textbf{The Inequality}
\end{center}

The derivation of one of the versions of the inequality is then standard and can be found in many articles and quantum mechanics textbooks. The straight forward line of argumentation emerges from the question: Suppose we explain the perfect (anti-)correlations of the measured two particle spins in the EPRB-experiment for the case of coinciding SGM-angles by common causes; is this pattern of explanation able to describe the spin-correlations for different settings of the SGM-angles? The ``predicted answer`` by quantum theory is \textit{``NO``}, and the experiments support the predictions of quantum theory\footnote{It is also possible to infer Bell's inequality without directly referring to Bell's formalization of local causality (and thereby without explicitly introducing some specification $\lambda$) but by just assuming the (pre-)existence of random-variables for the outcomes of the EPRB-experiment for various orientations of the SGMs. Then the violation of locality through a violation of the inequality is very obvious right from the start. See e.g.\! \cite{Detlef, Maudlin} and for a very nice and transparent illustration \cite{bricmont}.} -- given condition \eqref{nocons} is valid.

\begin{center}
\textbf{Summary}
\end{center}

Now let us summarize the facts: Bell gave a criterion to judge whether some arbitrarily given candidate theory is locally causal. From this criterion the separability condition \eqref{sep} follows immediately for the EPRB-setup. This condition together with condition \eqref{nocons} that ``nature is not conspiratorial`` yields some empirically testable prediction (Bell's inequality). Standard quantum mechanics, as well as Bohmian mechanics and GRW, violates the local causality condition and Bell's inequality. Experiments also show a violation of Bell's inequality \cite{Aspect} and the outcomes coincide very well with the predictions of quantum theory. Because Bell's locality criterion is a condition to check some given theory to be locally causal, the empirical violation of Bell's inequality means that no locally causal theory can make the right predictions for outcomes of a given experiment which are actually observed in the laboratories (if we take a non conspiratorial nature for granted). Thus nature can not be described satisfactory by a locally causal framework, i.e.\! \textit{nature is non-local}. And the impressive coincidence of experimental data with the predictions of quantum theory strongly suggest, that we somehow have to take the element of quantum theory serious, in which quantum nonlocality is encoded -- namely the wave-function on configuration space.

\subsection{\label{loccom}Local Commutativity and the Term \textit{Causality} in Relativistic Quantum Theory}

In contemporary physics the term ``causality'' caught on to describe a property of some given theory, which is -- as I will argue -- actually not very well covered by this term. In quantum theory the ``requirement of causality" (in QFT even more misleadingly also called "locality") is usually expressed by the mathematical requirement that self-adjoint operators, corresponding to measurements at space-like separation, should commute\footnote{In QFT a famous consequence of this requirement (for bilinear forms in the field-operators) is the celebrated 'spin statistics theorem'.}. And a straight forward physical consequence is that quantum nonlocality provides no strategies for an experimenter to inform another experimenter, space-like separated from the former, about any local actions she performed on a quantum system, i.e.\! it is not possible to ``send superluminal signals'' by ``external influence''. The ``pure fact of measurement`` in a space-time region $\mathcal{A}$ does not change the local probabilities of a measurement which is performed in space-like separated region $\mathcal{B}$. 

Now, it seems that vague and anthropocentric expressions like "measurement", ''external influence'' and "impossibility of sending superluminal signals/information" make the physical content of this requirement. This vagueness together with the claim of providing the fundamental causal structure of a theory might instigate us to investigate the implications of local commutativity more closely. Bell had similar concerns:
\begin{quotation}
 \textsl{Could it be that causal structure is something like a 'thermodynamic' approximation, where the notions 'measurement' and 'external field' become legitimate approximations? Maybe that is part of the story, but I do not think it can be all. Local commutativity does not for me have a thermodynamic air about it. It is a challenge now to couple it with sharp internal concepts, rather than vague external ones.} \cite{Bell} 
\end{quotation}

I will try to make a step in this direction in section \ref{meaningloccom}. I will use the concept of primitive ontology to give a surprisingly simple argument, why we should take local commutativity serious, in order to maintain a kind of ontological consistency within a relativistic framework. Since the starting point of the argument is commutativity of operators, and operator-formalism is measurement-formalism in quantum theory, I will have to start with considering quantum-measurement and to identify the primitive ontology with positions of pointers (for example). But nevertheless the conclusion will be transparent and the argument is less anthropocentric than the standard argument (for example it makes no use of the concept of controllability which is crucial for the standard argument). And we have available theories, like Bohmian mechanics or GRW, in which the operator-measurement-formalism of quantum theory can be derived and understood from an underlying precise physical description. But the task remains, to bring this argument down to the fundamental level of the theory, where we do not need pointer positions to identify the primitive ontology; and then to examine if it indeed provides contributions to constitute the causal structure of the theory. 

But first let me illustrate the standard line of argument which bases on the presumed impossibility of superluminal signals and then discuss briefly relativistic implications of the latter in section \ref{bitcausation}.

\subsubsection{\label{sectnosig}Local Commutativity, Signals \& Controllability}

\vspace{0.5cm}

\begin{center}
\textbf{Commutativity}
\end{center}

The mathematical fact that two self-adjoint operators (corresponding to quantum measurements) acting on some given Hilbertspace $\mathcal{H}$ commute, implies that the projection valued measures (PVM's), related to the operators by the spectral theorem, yield joint probability distributions for the outcomes of the two measurements. This is the key to understand why local commutativity excludes strategies to utilize the nonlocality of quantum measurement for superluminal signaling: 

Consider a quantum-mechanical system with wave-function $\Psi \in \mathcal{H}$ and two space-like separated space-time regions $\mathcal{A}$ and $\mathcal{B}$. Now suppose a quantum measurement is performed in each of this regions, where in $\mathcal{A}$ the physical quantity $\mathscr{A}$ is measured and the statistics of this measurement is described by the self-adjoint operator (acting on $\mathcal{H}$) 
\begin{equation}\label{opA}
\hat{A}=\sum_{\alpha} \sum_{i=1}^{\ell_{\alpha}} \alpha \mid \varphi^{(i)}_{\alpha} \rangle \langle \varphi^{(i)}_{\alpha} \mid   
\end{equation}
with eigenvalues $\alpha$ and eigenvectors $\mid \varphi^{(i)}_{\alpha} \rangle$ (the index $\ell_{\alpha}$ indicates the degree of degeneracy of eigenvalue $\alpha$) and analogously in $\mathcal{B}$ the quantity $\mathscr{B}$ is measured, described by the operator
\begin{equation}\label{opB}
 \hat{B}=\sum_{\beta} \sum_{j=1}^{\ell_{\beta}}\beta \mid \chi^{(j)}_{\beta} \rangle \langle \chi^{(j)}_{\beta} \mid \mbox{ .}
\end{equation}
Let us suppose now that $[\hat{A},\hat{B}]=0$.

Denote by $\mathcal{P}_{\alpha}:=\sum_{i=1}^{\ell_{\alpha}} \mid \varphi^{(i)}_{\alpha} \rangle \langle \varphi^{(i)}_{\alpha} \mid$ and by $\mathcal{P}_{\beta}:= \sum_{j=1}^{\ell_{\beta}} \mid \chi^{(j)}_{\beta} \rangle \langle \chi^{(j)}_{\beta} \mid$ the projection operators onto the eigenspace of $\hat{A}$ belonging to eigenvalue $\alpha$ and onto the eigenspace of $\hat{B}$ belonging to $\beta$, respectively. Then we can write the probability of some outcome, say $\alpha'$, of the $\mathscr{A}$-measurement as

\begin{equation}
\begin{gathered}\label{jointprob}
 \mathbb{P}^{\Psi}(\mathscr{A}=\alpha') = \langle \Psi \mid \mathcal{P}_{\alpha'} \mid \Psi \rangle = \langle \Psi \mid \mathds{1}_{\mathcal{H}} \mathcal{P}_{\alpha'} \mid \Psi \rangle \overset{\sum_{\beta} \mathcal{P}_{\beta} = \mathds{1}_{\mathcal{H}}}{=} \langle \Psi \mid \Big{[} \sum_{\beta} \mathcal{P}_{\beta} \Big{]} \mathcal{P}_{\alpha'} \mid \Psi \rangle \\
 = \sum_{\beta} \langle \Psi \mid \mathcal{P}_{\beta} \mathcal{P}_{\alpha'} \mid \Psi \rangle \overset{\mathcal{P}_{\beta}^2=\mathcal{P}_{\beta}}{=}  \sum_{\beta} \langle \Psi \mid \mathcal{P}_{\beta} \mathcal{P}_{\beta} \mathcal{P}_{\alpha'} \mid \Psi \rangle \overset{[\mathcal{P}_{\beta} , \mathcal{P}_{\alpha}]=0}{=} \sum_{\beta} \langle \Psi \mid \mathcal{P}_{\beta} \mathcal{P}_{\alpha'} \mathcal{P}_{\beta} \mid \Psi \rangle \\
 = \sum_{\beta} \bigg{[}\frac{\langle \Psi \mid \mathcal{P}_{\beta} \mathcal{P}_{\alpha'} \mathcal{P}_{\beta} \mid \Psi \rangle}{\langle \Psi \mid \mathcal{P}_{\beta} \mid \Psi \rangle} \bigg{]} \times \langle \Psi \mid \mathcal{P}_{\beta} \mid \Psi \rangle =\sum_{\beta}  \mathbb{P}^{\Psi}(\mathscr{A}=\alpha' \mid \mathscr{B}=\beta) \times \mathbb{P}^{\Psi}(\mathscr{B}=\beta) \\
 = \sum_{\beta} \mathbb{P}^{\Psi}(\{\mathscr{A}=\alpha'\} \cap \{\mathscr{B}=\beta\}) 
\end{gathered}
\end{equation}
where we have used standard properties of the spectral decompositions (completeness, projector property), the linearity of the scalar-product and the commutativity of projectors $\mathcal{P}_{\alpha}$ and $\mathcal{P}_{\beta}$, which follows from $[\hat{A},\hat{B}]=0$ by the spectral theorem. Thus we have a joint probability distribution for the outcomes of the two measurements. 

On the other hand, in order to calculate the conditional probability of outcome $\alpha'$ of the $\mathscr{A}$-measurement for the case that the $\mathscr{B}$-measurement was already performed, without specification of the actual outcome, we have to average over all possible results of the latter experiment:   

\begin{equation}
\mathbb{P}^{\Psi}(\mathscr{A}=\alpha' \mid \mathscr{B} \hspace{0.2cm} was \hspace{0.2cm} measured \hspace{0.2cm} in \hspace{0.2cm} \mathcal{B})=\sum_{\beta}  \mathbb{P}^{\Psi}(\mathscr{A}=\alpha' \mid \mathscr{B}=\beta) \times \mathbb{P}^{\Psi}(\mathscr{B}=\beta)
\end{equation}
which is obviously identical to the probability of the unconditioned measurement \eqref{jointprob}. 
\begin{equation}\label{nosupsig}
 \Longrightarrow \hspace{0.5cm} \mathbb{P}^{\Psi}(\mathscr{A}=\alpha' \mid \mathscr{B} \hspace{0.2cm} was \hspace{0.2cm} measured \hspace{0.2cm} in \hspace{0.2cm} \mathcal{B}) = \mathbb{P}^{\Psi}(\mathscr{A}=\alpha')
\end{equation}

It follows immediately from \eqref{nosupsig} that also variation of parameters of the device (e.g.\! a rotation of some SGM) at $\mathcal{B}$ does not change the local probabilities at $\mathcal{A}$ -- given that we do not care about the actual influence of the $\mathscr{B}$-device, of course. This is due to the fact that two different configurations of the $\mathscr{B}$-device yield both the same conditioned probability distribution at $\mathcal{A}$ which is according to \eqref{nosupsig} equal to the unconditioned distribution there.

All this does by no means mean that measurement-like events in the side-cone of a measurement event do not change the probabilities of outcomes of the latter -- given a sufficient specification of events in its past light cone -- i.e.\! it does not mean that Bell's local causality is recovered. This would be the case if and only if 
\begin{equation}
 \mathbb{P}^{\Psi}(\mathscr{A}=\alpha')=\mathbb{P}^{\Psi}(\mathscr{A}=\alpha' \mid \mathscr{B}=\beta) \hspace{0.5cm} \forall \beta 
\end{equation}
which is violated in quantum theory in general.\paragraph*{}

For example, in the EPRB-setup the two spin-measurements of the singlet particles are in the side-light-cone of one another if the setup is arranged in the true sense of the gedankenexperiment, and the two spin-operators corresponding to the two wings of the experiment, respectively, should indeed commute. But the fact that one of the measurements yields say ``z-spin-up'' forces the probability of the measurement-outcome: ``z-spin-up'' of the second particle to jump from $\frac{1}{2}$ to zero 

\begin{equation}
 \mathbb{P}^{\Psi_-} (\sigma^{(2)}_z = +1) = \frac{1}{2} \hspace{0.2cm} \boldsymbol{\neq} \hspace{0.2cm} 0 = \mathbb{P}^{\Psi_-} (\sigma^{(2)}_z = +1 \mid \sigma^{(1)}_z = +1) \mbox{ .}
\end{equation}

And Bell's theorem rules out all patterns of explanation of this perfect (anti-)correlations which draw upon events which lie in the intersection of the past light cones of the two measurements. 

But still the ``pure fact of measurement'' (i.e.\! if we do not condition on one specific outcome but just on the fact that the first measurement occurred) does not change these local probabilities (as it should be already clear from \eqref{nosupsig}):

\begin{equation}
\begin{gathered}\label{nosupsigeprb} 
 \mathbb{P}^{\Psi_-} (\sigma^{(2)}_z = +1 \mid \mbox{ \textsl{measurement of} } \sigma^{(1)}_z) = \sum_{\sigma^{(1)}_z = \pm 1} \mathbb{P}^{\Psi_-} (\sigma^{(2)}_z = +1 \mid \sigma^{(1)}_z) \times \mathbb{P}^{\Psi_-}(\sigma^{(1)}_z) \\ 
 =\sum_{\sigma^{(1)}_z = \pm 1} \mathbb{P}^{\Psi_-} (\{\sigma^{(2)}_z = +1\} \cap  \{\sigma^{(1)}_z\}) \quad =\quad \mid \langle \Psi_- \mid \uparrow \uparrow \rangle \mid^2 + \mid \langle \Psi_- \mid \downarrow \uparrow \rangle \mid^2 \\
 = \langle \Psi_- \mid \Big{[} (\mid \uparrow \rangle \langle \uparrow \mid)^{(1)} \otimes (\mid \uparrow \rangle \langle \uparrow \mid)^{(2)} \Big{]} \mid \Psi_- \rangle + \langle \Psi_- \mid \Big{[} (\mid \downarrow \rangle \langle \downarrow \mid)^{(1)} \otimes (\mid \uparrow \rangle \langle \uparrow \mid)^{(2)}\Big{]} \mid \Psi_- \rangle \\
=\langle \Psi_- \mid \Big{[} \Big{(} (\mid \uparrow \rangle \langle \uparrow \mid)^{(1)} + (\mid \downarrow \rangle \langle \downarrow \mid)^{(1)} \Big{)} \otimes (\mid \uparrow \rangle \langle \uparrow \mid)^{(2)} \Big{]} \mid \Psi_- \rangle =\langle \Psi_- \mid \Big{[}\mathds{1}_{\mathcal{H}_1} \otimes (\mid \uparrow \rangle \langle \uparrow \mid)^{(2)} \Big{]} \mid \Psi_- \rangle \\
  = \mathbb{P}^{\Psi_-} (\sigma^{(2)}_z = +1) \quad = \quad \frac{1}{2}
\end{gathered}
\end{equation}

Similar calculations can be found in different versions in various papers and textbooks, usually labeled by \textit{``impossibility of superluminal signaling''}. 

\begin{center}
\textbf{Signals}
\end{center}

To make this notion transparent it is necessary to give a transparent account of the notion of a \textit{signal}. In physics literature transmission of signals is sometimes characterized by energy or matter transmission. Since the utilization of quantum nonlocality for faster-than-light signaling (if it was possible) would in general not be related to such processes, I will pick up Maudlin's very general definition of a signal \cite{Maudlin} (which is also not necessarily related to energy or matter transport), which goes like this: \textit{A signal is a physical process which can be decomposed into two correlated parts: a controllable and an observable part}. 

The notions \textit{``controllable''} and \textit{``observable''} do not necessarily relate to human decisions and perceptions (respectively) in the physical world but (as explained above) rather ``controllable'' stands for some free parameter in the equations of the theory whose variation produces variation of predictions, which then stand for the observable part of the process. The justification of such free parameters was discussed at the end of chapter \ref{Einstein Bell} and in footnote \ref{extinf}.

%\newpage
\begin{center}
\textbf{Controllability}
\end{center}

So far, we have encountered two possible controllable processes provided by quantum theory: \paragraph*{}

\textit{i)} The choice of the theoretician (experimenter) to apply either measurement formalism or unitary state evolution (to perform the measurement or not). \paragraph*{}

\textit{ii)} The choice of the theoretician (experimenter) to vary an operator associated with a measurement (to vary the actual configuration of the measurement device). \paragraph*{}

And we have found that such controllable actions do not change local probabilities for measurement events at space-like separation, if the actual result of these actions is ignored \eqref{nosupsig}. This kind of controllable operations bases on the nonlinear time-evolution-principle of wave-function collapse. Quantum formalism actually provides a third possibility of controllable external influence based on unitary time evolution: \paragraph*{}

\textit{iii)} The choice of the theoretician (experimenter) to vary some parameter called \textit{''external field''}, which couples to some variable of the system in the equations of motion (to vary some part of the configuration of some device interacting with the system\footnote{What is the actual difference between the measuring-device of \textit{i)} and \textit{ii)} and the interacting device of \textit{iii)}? Ordinary quantum mechanics has no answer to that question, for the measurement-device is part of its axiomatic formulation, and a description of measurement process can not be inferred from fundamental interaction by fundamental postulates, but an extraordinary ``dynamics of measurement`` is itself a postulate. This is the root of the \textit{shifty split} \cite{Bell} between ordinary interaction (as a part of the unitary evolution) and measurement (causing collapse), which is vague and arbitrary.}):\paragraph*{}

So let us briefly check whether local commutativity also serves to leave local probabilities of outcomes of measurement-events in $\mathcal{A}$ invariant under controllable influence in $\mathcal{B}$ in the sense of \textit{iii)}. For this we couple an external field to a degree of freedom of the system associated to the physical quantity $\mathscr{B}$. The coupling is described by an interaction-Hamiltonian of the form

\begin{equation}
 \mathscr{H}_{int} = \tilde{k} \hat{B} \mbox{ ,}
\end{equation}
where $\tilde{k}$ contains functions associated with the external field (coupling constant, time dependence...). Since the operators $\hat{A}$ and $\hat{B}$ commute, we can find a basis in $\mathcal{H}$ of common eigenstates, which we will denote by $\{ \mid \alpha, \beta, i \rangle \}$, where the index i indicates possible degeneration, if $\hat{A}$ and $\hat{B}$ do not form a complete set. Then we can decompose an arbitrary incoming wave-function $\mid \Psi_{in} \rangle$ into a superposition of these states

\begin{equation}
 \mid \Psi_{in} \rangle = \sum_{\alpha, \beta, i} c_{\alpha \beta i} \mid \alpha, \beta, i \rangle
\end{equation}
with coefficients $c_{\alpha \beta i}$. 

After interaction with the external field the wave-function will be

\begin{equation}
 \mid \Psi_{out} \rangle = e^{-ik\hat{B}} \mid \Psi_{in} \rangle = \sum_{\alpha, \beta, i} c_{\alpha \beta i} e^{-ik\beta} \mid \alpha, \beta, i \rangle \mbox{ ,}
\end{equation}
where now $k = \int \tilde{k} dt$.

Then, with the projector $\mathcal{P}_{\alpha'} := \sum_{\beta,i} \mid \alpha', \beta, i \rangle \langle \alpha', \beta, i \mid$ onto the eigenspace of $\hat{A}$ belonging to eigenvalue $\alpha'$, the probability of the $\mathscr{A}$-measurement to yield $\alpha'$ is

\begin{equation}
 \begin{gathered}
\mathbb{P}^{\Psi_{in}}(\mathscr{A}=\alpha' \mid external \hspace{0.2cm} field \hspace{0.2cm} in \hspace{0.2cm} \mathcal{B}) = \mathbb{P}^{\Psi_{out}}(\mathscr{A}=\alpha') = \langle \Psi_{out} \mid \mathcal{P}_{\alpha'} \mid \Psi_{out} \rangle \\
= \bigg{(} \sum_{\alpha, \beta, i} \bar{c}_{\alpha \beta i} e^{ik \beta} \langle \alpha, \beta, i \mid \bigg{)} \mathcal{P}_{\alpha'} \bigg{(} \sum_{\tilde{\alpha}, \tilde{\beta}, \tilde{i}} c_{\tilde{\alpha} \tilde{\beta} \tilde{i}} e^{-ik\tilde{\beta}} \mid \tilde{\alpha}, \tilde{\beta}, \tilde{i} \rangle \bigg{)} \\
= \bigg{(} \sum_{\alpha, \beta, i} \bar{c}_{\alpha \beta i} e^{ik \beta} \langle \alpha, \beta, i \mid \bigg{)} \bigg{(} \sum_{\tilde{\beta}, \tilde{i}} c_{\alpha' \tilde{\beta} \tilde{i}} e^{-ik\tilde{\beta}} \mid \alpha', \tilde{\beta}, \tilde{i} \rangle \bigg{)} \\
=\sum_{\beta, i} \mid c_{\alpha' \beta i} \mid^2 = \mathbb{P}^{\Psi_{in}}(\mathscr{A}=\alpha') 
 \end{gathered}
\end{equation}
where we have used orthogonality of the eigenvectors. Thus with local commutativity, which is crucial for this result, also the imposition of external fields in $\mathcal{B}$ does not change local probabilities of measurement-outcomes in $\mathcal{A}$. \paragraph*{}

Now, given an actual wave-function, the outcomes of quantum-measurements are notoriously uncontrollable, even in deterministic Bohmian mechanics \cite{quantequi}. The only means of controllability in quantum theory arise from either of the three operations discussed: \textit{i)} performing measurements or not, \textit{ii)} variation of measurement settings or \textit{iii)} imposition of external fields. The actual effect of operations \textit{i)} and \textit{ii)} remains uncontrollable and operation \textit{iii)} does not change the relevant predictions anyway, given local commutativity. Thus, with local commutativity, controllable operations performed in $\mathcal{B}$ do not change the ''observable`` probabilities (relative frequencies) of measurement-events in $\mathcal{A}$, i.e.\! \textit{quantum measurement provides no strategies to send signals faster than light} :

\begin{equation}\label{nosig}
 \mathbb{P}(\mathscr{A}=\alpha \mid controllable \hspace{0.3cm} operations \hspace{0.3cm} performed \hspace{0.3cm} in \hspace{0.3cm} \mathcal{B}) = \mathbb{P}(\mathscr{A}=\alpha) 
\end{equation}
where it is not conditioned on the actual effect of the controllable operations. In the following I will refer to condition \eqref{nosig} as \textit{''no signaling``} condition\footnote{For a much more detailed analysis of no signalling conditions and their relation to local commutativity, see chapter 3 of \cite{diss}.}. \paragraph*{}

But did God invent this ``causality'' (local commutativity) just to prevent us from sending faster-than-light messages while nonlocal causal structures lie at the heart of the dynamics governing the world, or is there a more fundamental meaning? Does local commutativity provide serious contribution to the reconciliation of the space-time structure of special relativity with quantum nonlocality? 

I will propose a somewhat unusual view on local commutativity which might be appropriate to reveal such contribution in section \ref{meaningloccom}. But first, let me briefly discuss the (a bit anthropocentric) more common argument in support of local commutativity, which bases on the just discussed prohibition of superluminal signaling in order to preclude backwards-in-time causation and alleged resulting paradoxes. 

\subsubsection{\label{bitcausation}Faster-than-Light Signals \& Backwards-in-Time Causation}

It is often said that special relativity excludes superluminal signals. This is not true, from relativity principle only follows the existence of an invariant -- but not necessarily of a maximal velocity \cite{Sexl, Maudlin}. Tim Maudlin gives an extensive account of possible physical processes with signals faster than light, without touching relativistic space-time structure and the relativity principle (see chapter 4 in \cite{Maudlin}). Even tachyons can be easily implemented in a relativistic theory, i.e.\! even superluminal matter transport does not contradict relativity in principle (chapter 3 in \cite{Maudlin}).

More serious concerns arise from the directedness of signals. It is important to recognize that the definition of a signal, as given above, defines a directed causal process, for the controllable cause and the observable effect are well defined in each case and cannot being exchanged with one another in general. And if now a signal in the above sense could be ``superluminal'', i.e if cause and effect would lay in space-like separated space-time regions, there would always exist Lorentz-frames in which this signal travels backwards in time, i.e in which the effect would precede its cause. This would contradict very much the way in which we perceive the temporal structure of causality in the everyday life and, preferably, I would tend to avoid such rather strange elements in a physical theory, too. But at the end of the day I see neither physical nor logical necessity to do so.

\begin{figure}[htbp]
\centering
\includegraphics[scale=0.5]{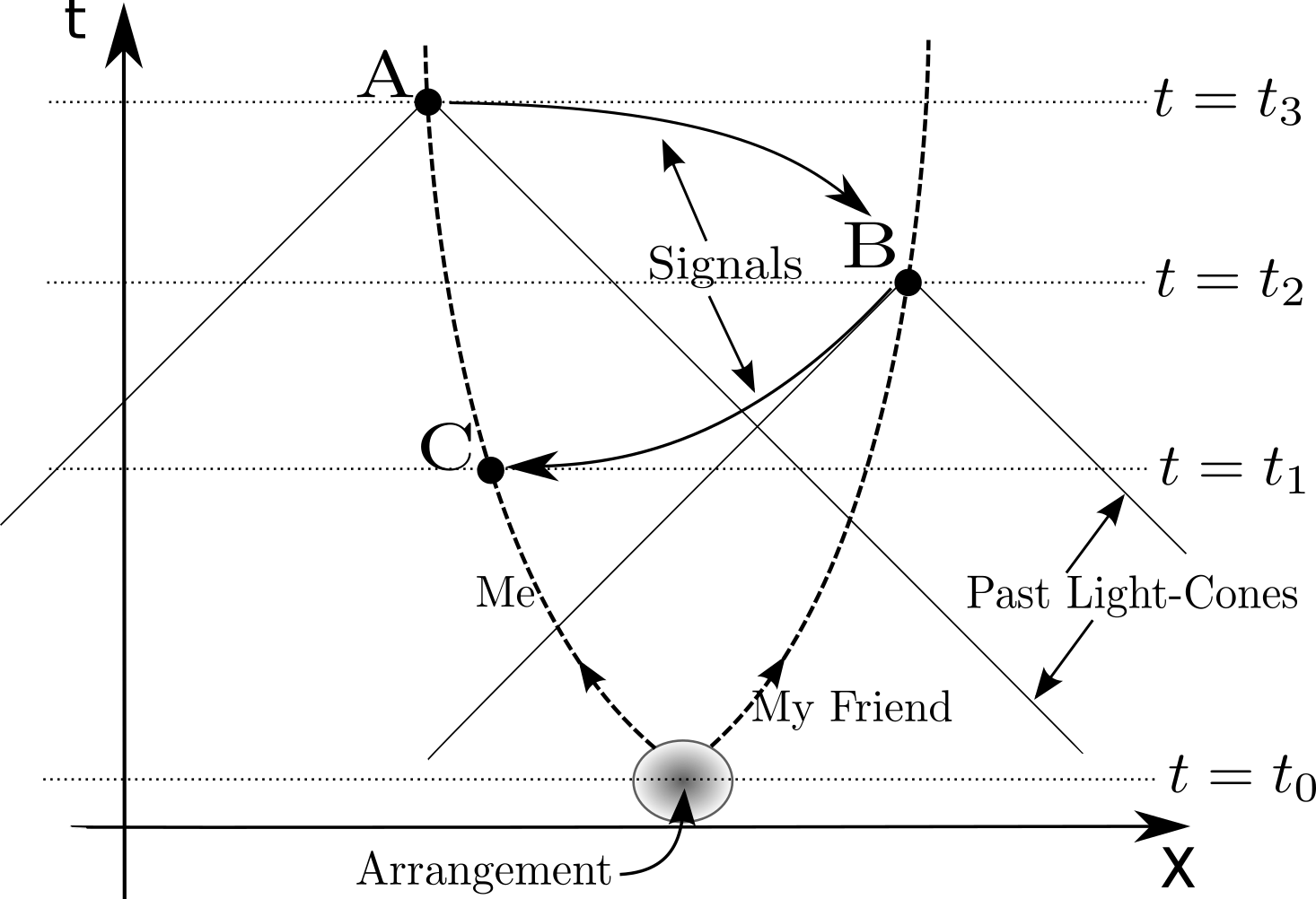}
\caption{\textbf{Causal Loop}: If it was possible to send signals faster than light, i.e.\! in the side light-cone, paradoxes might be constructed. Suppose that I make an arrangement with my friend: If he receives a signal from me at $B$, he will immediately send a signal back to space-time point $C$. If I receive a signal at $C$, I will omit to send a signal at $A$. Hence, if I send a signal from $A$ to $B$ I will omit to send a Signal from $A$ to $B$.}
\label{Loop}
\end{figure}

Various authors have indeed reasonable concerns about the logical implications of backwards-in-time causation. These concerns arise from the possibility of creating paradoxes -- basing on causal loops -- if we once allow for such kind of causality. If signals could be received in the side light-cone of the event of transmission, a closed causal chain might be constructed, e.g.\! between three space-time points A, B and C in figure \ref{Loop}. 

Suppose I make an arrangement with my friend: if he receives a signal he in turn will immediately send a signal back but if I receive a signal I will omit to send a signal. Then we separate and occupy different positions in space (see figure \ref{Loop}). Then, at space-time point A (time $t_3$), I send a signal to my friend which he receives at B (time $t_2$) in the side light-cone of A. And he in turn immediately sends a signal back which I receive at C (time $t_1$) prior to A which induces me to omit to send the signal at A. This is a contradiction as it seems to be contradictory in general that an event could be in its own range of causal influence. 

But there is one flaw in the argument, for it presupposes controllability of the device (which is used for signaling) at space-time points A and B. And it is rather obvious that the basis of the use of the concept ``controllability`` we worked out and worked with so far, does not work for closed causal chains in general. We justified a ''freedom of choice'' of the device settings within the procedure of calculating predictions for the devices' effect on the system. This freedom was compatible with physical description, even within a deterministic theory. We argued that device and system should be considered as independent physical systems before interaction and tacitly assumed no further fundamental constraints to be imposed by the theory on the device settings. But in the end, in a complete description the physics of the device should also be part of the physics described by the theory. And from the point of view of a theory which allows for closed causal chains, inconsistent loops as described above should not emerge as solutions of the theories equations if the dynamics of the theory is well defined. In such a case it might not be conspiratorial anymore to have restrictions to the device settings. Wheeler and Feynman \cite{wheelerfeynman} gave an example of a theory, in which backwards causation is a fundamental part of the dynamics and -- following from that -- in which events are in their own range of causal influence. In the chapter ``The Paradox of Advanced Actions`` they gave a paradoxical example analogous to the one I gave in the last paragraph and analyzed it carefully from the point of view of the theory they proposed. And it turned out that everything paradoxical vanishes as soon as the whole closed story is continuously described by the theory.

Thus a theory with a well defined dynamics can allow for backwards causation and no inconsistency should emerge. Of course, we might get into serious trouble if we allow for real metaphysical external influence on physical systems -- like me and my friend having a metaphysical free-will-choice to send superluminal signals or not. I do not want to reject the possibility of an existence of something like ''metaphysical free-will'' here. But for reasons of humility I plead in favor of avoiding the usage of free-will as justification of physical laws, and to look for physical alternatives instead.

\subsubsection{\label{meaningloccom}A Meaning of Local Commutativity: Ontological Consistency}

Here is now an argument why a violation of local commutativity would indeed have rogue and inconsistent consequences, which does not stick ultimately to the concept of controllability. To see this consider again an experiment consisting of two measurements performed on a quantum mechanical system at space-like separation and suppose the self-adjoint operators associated with these measurements do not commute -- we have seen above that this is a quantum mechanical formalization of ``possibility of superluminal signaling''. Assume further that the first measurement, e.g.\! of the physical quantity $\mathscr{A}$, takes place in space-time region $\mathcal{A}$ around time $t_1$ in the laboratory frame and the corresponding operator is $\hat{A}$ given by equation \eqref{opA}. Correspondingly $\mathscr{B}$ is measured in $\mathcal{B}$ at time $t_2 > t_1$ with corresponding operator $\hat{B}$ given by equation \eqref{opB}.          

If, before the $\mathscr{A}$-measurement, the system is prepared, say in the $\hat{A}$-eigenstate $\mid \varphi^{(\tilde{i})}_{\tilde{\alpha}} \rangle$, the outcome of this measurement will be $\tilde{\alpha}$ with certainty, e.g.\! there might be a pointer which points on the number $\tilde{\alpha}$. And if this procedure is performed on an ensemble of identically prepared systems (in the above sense) all the pointers will point on $\tilde{\alpha}$ after this sequence of experiments (there might be pairs $(t_1,t_2)_i$, one for each experiment).

Since the space-time regions $\mathcal{A}$ and $\mathcal{B}$ are space-like separated we can find a Lorentz-frame of reference in which the $\mathscr{A}$-measurement takes place at time $t'_1$ and the $\mathscr{B}$-measurement at time $t'_2$, but now $t'_2 < t'_1$ (for each experiment) and we can describe this ensemble of quantum mechanical processes from this point of view. In this frame the $\mathscr{B}$-measurements will precede the $\mathscr{A}$-measurements (respectively). And because $[\hat{A},\hat{B}] \neq 0$ the $\mathscr{B}$-measurements will destroy the ingoing states $\mid \varphi^{'(\tilde{i})}_{\tilde{\alpha}} \rangle$ (where the prime indicates the Lorentz transformation of the state) and produce a superposition 

\begin{equation}
 \sum_{\alpha,i} c_{\alpha,i} \mid \varphi^{'(i)}_{\alpha} \rangle \mbox{ ,}
\end{equation}
where in general $c_{\alpha,i} \neq 0$, also for $\alpha \neq \tilde{\alpha}$ (and $i \neq \tilde{i}$). Thus in this frame in our ensemble of experiments there will be cases in which the pointer points on numbers $\alpha \neq \tilde{\alpha}$ in the (subsequent) $\mathscr{A}$-measurement. 

Apparently -- given space-like separated measurements with associated non-commuting operators -- \textsl{the distribution of matter of some pointers would transform under a change of Lorentz-frame into distributions of matter of the pointers which are not the Lorentz-transform of the former distributions. Experimenters in the respective Lorentz-frames would not experience the respective Lorentz-transformed reality but a completely different reality.} 

We can look at it this way: This inconsistency emerges from the fact that in a reasonable physical theory the primitive ontology should be unique in some sense. In special relativity \textsl{a ruler transforms into the same ruler in a different frame}; although its shape might change according to Lorentz-contraction, every space-time point (as a Lorentz invariant locus in $\mathscr{M}$) associated with the ruler in one frame transforms into a space-time point associated with that ruler in another frame. And the same holds for pointers or any other distribution of matter, i.e.\! for the primitive ontology. We will call the requirement that distributions of matter in different Lorentz frames are the respective Lorentz transformed distributions of one another \textsl{ontological consistency}\footnote{The concept of ontological consistency in relativistic quantum theory is further developed in chapter 3 of \cite{diss}. In this reference, this concept is stated as a mathematical requirement under the name \textsl{relativistic consistency} and its relations to no signalling and local commutativity are analyzed in great detail.}. \paragraph*{}

Indeed, if Minkowski space-time structure is taken for granted, it is even technically impossible to construct non-commuting observables which correspond to measurements in space-like separated regions: there exists also always a frame in which $t_1=t_2$ and we all know that quantum theory excludes simultaneous measurements described by non-commuting operators. \paragraph*{}

In the following we will be concerned with measurement strategies for nonlocal physical quantities encoded in entangled wave-functions, and there it is a subtle business to identify the measurements which do not raise problems in a relativistic context. For example we will see, that for two spatially separated spin-$\frac{1}{2}$ particles each component of the total spin can be measured at a well defined time (in one frame) by purely local interactions without violating ``no signaling``-condition \eqref{nosig}, but the (square of the) total spin cannot. Both are nonlocal quantities of the composite system and the relativistic and quantum mechanical implications for the possibility or impossibility of such procedures are not obvious.

\subsubsection{Intermezzo: Relativistic Restrictions to the Set of Observables}

Consider the following scenario \cite{aa2, breuer, ghirardinonlocal}: A pair of spin-$\frac{1}{2}$-particles is prepared in the state 
\begin{equation}
\mid \Psi_{prep} \rangle = \mid \uparrow \uparrow \rangle 
\end{equation}
and subsequently the particles get spatially separated (far apart from each other). Now suppose it would be possible to measure the square of the total spin 
\begin{equation}
 (\boldsymbol{\sigma}^{tot})^2=(\boldsymbol{\sigma}^{(1)} + \boldsymbol{\sigma}^{(2)})^2
\end{equation}
 of this system at a well defined instant of time $t_0$ (in some frame \footnote{Obviously this measurement requires local interactions of both spatially separated particles with some measurement device. And to require both interactions to happen at the  same instant of time $t_0$ implies simultaneity of space-like separated events, which of course means the choice of a particular Lorentz-frame.}). The state  $\mid \uparrow \uparrow \rangle$ is an eigenstate of the operator $(\hat{\boldsymbol{\sigma}}^{tot})^2$ which describes the statistics of this kind of measurement and this implies (according to the rules of quantum theory) that the measurement does not disturb the state of the system. Thus if we perform a measurement of the z-component of the spin of particle(1) $\sigma^{(1)}_z$ right after the $(\boldsymbol{\sigma}^{tot})^2$-measurement -- say at time $t_0 + \epsilon$ -- it would yield the value +1 with certainty, the value -1 had probability zero.

\begin{figure}[htbp]
\centering
\includegraphics[scale=0.16]{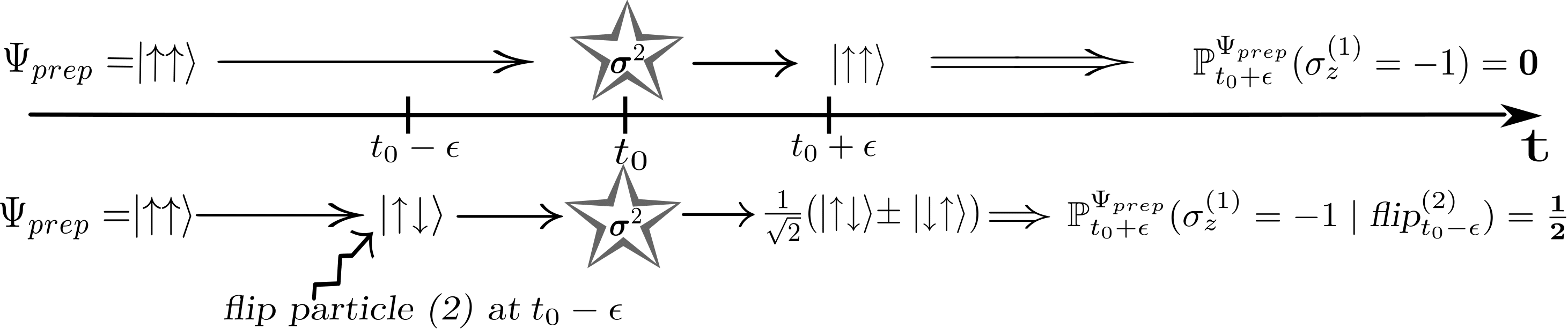}
\caption{If it was possible to measure the total (square of the) spin of a two particle spin-$\frac{1}{2}$-system at some well defined instant of time, controllable local operations in one space-time region (the one belonging to particle(2)) could be detectable in a space-like separated region (belonging to particle(1)), i.e the possibility of such a kind of measurement would give rise to violations of no-signaling condition \eqref{nosig}.}
\label{sigma^2}
\end{figure}

 Now imagine we made an arrangement with our friend before preparation: We will accompany particle(1) and he particle(2). If he is than in a good mood shortly before the $(\boldsymbol{\sigma}^{tot})^2$-measurement (say at time $t_0-\epsilon$), he will stay passive but if his mood is rather bad he will push the on-button of some magnetic device, which than interacts with particle(2) and flips the z-component of its spin to -1 (see figure \ref{sigma^2}). Thus if our friend is in black mood at $t_0-\epsilon$, the incoming state of the $(\boldsymbol{\sigma}^{tot})^2$-measurement will be   
\begin{equation}
 \mid \Psi_{in} \rangle = \mid \uparrow \downarrow \rangle
\end{equation}
 which is not an eigenstate of the operator $(\hat{\boldsymbol{\sigma}}^{tot})^2$ anymore, but rather a linear-combination of the eigenstates 
\begin{equation}
 \mid \Psi_{\pm}\rangle:=\frac{1}{\sqrt{2}}(\mid \uparrow \downarrow \rangle \pm \mid \downarrow \uparrow \rangle)\mbox{ .} 
\end{equation}

The measurement would then disrupt the incoming state and produce one of these two eigenstates (with probability $\frac{1}{2}$, respectively). The probability to measure $\sigma^{(1)}_z=-1$ for particle(1) at time $t_0 + \epsilon$ would not be zero anymore but $\frac{1}{2}$. Thus if we find the value of the z-component of the spin of particle(1) to be -1 at time $t_0 + \epsilon$ we immediately know that our friend has pushed the button and that we have to worry about his bad mental condition (if this value is +1 we cannot infer any information about our friend, of course).
Obviously some local operation onto particle(2) caused a sudden jump of the local probability of finding the z-component of the spin of particle(1) to be -1 from zero to $\frac{1}{2}$. Note that we may chose $\epsilon$ arbitrarily small in principle! Thus we have a violation of ''no signaling`` condition \eqref{nosig}:

\begin{equation}
\begin{gathered}
(\boldsymbol{\sigma}^{tot})^2 \mbox{-measurement at time } t_0 \hspace{1.0cm} \Longrightarrow \\
 \mathbb{P}^{\Psi_{prep}}_{t_0+\epsilon}(\sigma^{(1)}_z=-1 \mid \textsl{friend pushed a button near particle(2) at time } t_0-\epsilon ) = \frac{1}{2} \\
\neq 0 = \mathbb{P}^{\Psi_{prep}}_{t_0 + \epsilon}(\sigma^{(1)}_z=-1) \mbox{ .}
\end{gathered}
\end{equation}

Indeed, if it was possible to measure $(\boldsymbol{\sigma}^{tot})^2$ within a time interval smaller than $\Delta T = \frac{\Delta X}{c}$, where $\Delta X$ is the distance of the two particles at the time of measurement, it would be possible to send signals faster than light. \paragraph*{}

\textbf{Ontological Consistency:} Now, we have seen that the possibility of measuring the square of the total spin at some well defined instant of time, would violate the requirement of excluding superluminal signals. We encountered in the previous section, that this requirement is closely related with a maybe somewhat better justified requirement, namely ontological consistency. But it seems to be not that simple in this case to construct ontological contradictions out of the possibility of such measurements. We cannot simply transform the above scenario into a different frame, to see ontological inconsistency appearing\footnote{Suppose the story is described in a frame, in which our $\sigma^{(1)}_z$-measurement at particle(1) is prior to the spin-flip of particle(2). In this frame both interactions pertaining to the $(\boldsymbol{\sigma}^{tot})^2$-measurement do not occur simultaneously anymore, and -- as we will see in chapter \ref{AA} -- we can expect in general, that even an eigenstate of the operator $(\boldsymbol{\hat{\sigma}}^{tot})^2$ will be disturbed in the intermediate time between the two interactions. Thus even if the incoming state of the first interaction of the $(\boldsymbol{\sigma}^{tot})^2$-measurement at particle(1) is $\mid \uparrow \uparrow \rangle$, it is not contradictory to have a pointer at the $\sigma^{(1)}_z$-measurement device, pointing on the value -1 directly afterwards.}. 

Such a scenario requires local interactions of both particles with some devices. In the explicit quantum mechanical construction of such a process, operators are associated with these interactions (we will explicitly construct an analogous measurement procedure in chapter \ref{AA}). And the violation of condition \eqref{nosig} here, means that the operations performed on particle(1) and on particle(2) disturb one another, i.e.\! the operators associated with these operations do not commute (in calculation \eqref{jointprob} additional interference terms would appear if the projectors failed to commute, such that the ''joint probability property'' and thus the ''no signaling'' property \eqref{nosig} would be destroyed). Thus we can claim that such a procedure should be impossible if relativistic space-time is taken for granted and ontological consistency in the sense of the previous chapter is required.  \paragraph*{}      
   
It seems that there are certain restrictions to the set of physical quantities which can be measured at a well defined time within relativistic quantum mechanics. The consideration of such restrictions set the stage for a detailed investigation of the implications which Minkowski space-time has for wave-function collapse. Such investigations will be addressed in the following sections. 

\newpage

\section{\label{chap3}Relativistic State Description}

\subsection{Landau \& Peierls}

In 1931 Lew Landau and Rudolf Peierls \cite{landau} suggested a kind of relativistic precondition for quantum theory in the realm of relativity: there should be certain restrictions to the set of measurable physical quantities which are not obvious in the non-relativistic case. But the reasoning of Landau and Peierls suggested also tacitly the necessity to face ambiguities of wave-function collapse in a relativistic theory.

\begin{figure}[htbp]
\centering
\includegraphics[scale=0.158]{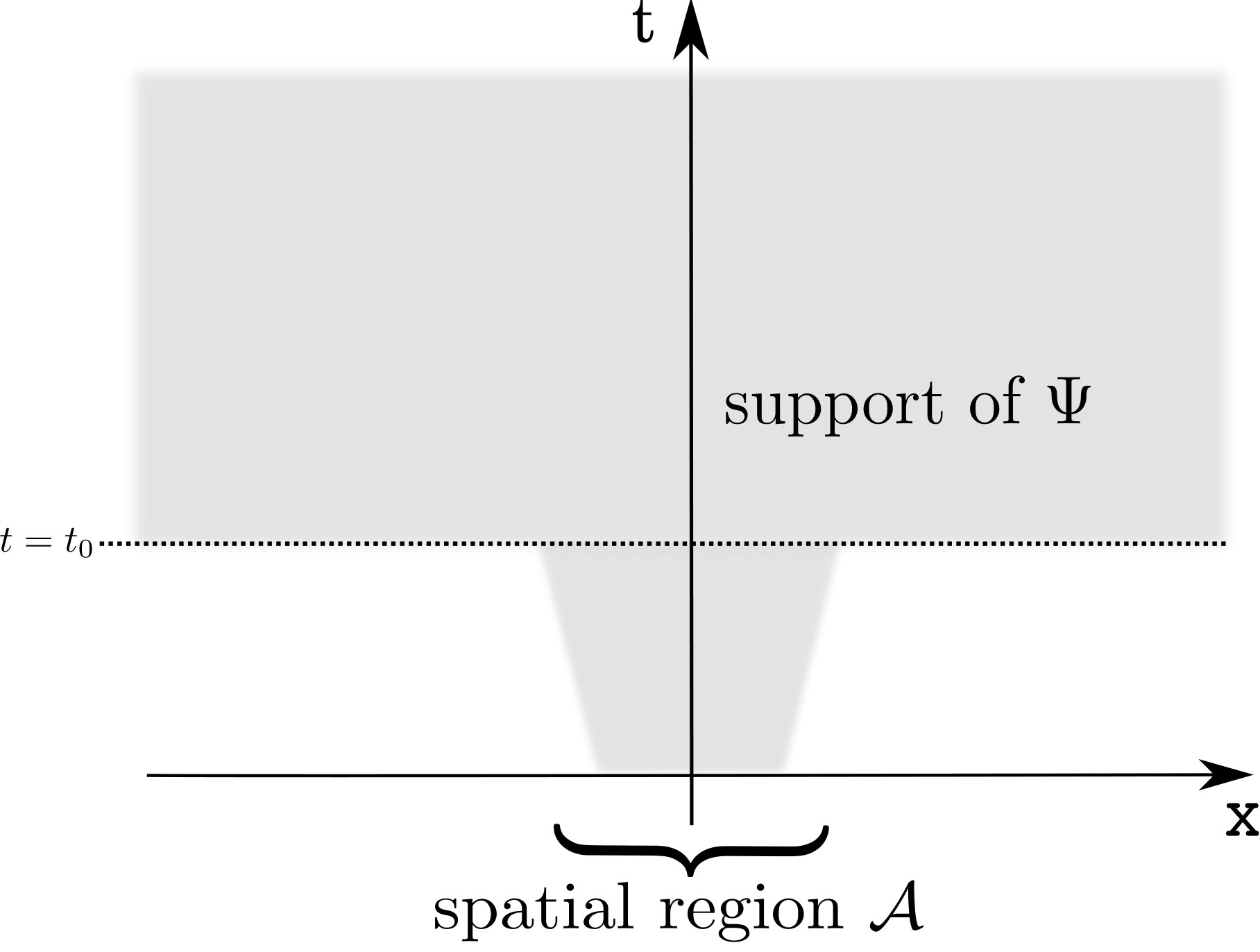}
\caption{\textbf{Localized State $\rightarrow$ Momentum Measurement:} Suppose a localized one particle state was prepared by a position measurement about space-time region $\mathcal{A}$, where the shaded region indicates the support of the wave-function. If it was possible to measure the momentum of this particle at some well defined instant of time $t=t_0$, there would be finite probability of finding the particle any desired distance far apart from $\mathcal{A}$ shortly after the detection about $\mathcal{A}$. Therefore, the probability of moving the particle with superluminal velocity would be nonzero.}
\label{LP}
\end{figure}

To follow their thought consider a one-particle system which is prepared with some initial wave-function $\Psi$ having support localized in some spatial region $\mathcal{A}$ (see figure \ref{LP}). At some time $t_0$ the momentum of the particle is measured and found to have the value $p$. After the measurement the wave-function will be a (non-normalizable) eigenstate of the momentum operator (in the idealized case) with eigenvalue $p$. The support of this wave-function is spread all over space. Thus the probability of finding the particle (by means of a position measurement) in some spatial region $\mathcal{B}$, far far apart from $\mathcal{A}$, jumps at time $t_0$ from zero to some finite value. This obviously means that according to quantum theory the described procedure yields a non vanishing probability that the particle is shifted from $\mathcal{A}$ to $\mathcal{B}$ with superluminal velocity (indeed the velocity would be infinite because the jump of the probability occurs at some definite instant of time $t_0$). 

Such reasoning motivated Landau and Peierls to infer relativistic restrictions to the set of quantum-mechanical observables. They argued that in order to prevent the theory to provide a mechanism for superluminal matter transport (or in the general case for arbitrary superluminal signaling) the above described procedure should be impossible, i.e.\! it should be impossible to measure the momentum of a particle at some well defined instant of time. 

They further generalized this requirement in the following way: If a measurement ``produces`` a wave-function with support of spatial extension $\Delta X$, the time $\Delta T$ of interaction between system and the measurement device should be restricted by\footnote{This restriction combined with the usual uncertainty principle led Landau an Peierls to propose a new relativistic uncertainty principle: ${{\Delta p} {\Delta T}} \gtrsim {\frac{\hbar}{c}}$} 

\begin{equation} \label{LPcondition}
\Delta T \gtrsim {\frac{\Delta X}{c}} \mbox{ .}
\end{equation}

However a closer look shows that condition \eqref{LPcondition} is not justified in this general form (with ``no signaling`` taken as the basis of justification) and further that it is indeed not consistent with the predictions of quantum mechanics: Following David Albert and Yakir Aharonov we will encounter quantum mechanical measurement strategies which give rise to violations of \eqref{LPcondition}, i.e.\! we will construct explicitly the description of a quantum measurement for which $\Delta X$ is arbitrarily big and $\Delta T$ arbitrarily small. But nonetheless this kind of measurement consists of purely local interactions between system and device and it does not give rise to violations of ''no signaling``-condition \eqref{nosig}.

\subsection{\label{chapHK}Hellwig \& Kraus} 

\begin{figure}[htbp]
\centering
\includegraphics[scale=0.51]{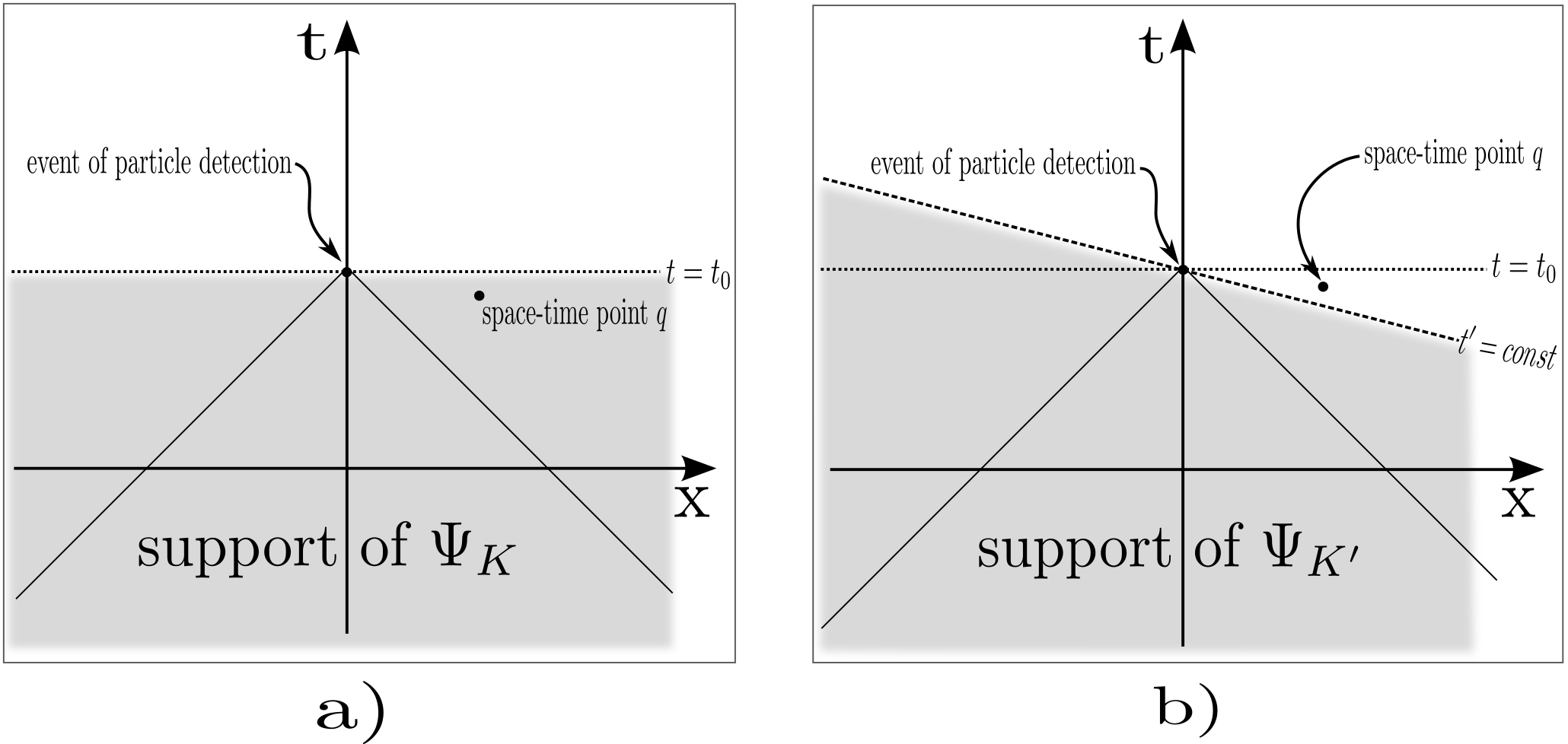}
\caption{\textbf{Spread State $\rightarrow$ Position Measurement:} $a)$ Consider an initial one particle wave-function with widely spread spatial support (shaded region), e.g.\! a momentum eigenstate. At time $t=t_0$ a position measurement is performed and the particle is found, say at $\v{x}=0$. $b)\mbox{ }$From the viewpoint of a different Lorentz-frame, the constant time hyperplane containing the event of detection is distinct from the $t=t_0$ hyperplane. Hence, if the wave-function collapses instantaneously in each frame, we have to account for two distinct state histories, which are apparently not the Lorentz-transform of one another. Has the wave-function non-vanishing support at space-time point $q$?}
\label{HK}
\end{figure}

Consider now again a one-particle system with initial wave-function $\Psi$ having support extended in space (e.g.\! some momentum eigenstate). Suppose at time $t_0$ a position measurement is carried out detecting the particle, say, at $x=0$. This measurement induces a collapse of the wave-function along the $t=t_0$ hyperplane in space-time (see figure \ref{HK}$a$): Subsequent detection of that particle by another detector far apart from the position detected in the first measurement must have probability zero.  

Transformed into a different Lorentz-frame, say $K'$ this scenario looks quite different: The $t'=const$  hyperplane through the space-time point of detection is not the same hyperplane anymore as in the un-primed frame $K$ (figure  \ref{HK}$b$). If we require the non-existence of a distinguished Lorentz-frame the probability of finding the particle above the $t'=const$ hyperplane of $K'$ (far apart from the point of detection) in a subsequent measurement must also vanish. And if we iterate this argument to cover all possible frames of reference we find that the probability of finding the particle anywhere forward to the surface of the backward light-cone of space-time point $(t_0,0)$ must be zero. 

Such reasoning is the content of a paper by Felix Bloch \cite{bloch}. Instigated by that paper Karl-Eberhard Hellwig and Karl Kraus \cite{hellwig} proposed a covariant law for wave-function collapse\footnote{Hellwig and Kraus actually proposed their model for local field observables in local QFT, but the model is in straight analogy to a consideration of common local quantum mechanical observables as done by Bloch and within this lines.}: Since every light-cone is a Lorentz invariant locus in $\mathscr{M}$ (a light-cone transforms into itself under Lorentz transformations) they proposed \textit{wave-function collapse to occur along the backward light-cone of each measurement event}.

\begin{figure}[htbp]
\centering
\includegraphics[scale=0.18]{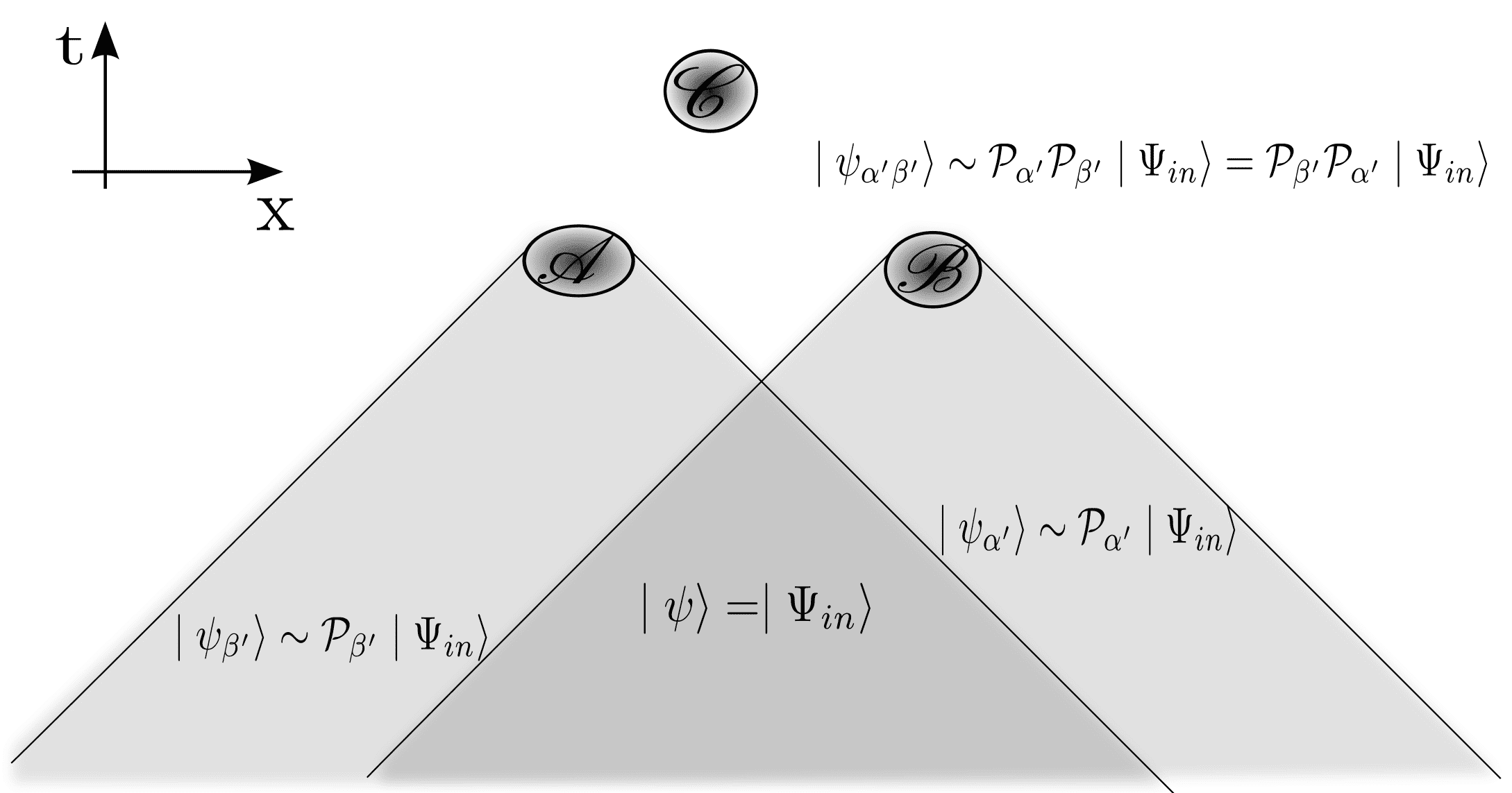}
\caption{\textbf{Hellwig-Kraus reductions for multiple local measurements:} Local measurements of physical quantities $\mathscr{A}$, $\mathscr{B}$ and $\mathscr{C}$ are performed in the indicated space-time regions. According to the proposal of Hellwig and Kraus the respective reduced states are to be taken forward to the backward light-cone of the respective measurements. In these regions the initial wave-function $\mid\Psi_{in}\rangle$ is projected (and renormalized) onto the subspace of Hilbertspace corresponding to the respective measurement outcomes. The projectors are denoted by $\mathcal{P}_i$.}
\label{HKpic}
\end{figure}

Local commutativity than guaranties the right and unique joint probability distributions, when sets of partly space-like separated local measurements are described by the Hellwig-Kraus formalism (see figure \ref{HKpic}): 

Suppose local measurements of physical quantities $\mathscr{A}$, $\mathscr{B}$ and $\mathscr{C}$ are performed in the indicated space-time regions of figure \ref{HKpic}, where the region belonging to the $\mathscr{A}$-measurement is space-like separated with respect to the region belonging to the $\mathscr{B}$-measurement. Then the collapsed wave-function due to the $\mathscr{A}$-measurement is to be taken in the part of space-time, forward to the surface of the past light-cone of the respective space-time region; and the same holds for the $\mathscr{B}$-measurement. The collapsed wave-function is in each case proportional to the projection of the initial wave-function $\mid \Psi_{in} \rangle$ onto some eigenspace (e.g.\! corresponding to eigenvalue/outcome $\alpha'$) of the corresponding operator. Then, according to local commutativity of these operators and according to joint-probability-calculation \eqref{jointprob}, the outcome statistics of the $\mathscr{B}$-measurement, for an ensemble of such scenarios, is unaltered by the fact of wave-function collapse according to the $\mathscr{A}$-measurement, and vice versa. Secondly, local commutativity of the corresponding operators and the resulting commutativity of the related projectors (spectral theorem), guaranties that the lack of a definite time order between the $\mathscr{A}$ and $\mathscr{B}$ measurement raises no problems to calculate a definite resulting wave-function of these two processes. Thus the local measurement of physical quantity $\mathscr{C}$ (in the indicated space-time region) has a unique initial wave-function. So far so good... 

\begin{figure}[htbp]
\centering
\includegraphics[scale=0.18]{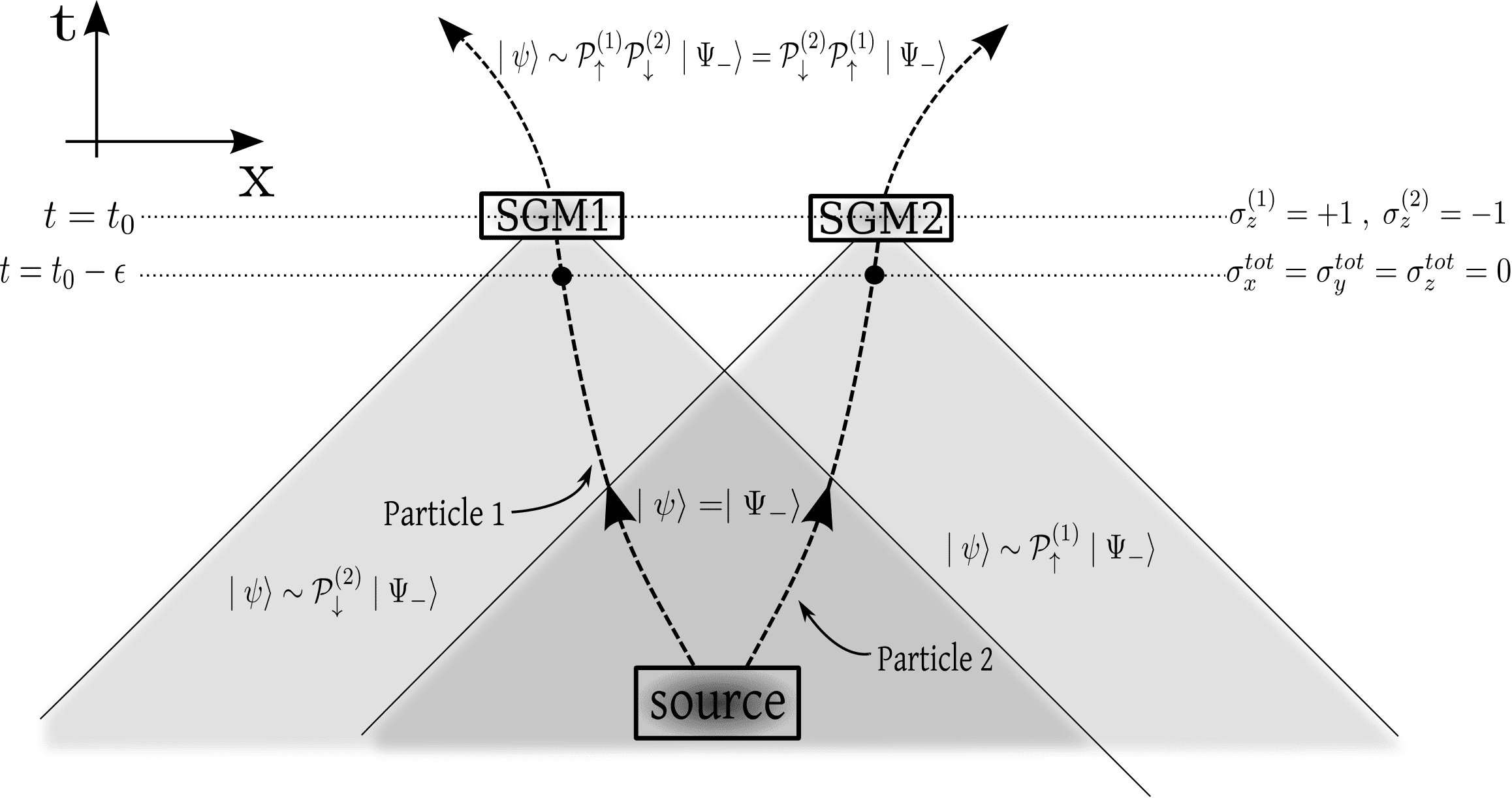}
\caption{\textbf{Contradiction:} EPRB-experiment at time $t=t_0$ with Hellwig-Kraus reductions and an Aharanov-Albert procedure performed at time $t=t_0 - \epsilon$ shortly before the one particle spin measurements of the EPRB-experiment. According to the Hellwig-Kraus proposal the wave-function is already reduced all along the $t=t_0-\epsilon$ hyperplane. This is contradictory to the possibility of verifying $\sigma^{tot}_x=\sigma^{tot}_y=\sigma^{tot}_z=0$ as a consequence of the initial singlet state at this time.}
\label{HKEPR}
\end{figure}

But apart from some oddities\footnote{The Hellwig-Kraus picture has the strange teleological feature that the history of the wave-function within the future light-cone of some event (say the big bang) is somehow determined in advance by ''measurement-events'' in the future. Also a strange and nontransparent kind of nonlocality is involved here: The above description justifies the right quantum mechanical statistics for ensembles of measurements, but it is not clear to me how e.g.\! the perfect (anti-)correlations in each single run of the EPRB-experiment find such justification (contemplate on figure \ref{HKEPR} and do not care about the Aharanov-Albert experiment at $t_0 - \epsilon$).} the formalism proves a failure as soon as measurements of physical quantities are considered, which are intrinsically encoded in nonlocal entangled wave-functions: Consider for example once again the EPRB-setup with simultaneous measurements of $\sigma^{(1)}_z$ and $\sigma^{(2)}_z$ at (or directly behind) the two SGMs, respectively, at some time $t_0$ (see figure \ref{HKEPR}). Now suppose it would be possible to perform a measurement-procedure arbitrarily short before that time (say at time $t_0 - \epsilon$), which verifies that the total spin of the two particles is zero: $\sigma^{tot}_x=\sigma^{tot}_y=\sigma^{tot}_z=0$. This is true for the singlet state

\begin{equation}
 \hat{\sigma}_x^{tot} \mid \Psi_- \rangle = \hat{\sigma}_y^{tot} \mid \Psi_- \rangle = \hat{\sigma}_z^{tot} \mid \Psi_- \rangle = 0 \mbox{ ,}
\end{equation}

but neither for the state $\mid \uparrow \downarrow \rangle$ nor for $\mid \downarrow \uparrow \rangle$ (indeed in each of the latter two states it is even not possible to assign definite values to all three components of the total spin\footnote{The impossibility of such assignments is actually a basic content of no-go theorems like the one of Kochen and Specker and also of Bell's theorem. Such view on these theorems is usually expressed by a requirement of \textit{``contextuality``} for variables like spin.}). But in the Hellwig-Kraus picture one of the latter two states is already realized in \textit{all} space-time regions which are crossed by the constant-time $t_0-\epsilon$ slice.

Such a measurement-scheme was explicitly constructed by Yakir Aharanov and David Albert \cite{aa1, aa2} (and later in a more realistic version by Gian-Carlo Ghirardi \cite{ghirardinonlocal}) and thereby the Hellwig-Kraus proposal is refuted. But before we come to this scheme one last remark on the Hellwig-Kraus formalism.         

\vspace{1cm}

\begin{center}
\textbf{Wave-Function as a Functional on the Set of Space-Time Points}
\end{center}

It was pointed out by Tumulka \cite{pointproc} that Hellwig and Kraus actually (and tacitly) proposed to take the wave-function as a functional on the set of points of space-time:

Denote by $\mathscr{F}(x)$ the (absolute) future of space-time point $x$ (as formalized at the very beginning). Now consider the set of space-time points $\{X_1,X_2,...,X_N\}$ where measurements of physical quantities $\mathscr{A}_{X_i}$ ($i \in \{1,...,N\}$) are performed and which shall thus be taken as the vertices of Hellwig-Kraus reductions. The outcome of measurement $i$ of quantity $\mathscr{A}_{X_i}$ at point $X_i$ shall be denoted by $\alpha_i$ and the projector onto the corresponding subspace $\mathcal{H}_{\alpha_i}$ of Hilbertspace $\mathcal{H}$ by $\mathcal{P}_{\alpha_i}$. The ''initial'' wave-function (to be taken in the area of space-time where all past light-cones of the points $X_i$ intersect) is denoted by $\mid \Psi_{in} \rangle$. 

Then we can associate a Hellwig-Kraus wave-function $\mid \psi_x \rangle$ with every space-time point $x$ by\footnote{The space-time index $x \in \mathbb{R}^4$ is not to be confused with the argument of the wave-function (in position representation), which still lives on configuration space(-time), of course. }   
\begin{equation}
 \mid \psi_x \rangle = \frac{ \Big{(} \raisebox{0.15cm}{$\tilde{\prod \limits_{ i: X_i \notin \mathscr{F}(x) }{}}$} \mathcal{P}_{\alpha_i} \Big{)} \mid \Psi_{in} \rangle}{\Big{\|} \Big{(} \raisebox{0.15cm}{$\tilde{\prod \limits_{ i: X_i \notin \mathscr{F}(x) }{}}$} \mathcal{P}_{\alpha_i} \Big{)} \mid \Psi_{in} \rangle \Big{\|}} \in \mathcal{H} \mbox{ ,}
\end{equation}
where the tilde on the product shall indicate a (partial) ``chronological order'' of the factors: whenever $X_i \in \mathscr{F}(X_j)$ the projector $\mathcal{P}_{\alpha_j}$ acts on the wave-function first, i.e.\! $\mathcal{P}_{\alpha_j}$ stands to the right of $\mathcal{P}_{\alpha_i}$. If $X_i\in \mathscr{F}(x)$ the corresponding projector $\mathcal{P}_{\alpha_i}$ is left out of the product.

But it turns out that a much more accurate description of wave-functions in a theory with Minkowski space-time background is to define them as functionals on the set of space-like hyper-surfaces of $\mathscr{M}$. In the case of unitarily evolving wave-functions this was already realized in 1946 by Tomonaga \cite{Tomonaga} (see also Schwinger \cite{Schwinger}). But also -- and in particular -- when sets of partially space-like separated measurements (or better and more general: collapse events) are considered, a transparent Lorentz invariant description of wave-function evolution can be built that way \cite{aa3, rGRWf, pointproc}. We will derive such a description in section \ref{functional} and apply it within the framework of a complete Lorentz invariant (but non-local) theory of wave-function collapse in section \ref{rgrwf}.

But first let us investigate the mentioned measurement strategy proposed by Aharonov and Albert.

\subsection{Aharanov \& Albert \label{AA}}

\vspace{0.5cm}

\begin{center}
 \textbf{Preliminary Remarks}
\end{center}

The result of the procedure proposed by Aharonov and Albert \cite{aa1, aa2}, which we shall develop and investigate now, will be (at least) the following: Suppose a two-particle spin-$\frac{1}{2}$-system is in the singlet state $\mid \Psi_- \rangle$; then suitable designed simultaneous (and arbitrarily short) local interactions of the singlet-particles with some probe system will enable us to verify that all three components of the total spin of the singlet particles are zero. In addition the procedure constitutes a non-demolition measurement\footnote{A non-demolition measurement \cite{braginsky1, braginsky2}) is at the end of the day an ideal projective quantum-measurement, i.e.\! one which leaves the state in an eigenstate of the operator corresponding to the measurement. Most realistic measurements are not of that kind, since e.g.\! particles are absorbed by detectors during measurement. The necessary and sufficient condition for a measurement of physical quantity $q$ to be non-demolition is that $[\mathscr{H},\hat{q}] \mid \Psi \rangle = 0$ holds, where $\mathscr{H}$ is the joint Hamiltonian of measuring device (e.g.\! probe particle) and object under study, $\hat{q}$ is the self-adjoint operator associated with quantity $q$ and $\mid \Psi \rangle$ is the initial wave-function (the usually applied condition $[\mathscr{H},\hat{q}]=0$ is only sufficient, not necessary for a quantum non-demolition measurement).} and it will leave the singlet-state untouched. \paragraph*{}

Because of the non-demolition property and the fact that the condition $\sigma^{tot}_x=\sigma^{tot}_y=\sigma^{tot}_z=0$ is a necessary and sufficient condition for the singlet-state, the authors speak of ``state verification of the singlet-state at some well defined instant of time''. Since we will encounter possible subtle circumstances under which despite the right pointer positions at the end of this procedure it will be difficult to conclude that the system is in that state (see chapter \ref{distfol}), I will try to be careful when using such expressions.    \paragraph*{}

Also the authors and others use often the phrase \textsl{``nonlocal measurement by purely local interactions''} \cite{breuer} or ``local measurement of non-local observables`` \cite{ghirardinonlocal}. I am ok with this formulation but it seems to me that they need some disambiguation: For example the total (classical) momentum of two spatially separated flying stones is simply the sum of the respective momenta and if we (classically) measure the momentum of each stone and combine the results we have performed a measurement of the nonlocal property ``total momentum'' of the system ``two stones``. The case we shall investigate in a moment will be similar in some sense, but the stones are now singlet particles and the measured quantity is the total spin; and there appears an important conceptual difference: This system is not separable (in contrast to the stones), i.e.\! we cannot simply measure all three components of the spin of each particle (that is indeed impossible at some well defined instant of time and performed in a row it would destroy the singlet state) and sum them up. The phrase ''nonlocal measurement'' or ''measurement of a nonlocal observable'' here shall capture exactly this distinctive feature: It provides a strategy to get access to a physical quantity which is intrinsically encoded in an entangled nonlocal wave-function and which cannot be reduced to the composition of local properties of the two particles. It might be interesting to anticipate already, that in the procedure we will construct now, the readout on one wing of the experiment will provide no relevant information whatsoever about the spin-$\frac{1}{2}$-system, only the composition of both results carries real physical meaning (in contrast to the stones...). \paragraph*{}

The procedure is a bit artificial, since the probe-system needs to be prepared in a non-normalizable state, a state of the kind considered by Einstein, Podolsky and Rosen in their famous paper \cite{EPR} (see footnote \ref{EPRstates}). But that is no big deficiency, since the conceptual implications of the result do not depend on artificial states. Essentially the same indirect measurement, only with normalizable probe-particle states (where each has three degrees of freedom), has been constructed by Ghirardi \cite{ghirardinonlocal}.    

 \begin{center}
 \textbf{Step \RM{1}: $\sigma^{(j)}_z$-measurement}
\end{center}

Consider a system of two spin-1/2-particles (j) (j=1,2) and for now suppose the particles are in one of the common eigenstates of the operators $\hat{\sigma}^{(j)}_z$. First we design a non-demolition measurement of the z-component of the Spin $\sigma^{(j)}_z$ of either of the particles. Think of the interaction with the measuring device as a (unitary) interaction between particle(j) with some probe particle (indirect measurement) which immediately after this interaction interacts again with some macroscopic apparatus in a specific way to yield the definite outcome of the measurement (position of a pointer...). The quantity of the probe to be measured will be $\pi_j$, which is the canonically conjugate (generalized) momentum of some internal variable $q_j$ of the probe. $q_j$ shall couple to the z-component of the spin of particle(j). Thus the interaction will be described by a Hamiltonian of the following form: 

\begin{equation} \label{AAHam}
\mathscr{H}^{\mbox{\tiny{\textsl{int}}}}_j=f_j(t) \: \hat{q}_j\hat{\sigma}^{(j)}_z \hspace{0.1cm} , \hspace{0.1cm} j\in\{1,2\} \hspace{0.3cm} ; \hspace{0.3cm} [\hat{q}_j,\hat{\pi}_j]=i 
\end{equation}
where $f_j(t) \thicksim \tilde{f}_j(t) \cdot \mathds{1}_{\{[t^{(j)}_1,t^{(j)}_2]\}}(t)$ (with indicator-function $\mathds{1}_{\{\Delta\}}(t)$ of interval $\Delta$ and an arbitrary (bounded) function $\tilde{f}_j(t)$ ) is a function which describes the time dependence and strength of the interaction; it has non-vanishing values only within the (arbitrarily short) time interval $[t^{(j)}_1,t^{(j)}_2]$. 

$\hat{\pi}_j$ is the self-adjoint operator corresponding to the physical quantity $\pi_j$. We assume that the free part of the Hamiltonian $\mathscr{H}_j=\mathscr{H}^{\mbox{\tiny{\textsl{free}}}}_j + \mathscr{H}^{\mbox{\tiny{\textsl{int}}}}_j$ does not contain operators which do not commute with $\hat{\pi}_j$, $\hat{q}_j$ or $\hat{\sigma}^{(j)}_z$, or at least that we can neglect the free dynamics (with these requirements the non-demolition measurement admittedly might have some contrived air about it).

In the Heisenberg-picture the time evolution of the operator $\hat{\pi}_j$ is then given by 
\begin{equation}\label{dpidt}
\frac{\partial{\hat{\pi}_j}}{\partial{t}}=\frac{1}{i}[\hat{\pi}_j,\mathscr{H}_j]=-f_j(t)\hat{\sigma}^{(j)}_z 
\end{equation}
 and for $\hat{\sigma}^{(j)}_z$ we have the desired non-demolition property
\begin{equation}\label{nondemz}
 \frac{\partial{\hat{\sigma}^{(j)}_z}}{\partial{t}}=\frac{1}{i}[\hat{\sigma}^{(j)}_z,\mathscr{H}_j]=0 \mbox{ .}
\end{equation}
Now we can easily integrate Heisenberg-equation \eqref{dpidt} and solve it for $\hat{\sigma}^{(j)}_z$. Therefore, given actual values $\tilde{\pi}_j$ (of the quantity $\pi_j$ corresponding to the operator $\hat{\pi}_j$) at some time $t<t^{(j)}_1$ (given by preparation) and at some time $t>t^{(j)}_2$ (given by measurement), we obtain the value of the z-component of the spin of particle(j):
\begin{equation}\label{sigmajz}
{\sigma}^{(j)}_z=\frac{{\pi}_j(t < t^{(j)}_1) - {\pi}_j(t > t^{(j)}_2)}{\int_{t^{(j)}_1}^{t^{(j)}_2}dt f_j(t)} \mbox{ .}
\end{equation}

%\vspace{0.5cm}

\begin{center}
 \textbf{Step \RM{2}: $\sigma^{(tot)}_z$-measurement}
\end{center}

Let`s now combine the two devices. The quantity we are interested in now is the z-component of the total spin $\sigma^{tot}_z=\sigma^{(1)}_z + \sigma^{(2)}_z$. Within this chapter we set $t^{(1)}_i\equiv t^{(2)}_i:=t_i, \quad i=1,2$ and $f_1(t)=f_2(t) := f(t)$, i.e.\! the two interactions of the spin-$\frac{1}{2}$-particles with the probe-particles have the same time-dependence and strength. As ``laboratory-frame`` we have obviously chosen a Lorentz-frame in which the two measurements are performed simultaneously. The Hamiltonian now reads
\begin{equation}\label{AAhamtot}
 \mathscr{H}=\mathscr{H}_1+\mathscr{H}_2 \mbox{ .}
\end{equation}

\textsl{And here comes the trick:} In order to ''detect'' the actual value of $\sigma^{tot}_z$ we bring the two probes together before the interaction with the spin-$\frac{1}{2}$-particles at some time $t_0 < t_1$ and prepare them in an (entangled) state in which the $q$'s and $\pi$'s take values such that\footnote{\label{EPRstates}Equations \eqref{prepAA} might look a bit strange at first glance but they correspond exactly to a preparation of the probe in a kind of state which was discussed in detail by Einstein, Podolski and Rosen in the famous EPR-paper: The (non-normalizable) state $\Psi(x_1,x_2)=\int dp e^{i(x_1-x_2+x_0)p}$, for which the values of position and momentum must fulfill constraints $i)$ $x_2=x_1+x_0$ and $ii)$ $p_1=-p_2$.}
\begin{equation}
\label{prepAA}
 q_1(t_0)-q_2(t_0)=0  \hspace{1cm} \mbox{ and } \hspace{1cm} \pi_1(t_0)+\pi_2(t_0)=0 \mbox{ .}
\end{equation}
If now after the interactions described by \eqref{AAHam} and \eqref{AAhamtot} at some time $t>t_2$ the actual values $\tilde{\pi}_1$ and $\tilde{\pi}_2$ of the generalized momenta are measured, equation \eqref{sigmajz} yields the value for $\sigma^{tot}_z$:
\begin{equation}\label{AAsigmaz}
 \sigma^{tot}_z=\sigma^{(1)}_z + \sigma^{(2)}_z=-\frac{\tilde{\pi}_1(t > t_2) + \tilde{\pi}_2(t > t_2)}{\int_{t_1}^{t_2}dt f(t)} \mbox{ .}
\end{equation}
Thus, finding the values of $\pi_1$ and $\pi_2$ after the interaction (by subsequent measurements of the two probes-particles) reveals the value of the total z-spin of the spin-$\frac{1}{2}$-system (when combined later). 

Note that (as already mentioned) either of the measured values $\tilde{\pi}_j$ alone does not provide any information about the total z-spin or about the z-spin of either of the two spin-$\frac{1}{2}$-particles: According to \eqref{sigmajz} (for the latter) it would be necessary to have the value of $\pi_j(t<t_1)$ (before the measurement) available to infer such information. But at this time the value of $q_1-q_2$ is already fixed to be zero and $\hat{\pi}_j$ does not commute with $\hat{q}_1-\hat{q}_2$, such that it is impossible to fix some value of $\pi_j(t<t_1)$. Only the combined measured values from both wings of the experiment can reveal something about the spin-$\frac{1}{2}$-system.  

If now the initial wave-function is the singlet state $\mid \Psi_- \rangle$ the above procedure will leave the system in that state, for the non-demolition property $[\mathscr{H},\hat{\sigma}^{tot}_z]=0$ is valid and $\mid \Psi_- \rangle$ is an eigenstate of $\hat{\sigma}^{tot}_z$. 

\begin{center}
 \textbf{Step \RM{3}: Verification of $\sigma^{tot}_x=\sigma^{tot}_y=\sigma^{tot}_z=0$}
\end{center}

Suppose the spin-$\frac{1}{2}$-system is in the singlet state $\mid \Psi_- \rangle$. Now we can extend the above procedure straight forwardly to get access to the other two components of the total spin. 

We need three pairs of probe-particles, say one with conjugated variables $q^x_j$ and $\pi^x_j$ associated with $\sigma^{(j)}_x$, one with conjugated variables $q^y_j$ and $\pi^y_j$ associated with $\sigma^{(j)}_y$ and one with conjugated variables $q^z_j$ and $\pi^z_j$ associated with $\sigma^{(j)}_z$ (''associated'' means that the $q$'s couple to the $\sigma$'s in the sense of \eqref{AAHam}). These variables should be prepared in initial-states analogous to \eqref{prepAA}, i.e.\! states determined by values which fulfill:
\begin{equation}\label{prepxyz}
 q^x_1(t_0)-q^x_2(t_0) \quad = \quad q^y_1(t_0)-q^y_2(t_0) \quad = \quad q^z_1(t_0)-q^z_2(t_0)\quad = \quad 0 
\end{equation}
 \begin{center}
  and
 \end{center} 
\begin{equation}
 \pi^x_1(t_0)+\pi^x_2(t_0) \quad = \quad \pi^y_1(t_0)+\pi^y_2(t_0) \quad = \quad \pi^z_1(t_0)+\pi^z_2(t_0) \quad = \quad 0 \mbox{ .}
\end{equation}
That this leads to a non-demolition measurement and that it leaves the singlet-state untouched is heuristically easy to see: $\mid \Psi_- \rangle$ is an common eigenstate of each of the operators $\hat{\sigma}^{tot}_x$, $\hat{\sigma}^{tot}_y$ as well as $\hat{\sigma}^{tot}_z$ (with eigenvalue zero in each case), such that each procedure alone (corresponding to one pair of interactions) leaves the singlet state undisturbed. Now we can simply perform the three pairs of interaction in a row (each will preserve the singlet state) and make the intermediate time arbitrarily short, such that in the limit of vanishing intermediate time we have a simultaneous measurement of $\sigma^{tot}_x$, $\sigma^{tot}_y$ and $\sigma^{tot}_z$ (with outcome zero in each case) which preserves the singlet state.

More rigorously we can observe that 
\begin{equation}
\frac{\partial{}}{\partial{t}} \: \hat{q}^{x_i}_j =\frac{1}{i}[\hat{q}^{x_i}_j,\mathscr{H}]=0, \hspace{0.5cm} x_i=x,y,z \quad ; \quad j=1,2
\end{equation}
and therefore we can use the constraints \eqref{prepxyz} to modify the corresponding operators in the interaction Hamiltonian
\begin{equation}\label{AAhamxyz}
 \begin{gathered}
  \mathscr{H}^{\mbox{\tiny{\textsl{int}}}} = f(t) \cdot \sum_{i=1}^3 \Big{(} \hat{q}^{x_i}_1\hat{\sigma}^{(1)}_{x_i}
+ \hat{q}^{x_i}_2\hat{\sigma}^{(2)}_{x_i} \Big{)} = \\ 
f(t) \cdot \sum_{i=1}^3 \frac{1}{2} \Big{(} (\hat{q}^{x_i}_1 + \hat{q}^{x_i}_2) \cdot (\hat{\sigma}^{(1)}_{x_i} + \hat{\sigma}^{(2)}_{x_i}) + (\hat{q}^{x_i}_1 - \hat{q}^{x_i}_2) \cdot (\hat{\sigma}^{(1)}_{x_i} - \hat{\sigma}^{(2)}_{x_i}) \Big{)}\overset{\eqref{prepxyz}}{=} \\ 
f(t) \cdot \sum_{i=1}^3 \frac{1}{2} \Big{(} (\hat{q}^{x_i}_1 + \hat{q}^{x_i}_2) \cdot (\hat{\sigma}^{(1)}_{x_i} + \hat{\sigma}^{(2)}_{x_i}) \Big{)} = \\
\frac{f(t)}{2}  \Big{(} (\hat{q}^x_1+\hat{q}^x_2) \cdot \hat{\sigma}^{tot}_x
+ (\hat{q}^y_1+\hat{q}^y_2) \cdot \hat{\sigma}^{tot}_y + (\hat{q}^z_1+\hat{q}^z_2) \cdot \hat{\sigma}^{tot}_z \Big{)} \mbox{ .}
 \end{gathered}
\end{equation}
Now (given that the free part of the Hamiltonian contains no \textsl{``bad``} operators) with \eqref{AAhamxyz} and with the help of the commutator-relations
\begin{equation}\label{commsigma}
 [\hat{\sigma}^{tot}_{x_i} , \hat{\sigma}^{tot}_{x_j}] = \frac{i}{2} \varepsilon_{ijk} \hat{\sigma}^{tot}_{x_k}
\end{equation}
 it is easy to see that 
\begin{equation}\label{nondem}
 [\hat{\sigma}^{tot}_x , \mathscr{H}] \mid \Psi_- \rangle = [\hat{\sigma}^{tot}_y , \mathscr{H}] \mid \Psi_- \rangle = [\hat{\sigma}^{tot}_z , \mathscr{H}] \mid \Psi_- \rangle = 0 \mbox{ .}
\end{equation}
Thus, given an initial singlet state, interaction \eqref{AAhamxyz} gives rise to a non-demolition measurement of all three components of the total spin $\boldsymbol{\sigma}^{tot}$. When the $\pi$'s are measured after the interaction, the actual values $\tilde{\pi}^{x_i}_j$ will yield
\begin{equation}
 \sigma^{tot}_{x_i} = -\frac{\tilde{\pi}^{x_i}_1(t > t_2) + \tilde{\pi}^{x_i}_2(t > t_2)}{\int_{t_1}^{t_2}dt f(t)} = 0 
\end{equation}
for all three components $x_i=x,y,z$.

\subsection{\label{timedisp}The Aharonov-Albert Procedure on $\mathscr{M}$} 
\vspace{.5cm}

\begin{center}
\textbf{State Description}
\end{center}

In order to see how the Aharonov-Albert measurement looks like from the viewpoint of a different Lorentz-frame we have to develop the description in the Schr\"odinger-picture first: Let us go back to the $\sigma^{tot}_z$-measurement (and drop again the index $z$ at the $\pi$'s and $q$'s) and consider the state of the combined probe-singlet system living on Hilbertspace
\begin{equation}
 \mathcal{H}=\mathcal{H}_{S_1} \otimes \mathcal{H}_{S_2} \otimes \mathcal{H}_{P_1} \otimes \mathcal{H}_{P_2} \mbox{ .}
\end{equation}
$\mathcal{H}_{P_i}$ denotes the Hilbertspace of probe-particle(i) and $\mathcal{H}_{S_i}\cong\mathbb{C}^2$ is the spin part of the Hilbert-space of singlet-particle(i) (as argued at the very beginning we neglect the spatial $L^2$-part of the singlet particles' Hilbertspace and pretend to consider particles on classical trajectories, which -- at this stage -- does not matter for the results we shall derive). 

\begin{center}
\textbf{The Probe}
\end{center}

Let us define the operators  

\begin{equation}
 \hat{q}_- := \hat{q}_1 - \hat{q}_2  \hspace{1cm} \mbox{ and } \hspace{1cm} \hat{\pi}_+ := \hat{\pi}_1 + \hat{\pi}_2 \mbox{ .}
\end{equation}
acting on $\mathcal{H}_P=\mathcal{H}_{P_1} \otimes \mathcal{H}_{P_2}$. Since these operators commute we can find a basis of $\mathcal{H}_P$ consisting of joint eigenstates of $\hat{q}_-$ and $\hat{\pi}_+$. Let us denote these states by $\mid q_-;\pi_+ \rangle$ defined by
\begin{equation}\label{pi+-}
 \hat{q}_-\mid q_-;\pi_+ \rangle = q_-\mid q_-;\pi_+ \rangle  \hspace{1cm} \mbox{ and } \hspace{1cm}  \hat{\pi}_+\mid q_-;\pi_+ \rangle=\pi_+\mid q_-;\pi_+ \rangle \mbox{ .}
\end{equation}
In this notation the initial-state of the probe defined by conditions \eqref{prepAA} can be written as 
\begin{equation}
 \mid \Phi_0 \rangle = \mid q_-=0 ; \pi_+=0 \rangle \mbox{ .}
\end{equation}
Another useful representation will be given by the joint eigenstates of the commuting operators $\hat{\pi}_1$ and $\hat{\pi}_2$: $\mid \pi_1;\pi_2 \rangle$ defined by
\begin{equation}
 \hat{\pi}_1 \mid \pi_1;\pi_2 \rangle = \pi_1 \mid \pi_1;\pi_2 \rangle  \hspace{1cm} \mbox{ and } \hspace{1cm}  \hat{\pi}_2 \mid \pi_1;\pi_2 \rangle = \pi_2 \mid \pi_1;\pi_2 \rangle \mbox{ .}  
\end{equation}
With the help of this representation we will be able to calculate the effect of unitary time-evolution operators due to interaction Hamiltonians of the form \eqref{AAHam}. 

Armed with this machinery we can write the initial state of the probe a bit more complicated as
\begin{equation}\label{ev}
 \mid \Phi_0 \rangle = \int d\pi_1 \int d\pi_2 \mid \pi_1 ; \pi_2 \rangle \langle \pi_1 ; \pi_2 \mid q_-=0 ; \pi_+=0 \rangle
\end{equation}
with the following selection rule for the matrix-elements appearing in the evolution:
\begin{equation}\label{select}
 \langle \pi_1 ; \pi_2 \mid q_-=0 ; \pi_+=0 \rangle = 0  \hspace{1cm} \mbox{ for } \hspace{1cm}  \pi_1 + \pi_2 \neq 0 \mbox{ .}
\end{equation}

\begin{center}
\textbf{Time-Evolution}
\end{center}

Since $[\mathscr{H}(t),\mathscr{H}(t')]=0$ the unitary time-evolution generated by interaction-Hamiltonian \eqref{AAHam} will act on the state by the operation 
\begin{equation}\label{AAint}
 \mathcal{U}_j=e^{-i F_j \hat{q}_j \hat{\sigma}^{(j)}_z}
\end{equation}
for times $t > t^{(j)}_2$, where $F_j:=\int_{t^{(j)}_1}^{t^{(j)}_2} f_j(t) dt$. In the following $f_1(t)=f_2(t)$ will not be true anymore, but we shall set $F_1=F_2=:F$. 
 
Since the time-interval of interaction $\epsilon= t^{(j)}_2 - t^{(j)}_1$ may be chosen arbitrarily small, we will pretend in the following that such interactions take place at some well defined instant of time, e.g.\! at time $t_{\alpha}$.  

A helpful relation for evaluating the effect of time-evolution operators as considered above, is the fact, that one variable out of a pair of canonically conjugated variables is the generator of translations with respect to the other one, respectively: For eigenstates $\mid \pi \rangle$ corresponding to eigenvalue $\pi$ of some operator $\hat{\pi}$ we have
\begin{equation}\label{trans}
 [\hat{q},\hat{\pi}]=i \hspace{1cm} \Longrightarrow \hspace{1cm} e^{\pm i \xi \hat{q}} \mid \pi \rangle = \mid \pi \pm \xi \rangle \mbox{ .} 
\end{equation}

\begin{center}
\textbf{Time-displaced Interactions}
\end{center}

\begin{figure}[htbp]
\centering
\includegraphics[scale=0.16]{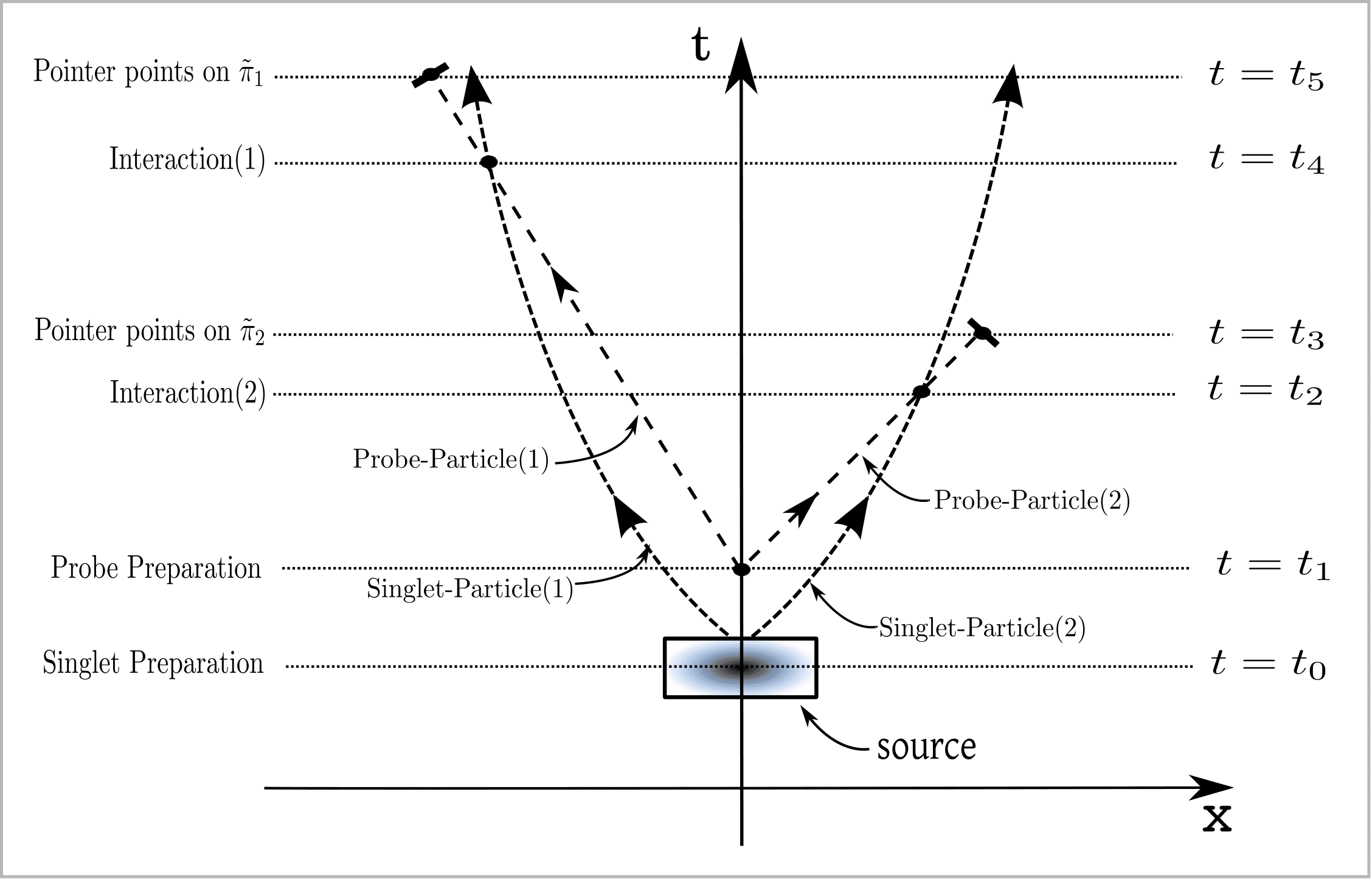}
\caption{\textbf{The Aharonov-Albert procedure with time displaced interactions}}
\label{stateAA}
\end{figure}

As in a different frame the two interactions of the probe with the singlet-particles (and the subsequent measurement of the generalized momenta of the probe-particles) will not be simultaneous anymore we shall now describe the measurement with time-displaced interactions as illustrated in figure \ref{stateAA}. Primary we are interested in the state between time $t_3$ of measurement of probe-particle(2) (after interaction with singlet-particle(2) ) and time $t_4$ of interaction of probe-particle(1) with singlet-particle(1). The initial state of the complete system after preparation (i.e.\! for times $t_1<t<t_2$) is the product wave-function 

\begin{equation}
 \mid \Psi_{in} \rangle = \mid \Psi_{(t_1<t<t_2)} \rangle = \mid \Phi_0 \rangle \otimes \mid \Psi_- \rangle \quad \in \quad \mathcal{H} \mbox{ .}
\end{equation}
 
According to \eqref{AAint} the interaction $\mathscr{H}^{\mbox{\tiny{\textsl{int}}}}_2$ given by \eqref{AAHam} will transform this state into
\begin{equation}
\begin{gathered}\label{t2t3}
 \mid \Psi_{(t_2<t<t_3)} \rangle = \; \mathcal{U}_2 \mid \Psi_{in} \rangle= \frac{1}{\sqrt{2}} \bigg{(} e^{-i F \hat{q}_2 \hat{\sigma}^{(2)}_z} \mid \Phi_{0} \rangle \otimes \mid \uparrow \downarrow \rangle - e^{-i F \hat{q}_2 \hat{\sigma}^{(2)}_z} \mid \Phi_{0} \rangle \otimes \mid \downarrow \uparrow \rangle \bigg{)} = \\
\frac{1}{\sqrt{2}}\bigg{(} \int d\pi_1 \int d\pi_2 \; e^{i F \hat{q}_2} \mid \pi_1 ; \pi_2 \rangle \langle \pi_1 ; \pi_2 \mid q_-=0 ; \pi_+=0 \rangle \otimes \mid \uparrow \downarrow \rangle - \\ 
\int d\pi_1 \int d\pi_2 \; e^{-i F \hat{q}_2} \mid \pi_1 ; \pi_2 \rangle \langle \pi_1 ; \pi_2 \mid q_-=0 ; \pi_+=0 \rangle \otimes \mid \downarrow \uparrow \rangle \bigg{)} \overset{\eqref{trans}}{=} \\
\frac{1}{\sqrt{2}}\bigg{(} \int d\pi_1 \int d\pi_2 \; \mid \pi_1 ; \pi_2 + F \rangle \langle \pi_1 ; \pi_2 \mid q_-=0 ; \pi_+=0 \rangle \otimes \mid \uparrow \downarrow \rangle - \\ 
\int d\pi_1 \int d\pi_2 \; \mid \pi_1 ; \pi_2 - F \rangle \langle \pi_1 ; \pi_2 \mid q_-=0 ; \pi_+=0 \rangle \otimes \mid \downarrow \uparrow \rangle \bigg{)}
\end{gathered}
\end{equation}
and with substitution of integration-variables we can write this as
\begin{equation}
 \begin{gathered}\label{t23}
  \mid \Psi_{(t_2<t<t_3)} \rangle = \frac{1}{\sqrt{2}}\bigg{(} \int d\pi_1 \int d\pi_2 \; \mid \pi_1 ; \pi_2 \rangle \langle \pi_1 ; \pi_2 - F \mid q_-=0 ; \pi_+=0 \rangle \otimes \mid \uparrow \downarrow \rangle - \\ 
\int d\pi_1 \int d\pi_2 \; \mid \pi_1 ; \pi_2 \rangle \langle \pi_1 ; \pi_2 + F \mid q_-=0 ; \pi_+=0 \rangle \otimes \mid \downarrow \uparrow \rangle \bigg{)}
 \end{gathered}
\end{equation}
Now at time $t_3$ a measurement of probe-quantity $\pi_2$ is performed (see figure \ref{stateAA}) and the readout might be the value $\tilde{\pi}_2$. If we project the state \eqref{t23} onto the state $\mid \tilde{\pi}_2 \rangle \in \mathcal{H}_{P_2}$ we find with the help of selection rule \eqref{select} and correct normalization the state of the remaining particles of interest $\mid \Psi_{(t_3<t<t_4)} \rangle \in \mathcal{H}_{S_1} \otimes \mathcal{H}_{S_2} \otimes \mathcal{H}_{P_1}$ to be 
\begin{equation}\label{t3t4}
\mid \Psi_{(t_3<t<t_4)} \rangle = \frac{1}{\sqrt{2}}\bigg{(}  \mid \pi_1 = -\tilde{\pi}_2 + F \rangle \otimes \mid \uparrow \downarrow \rangle - \mid \pi_1 = -\tilde{\pi}_2 - F \rangle \otimes \mid \downarrow \uparrow \rangle \bigg{)} \mbox{ .}
\end{equation}
Obviously the singlet-system is not in the singlet state in the intermediate time between the interactions, but in an entangled state with the probe!

At time $t_4$ then singlet-particle(1) interacts with probe-particle(1) and the resulting state will be
\begin{equation}
\begin{gathered}
 \mid \Psi_{t_4<t<t_5} \rangle = \mathcal{U}_1 \mid \Psi_{(t_3<t<t_4)} \rangle = \\
\frac{1}{\sqrt{2}}\bigg{(} e^{-i F \hat{q}_1} \mid \pi_1 = -\tilde{\pi}_2 + F \rangle \otimes \mid \uparrow \downarrow \rangle - e^{i F \hat{q}_1} \mid \pi_1 = -\tilde{\pi}_2 - F \rangle \otimes \mid \downarrow \uparrow \rangle \bigg{)} \\
=\frac{1}{\sqrt{2}} \bigg{(} \mid \uparrow \downarrow \rangle - \mid \downarrow \uparrow \rangle \bigg{)} \otimes \mid \pi_1 = -\tilde{\pi}_2 \rangle = \\ 
\mid \Psi_- \rangle \otimes \mid \pi_1 = -\tilde{\pi}_2 \rangle
\end{gathered}
\end{equation}
 
Thus when at time $t_5$ the generalized momentum of probe-particle(1) is measured it will be found to have the value $\pi_1=-\tilde{\pi}_2$; and therefore with \eqref{AAsigmaz} the experimenter has found that $\sigma^{tot}_z=0$. The resulting state for times $t>t_5$ (or actually $t>t_4$) is the singlet-state again. But the measurement does not leave the singlet state untouched anymore: \textsl{The singlet-state is disturbed in the intermediate time interval $t_2<t<t_4$ between the interactions with the probe}. During that interval the actual state is characterized by entanglement between the two spin-$\frac{1}{2}$-particles and the probe. 

Again this procedure can be combined with analogous procedures for the other two components of the total spin and the finding will be the same: Between the relevant interactions the state will be disturbed and entangled with the probe, the readout of the probe measurements will yield that $\sigma^{tot}_z=\sigma^{tot}_x=\sigma^{tot}_y=0$ and after the interactions the spin-$\frac{1}{2}$-particles will be in the singlet state again (see \cite{breuer}).

\begin{center}
\textbf{Relativity}
\end{center}
      
Let us relate now these calculations to the title of this section. As mentioned already, in some Lorentz-transform of the Aharonov-Albert measurement-scenario of section \ref{AA} the simultaneous interactions will not be simultaneous anymore. So except from the fact that we did not consider Lorentz-transformed states, the just investigated process (displayed in figure \ref{stateAA}) corresponds to the Lorentz-transformed process of section \ref{AA} (with respect to some particular frame). If the reader is afraid of possible conceptual relevance of the transformed states, she should notice that we (theoretically) can also perform this procedure with states, which do not transform under Lorentz transformations, e.g.\! by exchanging the singlet spin-$\frac{1}{2}$-system with a corresponding singlet isospin-$\frac{1}{2}$-system \cite{ghirardilessons, ghirardinonlocal}.

Thus we can literally translate the relevant features of our finding for the Aharonov-Albert procedure for time-displaced interactions to some Lorentz-transform of the original procedure of section \ref{AA}. In particular in a different Lorentz-frame the relevant simultaneous interactions will not be simultaneous anymore, the singlet-state will be disturbed in the intermediate time range and the out-coming state will be the singlet-state again.   

\subsection{Suggestion: No Single Covariant State History}

In the following two sections we shall infer some consequences from the possibility of ``non-local measurements'' of the kind we have investigated in the last sections, for theories which claim to get along without employing distinguished space-like structures on Minkowski space-time. 

In the case of such theories (which apparently have a \textsl{high degree of relativistic compatibility}) some simultaneity-slice ($t=const$) in the laboratory frame -- in which the pairs of interactions of the Aharonov-Albert procedure are simultaneous -- has the same physical significance as any other space-like hyper-surface. We can assign a wave-function to each simultaneity-slice within this frame (as we can assign a wave-function to each space-like hyper-surface) and the \textsl{instantaneous}\footnote{The term \textit{``instantaneous``} is very important here: We will see in section \ref{distfol} that the right pointer-positions at the end of the Aharonov-Albert experiment (i.e.\! $\tilde{\pi}^{x_i}_1=-\tilde{\pi}^{x_i}_2$) do not allow to conclude in general, that the system is in the singlet state subsequent to the measurement, if the pairs of interactions are not simultaneous. This holds also for theories without distinguished space-like structures on space-time. In theories with a preferred foliation even \textsl{instantaneous} verification of $\boldsymbol{\sigma}^{tot}=0$ does not allow such conclusion.} verification of $\sigma^{tot}_x=\sigma^{tot}_y=\sigma^{tot}_z=0$ for the total spin of the two-particle spin-$\frac{1}{2}$-system can indeed be seen as a state-verification of the singlet state (which is uniquely determined by these relations), at least on the simultaneity-slices of the laboratory-frame subsequent to the interactions. This fact opens the door for a special kind of continuous measurement, which is called \textsl{``monitoring of the state history''} and which is extensively discussed by Aharonov and Albert. I will now give a survey of this issue and then discuss the consequences. 

\begin{center}
 \textbf{Monitoring}
\end{center}

The monitoring procedure is simply the successive performance of one and the same non-demolition measurement, which will (according to the non-demolition property) yield the same outcome in every run. The intermediate time between to runs of the experiment can be chosen arbitrarily short, in principle. Thus, the succession of ``photographs of the singlet state'' (in the sense described in the last paragraph) turn into a film of the (not very thrilling) story of the singlet state in the limit of vanishing intermediate time.   

\begin{figure}[htbp]
\centering
\includegraphics[scale=0.6]{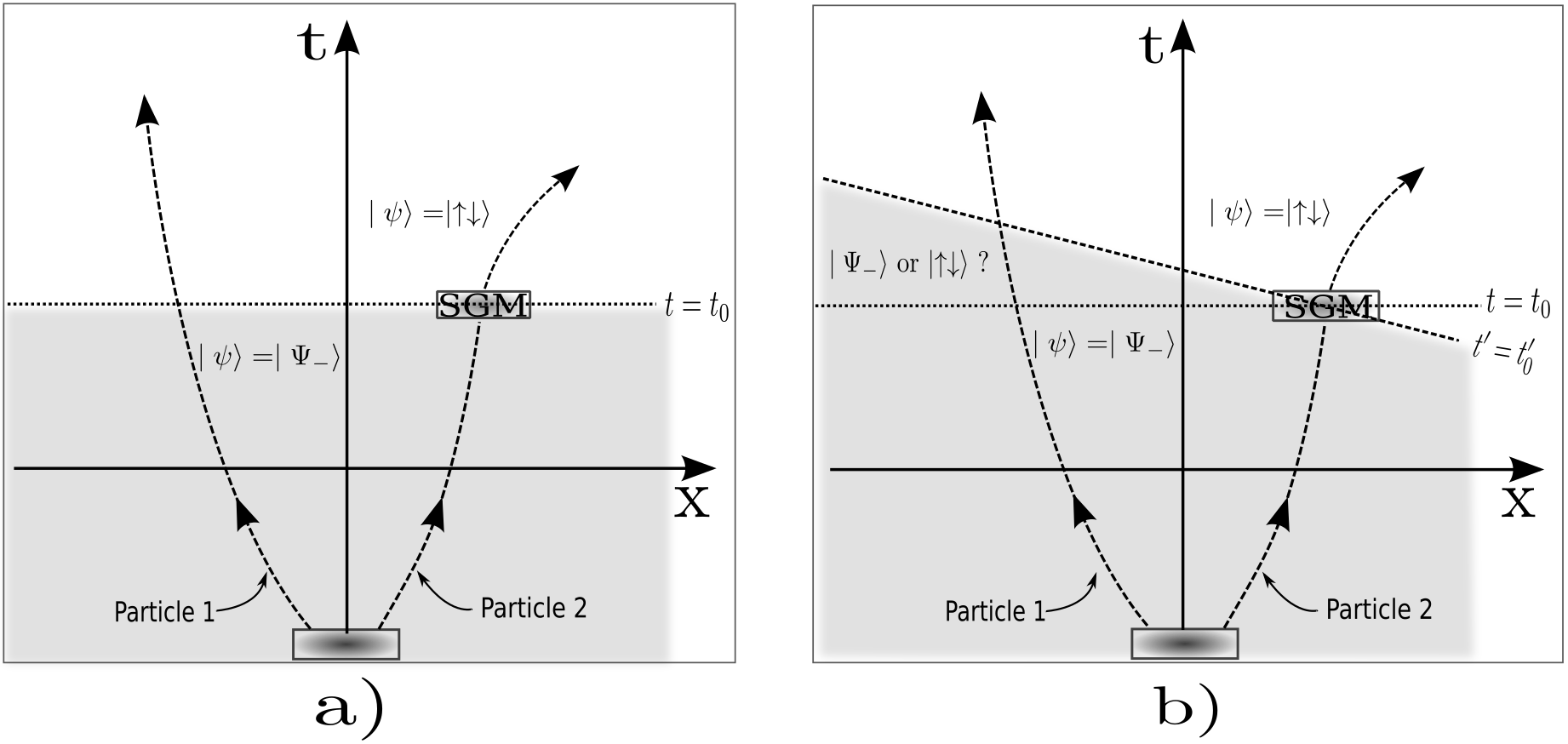}
\caption{\textbf{Inconsistent State Histories \RM{1}:} $\boldsymbol{a)}$ Collapse of the singlet state from the viewpoint of Lorentz-frame $K$. $\boldsymbol{b)}$ The hyperplane along which the wave-function collapses in frame $K'$ (again from the viewpoint of frame $K$). In each frame the state history can be monitored in principle, but the two state histories are not consistent with each other.}
\label{nosinglehist1}
\end{figure}	 

\begin{center}
 \textbf{Monitoring in Different frames}
\end{center}

Now suppose at some time $t=t_0$ the z-component of the spin of particle(2) is measured (e.g.\! at some SGM) and found to have the value -1. This means that the singlet state $\mid \Psi_- \rangle$ collapses at time $t=t_0$ and the resulting state will be $\mid \uparrow \downarrow \rangle$. If the history of the singlet state is monitored, the collapse along the $t=t_0$-hyperplane (at least at the two relevant (distinct) points laying on it, which is sufficient) of the ``laboratory frame`` $K$ can be recorded in principle\footnote{There is a certain uncertainty in recording the collapse, for there is non-vanishing probability to produce the singlet state again out of the state $\mid \uparrow \downarrow \rangle$ and thereby get the right pointer positions, i.e.\! $\boldsymbol{\sigma}^{tot}=0$. But of conceptual relevance is only the fact, that it is possible in principle to record the collapse non-locally and it is not essential that the recording might fail sometimes.} (see figure \ref{nosinglehist1} $a)$). If the state history is not monitored in frame $K$, but in another frame $K'$, the collapse can be recorded in that frame, say along the $t'=t'_0$-hyperplane, which crosses the measurement-event and which is different from the $t=t_0$-hyperplane as illustrated in figure \ref{nosinglehist1} $b)$. If we compare these two histories, this means in particular that there is a region in which particle(1) is already in the collapsed state $\mid \uparrow \downarrow \rangle$ according to the state history which might be monitored in the $K$-frame, whereas it is still in the singlet state $\mid \Psi_- \rangle$ according to the history which might be recorded in frame $K'$. 

According to the type of theory we are considering right now there is no distinguishing structure on space-time which prefers one of the two respective contradicting state histories with respect to the other one. So the urgent question arises what might happen if the state history is monitored in both frames at once. 

\begin{center}
 \textbf{Monitoring in Different Frames at Once}
\end{center}

\begin{figure}[htbp]
\centering
\includegraphics[scale=0.17]{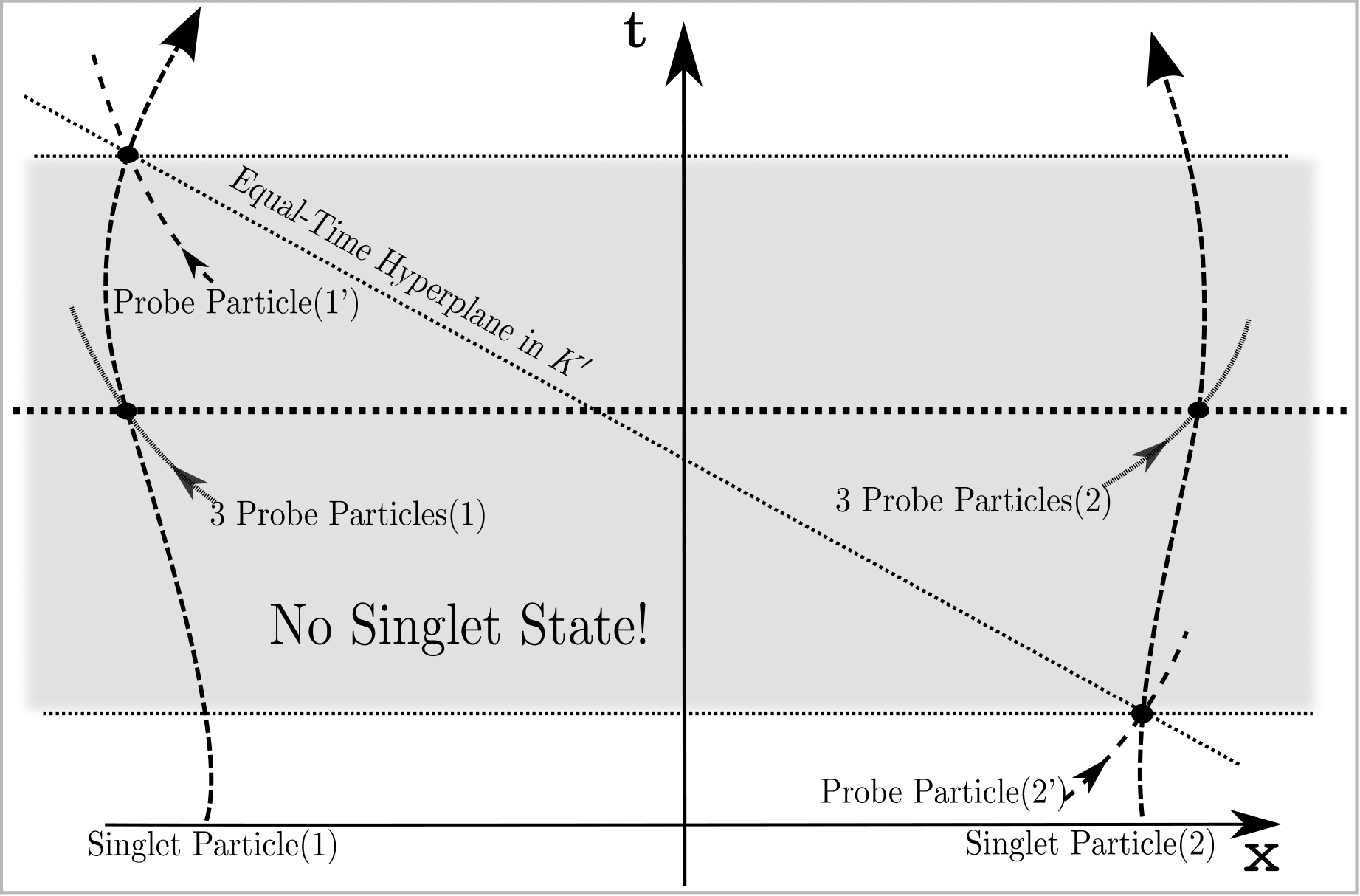}
\caption{\textbf{Disturbance:} The Aharonov-Albert experiment performed in frame $K'$ disturbs the singlet state in frame $K$. In the latter frame the wave-function is entangled with the $K'$-probe-particles in the intermediate time (shaded region) between the interactions of the spin-$\frac{1}{2}$-particles with the primed probe particles. A second Aharonov-Albert experiment performed in $K$ will not be a non-demolition measurement in the disturbed state anymore. Hence, the singlet-state history cannot be monitored in two different Lorentz-frames at once. [For simplicity we consider only the $\sigma^{tot}_z$-measurement in $K'$ (two primed probe particles), but the full procedure in $K$ given by Hamiltonian \eqref{AAhamxyz} (six probe particles).]}
\label{simmonitor}
\end{figure}

Heuristically it is easy to see that this is actually impossible: Consider one run of the Aharonov-Albert experiment performed in frame $K'$ from the viewpoint of frame $K$ (figure \ref{simmonitor}). For simplicity consider only the $\sigma^{tot}_z$-measurement performed in $K'$ (in case of the whole $\boldsymbol{\sigma}^{tot}$ measurement the entangled state would only be more complicated and the result is essentially the same), such that we can make use of the detailed analysis we made in the last chapter. As encountered there, the two interactions of the probe-particles with the spin-$\frac{1}{2}$-particles are not simultaneous in $K$ and the singlet state is disturbed in the intermediate time interval between the interactions. During that time the wave-function is an entangled state of system and probe, given by \eqref{t3t4}. If now the Aharonov-Albert procedure (consider now the full procedure given by interaction-Hamiltonian \eqref{AAhamxyz}) is performed also in $K$ within that time interval (see figure \ref{simmonitor}), non-demolition conditions \eqref{nondem} will not be fulfilled anymore for the initial state of that measurement: If for example the commutator $[\hat{\sigma}^{tot}_y , \mathscr{H}] \overset{\eqref{AAhamxyz} \& \eqref{commsigma}}{\thicksim} a \cdot \hat{\sigma}^{tot}_x + b \cdot \hat{\sigma}^{tot}_z$ (where $a$ and $b$ contain operators belonging to the probe-system of frame $K$) acts on state \eqref{t3t4}, the result will not be zero anymore:

\begin{equation}
\begin{gathered}
{[\hat{\sigma}^{tot}_y , \mathscr{H} ]} \mid \Psi_{(t_3<t<t_4)} \rangle \thicksim \\
\Big{(} a \cdot \hat{\sigma}^{tot}_x + b \cdot \hat{\sigma}^{tot}_z \Big{)} \frac{1}{\sqrt{2}}\bigg{(}  \mid \pi'_1 = -\tilde{\pi}'_2 + F \rangle \otimes \mid \uparrow \downarrow \rangle - \mid \pi'_1 = -\tilde{\pi}'_2 - F \rangle \otimes \mid \downarrow \uparrow \rangle \bigg{)} = \\
 \frac{a}{\sqrt{2}}\bigg{(}  \mid \pi'_1 = -\tilde{\pi}'_2 + F \rangle \otimes \hat{\sigma}^{tot}_x \mid \uparrow \downarrow \rangle - \mid \pi'_1 = -\tilde{\pi}'_2 - F \rangle \otimes \hat{\sigma}^{tot}_x \mid \downarrow \uparrow \rangle \bigg{)} \overset{\eqref{sigmatrans} \mbox{\tiny{ with }} \varphi=0, \theta=\frac{\pi}{2}}{=} \\
\frac{a}{\sqrt{2}}\bigg{(}  \mid \pi'_1 = -\tilde{\pi}'_2 + F \rangle \otimes \hat{\sigma}^{tot}_x \Big{(} \mid \uparrow \uparrow \rangle_x - \mid \downarrow \uparrow \rangle_x + \mid \uparrow \downarrow \rangle_x -\mid \downarrow \downarrow \rangle_x \Big{)} - \\
\mid \pi'_1 = -\tilde{\pi}'_2 - F \rangle \otimes \hat{\sigma}^{tot}_x \Big{(} \mid \uparrow \uparrow \rangle_x + \mid \downarrow \uparrow \rangle_x - \mid \uparrow \downarrow \rangle_x -\mid \downarrow \downarrow \rangle_x \Big{)} \bigg{)} = \\
\frac{2a}{\sqrt{2}} \bigg{(}  \mid \pi'_1 = -\tilde{\pi}'_2 + F \rangle \otimes \Big{(} \mid \uparrow \uparrow \rangle_x + \mid \downarrow \downarrow \rangle_x \Big{)} - \mid \pi'_1 = -\tilde{\pi}'_2 - F \rangle \otimes \Big{(} \mid \uparrow \uparrow \rangle_x + \mid \downarrow \downarrow \rangle_x \Big{)} \bigg{)} = \\
\sqrt{2} a \Big{(}  \mid \pi'_1 = -\tilde{\pi}'_2 + F \rangle  - \mid \pi'_1 = -\tilde{\pi}'_2 - F \rangle \Big{)} \otimes \Big{(} \mid \uparrow \uparrow \rangle_x + \mid \downarrow \downarrow \rangle_x \Big{)} \neq 0 \mbox{ ,}
\end{gathered}
\end{equation}
where the primed $\pi$'s are the generalized momenta of the probe-particles belonging to the measurement in the primed frame.      

This suggests that it is indeed not possible to monitor the state history in two different frames at once: The measurements performed in two different frames at once would disturb one another and thereby destroy the singlet state. If the state evolution corresponding to the scheme depicted in figure \ref{simmonitor} is explicitly calculated (which is a bit more extensive, but straight forward) the destruction of the singlet state (in general, i.e.\! apart from some finite probability of ``no destruction'') appears explicitly, of course. 

Thus the problem we are faced with is the following: \textsl{If we want to omit the utilization of distinguished intrinsic space-time structures in order to explain (theoretically) possible experimental results (see section \ref{distfol}) we have to account for respective contradicting state histories, where each can be verified experimentally in principle, but not both at once.} 

\begin{center}
 \textbf{Illustration of the Problem}
\end{center}

\begin{figure}[htbp]
\centering
\includegraphics[scale=0.15]{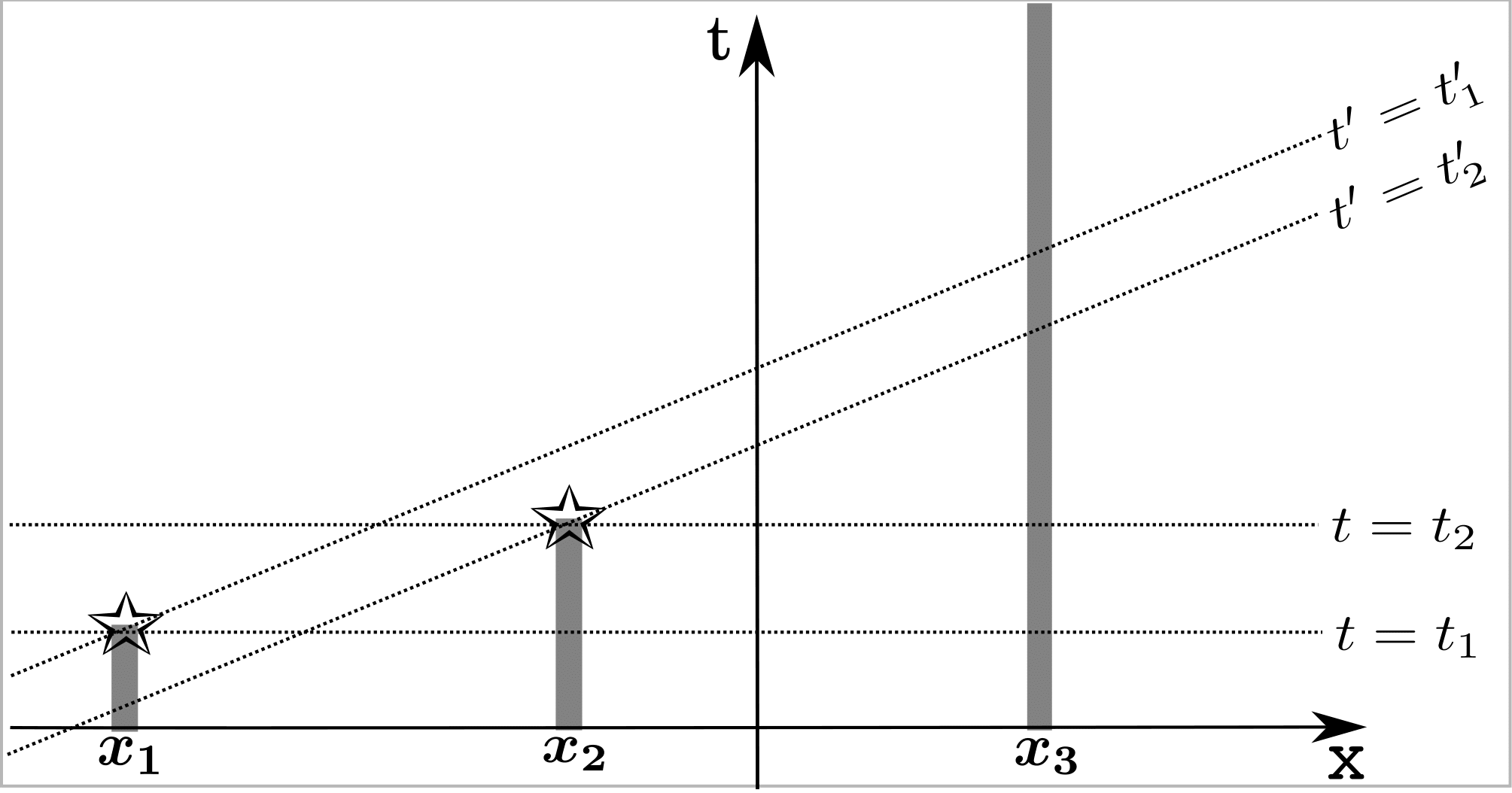}
\caption{\textbf{Inconsistent State Histories \RM{2}:} State history of a particle whose initial wave-function has support only in three spatial regions about the locations $\boldsymbol{x_1}$, $\boldsymbol{x_2}$ and $\boldsymbol{x_3}$. At time $t_1$ some particle-detector at $\boldsymbol{x_1}$ yields a negative result and collapses the wave-function; at time $t_2$ the same happens at $\boldsymbol{x_2}$. This state history is not consistent with the state history from the viewpoint of the indicated primed frame.}
\label{nosinglehist2}
\end{figure}

In order to highlight this point in a picturesque way, let me give one last simple example \cite{aa3}: Consider a one-particle wave-function, where the potential in the Hamiltonian allows only non-vanishing $\Psi$ within three small (but still bigger than the Compton-wavelength) distinct regions regions of space, say near the points $\boldsymbol{x_1}$, $\boldsymbol{x_2}$ and $\boldsymbol{x_3}$, respectively (see figure \ref{nosinglehist2}). Suppose now it would be possible to monitor the state history of this state.

Say at some time $t=t_1$ in the depicted frame $K$ a particle detector crosses the region around $\boldsymbol{x_1}$ and the detection is negative such that the support of the wave-function vanishes at $\boldsymbol{x_1}$. The same goes for the region around $\boldsymbol{x_2}$ at time $t=t_2>t_1$, where $(t_1,\boldsymbol{x_1})$ is supposed to be space-like with respect to $(t_2,\boldsymbol{x_2})$. Thus we can find some frame $K'$ in which the wave-function around $\boldsymbol{x_2}$ vanishes prior to the vanishing of the wave-function around $\boldsymbol{x_1}$, i.e in which the Lorentz-transformed (negative) measurement-events $(t'_1,\boldsymbol{x_1}')$ and $(t'_2,\boldsymbol{x_2}')$ fulfill $t'_2<t'_1$ (also indicated in figure \ref{nosinglehist2}).

Let us denote by $\mid \boldsymbol{x_i} \rangle$ the wave-function with support only in a small region around $\boldsymbol{x_i}$. Then the state-history which might be monitored in $K$ would look like (for simplicity let us drop the normalization)
\begin{align}
 & \mid \Psi_{(t<t_1)} \rangle =  \mid \boldsymbol{x_1} \rangle + \mid \boldsymbol{x_2} \rangle + \mid \boldsymbol{x_3} \rangle \\
& \mid \Psi_{(t_1<t<t_2)} \rangle  =  \mid \boldsymbol{x_2} \rangle + \mid \boldsymbol{x_3} \rangle \\
& \mid \Psi_{(t>t_3)} \rangle = \mid \boldsymbol{x_3} \rangle \mbox{ .}
\end{align}
On the other hand, monitored in $K'$ the story would look like
\begin{align}
 & \mid \Phi_{(t'<t'_2)} \rangle =  \mid \boldsymbol{x_1}' \rangle + \mid \boldsymbol{x_2}' \rangle + \mid \boldsymbol{x_3}' \rangle \\
& \mid \Phi_{(t'_2<t'<t'_1)} \rangle  =  \mid \boldsymbol{x_1}' \rangle + \mid \boldsymbol{x_3}' \rangle \\
& \mid \Phi_{(t'>t'_1)} \rangle = \mid \boldsymbol{x_3}' \rangle \mbox{ .}
\end{align}	

Observe that ``the system will realize the state`` $\mid \boldsymbol{x_2} \rangle + \mid \boldsymbol{x_3} \rangle$ sometime within its history in $K$ but apparently the Lorentz-transform of that state never appears in the history which might be monitored in $K'$. And the same goes the other way around for the state $\mid \boldsymbol{x_1}' \rangle + \mid \boldsymbol{x_3}' \rangle$ which appears in the state history in $K'$ but the transformed state never appears in $K$. 

To begin with, this strongly suggests a violation of fundamental principles of relativity (Lorentz invariance of the state history). And of course, the theoretical description alone suggests such violation, regardless of whether the histories can be monitored experimentally. But the (theoretical) possibility of monitoring inconsistent histories of non-local states experimentally (as we have extensively argued for the singlet state) highlights the necessity of reasoning a little bit deeper, whether a description can be given, which reconciles the (experimentally well confirmed) non-local action at a distance of quantum theory with the (experimentally well confirmed) relativistic space-time structure.

\begin{center}
 \textbf{Conclusion}
\end{center}

The above examples have shown that it is not possible in general to assign a single Lorentz invariant state history to a quantum mechanical system if we are adamant in avoiding the utilization of preferred structures of space-time. Therefore we have two possibilities at hand to come to a reconciliatory solution: 

Either we develop a consistent description of state evolution which also accounts for the possible coexistence of inconsistent state histories in different frames. This implies to assign some strange and unfamiliar properties to the wave-function and its evolution; in particular ontological interpretation of the wave-function (as something which is there...) becomes impossible. Its status must be reduced then to a purely nomological object (a part of the law which describes the dynamics of the primitive ontology). We shall call such state description \textit{Solution \RM{1}} which will be developed in the next chapter.

Or we employ distinguished space-like structures to accounts for the non-local connections of space-like separated events inherent in quantum theory. We shall work out in section \ref{distfol} that such a solution (which I'll call \textit{Solution \RM{2}}) is indeed well designed to solve the problem we are faced with.

\subsection{\label{functional}\textit{Solution \texorpdfstring{\RM{1}}{}}: Wave-Functionals \& Collapse Along Arbitrary Space-Like Hypersurfaces}

After their first two papers \cite{aa1, aa2} on that issue, in which Aharonov and Albert investigated the problems (without finding a solution) we have developed so far, they proposed a new kind of state description in a third paper \cite{aa3}, which is appropriate to account for wave-function collapse in Minkowski space-time in an unambiguous and Lorentz invariant way. In 2006 -- apparently independent of the investigations of Aharonov and Albert -- Roderich Tumulka developed essentially the same relativistic description of wave-function collapse in order to define a relativistic law for the dynamics of the primitive ontology in a relativistic collapse model \cite{rGRWf}.

The key is to treat wave-functions as functionals on the set of space-like hyper-surfaces of space-time. An appropriate dynamics has to be developed, such that ''collapse causing events`` enter the description in a transparent, Lorentz invariant and consistent way. But before we come to collapse let us briefly develop an appropriate description for unitary time-evolution.  

\begin{center}
 \textbf{No Collapse: Unitary Evolution}
\end{center}

Aharonov and Albert \cite{aa3} used the formalism developed by Tomonaga \cite{Tomonaga} to account for unitary time evolution. In Tumulka's relativistic GRW model \cite{rGRWf} the underlying equation is the multi-time Dirac-equation \cite{Dirac} such that it seems natural there to go a somewhat more direct way. But the result will be essentially the same: Unitary time evolution is generalized, such that the wave-function can be treated as a functional on the set of space-like hyper-surfaces. The description enables us to calculate the wave-function associated with some arbitrary space-like hyper-surface $\Sigma$, given some ''initial'' wave-function associated with another arbitrary space-like hyper-surface $\Sigma_0$, and given that no events causing wave-function reduction (like measurement events) lie in the space-time volume enclosed by $\Sigma_0$ and $\Sigma$. \paragraph*{} 

\textbf{Tomonaga} \cite{Tomonaga}: In order to build up the ``quantum theory of wave fields`` on a basis which does not rely on the choice of a particular frame of reference, Tomonaga generalized the description of time evolution of the wave-function $\Psi$. The time evolution is generated by some Hamilton density $\mathscr{H}(x)$ (the interaction part in the interaction picture). By this time evolution the wave-function is a function of an absolute time $\Psi=\Psi(t)$. Motivated by Dirac's multi time theory \cite{Dirac}, Tomonaga generalized $\Psi$ -- as a function of $t$ -- to a functional $\Psi[t(\v{x})]$ of functions $t(\v{x})$ which constitute arbitrary space-like hypersurfaces on space-time.     

Consider an arbitrary function $\tilde{t}:\mathbb{R}^3\rightarrow\mathbb{R}$ of space with the only constraint that each pair of space-time points $x=(t,\boldsymbol{x})$ and $y=(s,\boldsymbol{y})$ with the properties $t=\tilde{t}(\boldsymbol{x})$ and $s=\tilde{t}(\boldsymbol{y})$ is space-like separated\footnote{To be exact, we have to require in addition the correct transformation under Lorentz transformations for $\tilde{t}$ , i.e.\! it transforms such that $(\tilde{t}(\boldsymbol{x}),\boldsymbol{x})$ is a Lorentz-vector.}. $\tilde{t}(\boldsymbol{x})$ was called \textit{``local time``} by Stueckelberg \cite{Stueckelberg}. We shall call the graph $(\tilde{t}(\boldsymbol{x}),\boldsymbol{x})$ of such a function a space-like leaf $\Sigma$ on space-time. Now consider the functional $\Psi[\tilde{t}(\boldsymbol{x})]$ which is a solution to the (infinitely many) variational equations
\begin{equation}\label{tomo1}
 i \frac{\delta}{\delta{\tilde{t}(\boldsymbol{x}_0)}}\Psi[\tilde{t}(\boldsymbol{x})]=\mathscr{H}\Psi[\tilde{t}(\boldsymbol{x})] \mbox{ ,}
\end{equation}
where the local Hamilton-density operator $\mathscr{H}=\mathscr{H}(x)$ is restricted to the subset of space-time given by $t=\tilde{t}(\boldsymbol{x})$. The functional derivative $\frac{\delta}{\delta{\tilde{t}(\boldsymbol{x})}}$ is defined in the following way
\begin{equation}\label{varderi1}
 \frac{\delta \Psi[\tilde{t}(\boldsymbol{x})]}{\delta{\tilde{t}(\boldsymbol{x}_0)}}=\lim_{\substack{\varepsilon\rightarrow0\\V_0\rightarrow0}}\frac{\Psi[\tilde{t}(\boldsymbol{x}) + \varepsilon f_{\boldsymbol{x}_0}(\boldsymbol{x})]-\Psi[\tilde{t}(\boldsymbol{x})]}{\int \varepsilon f_{\boldsymbol{x}_0}(\boldsymbol{x}) \: d^3x} \mbox{ ,}
\end{equation}
where the test-function  $f_{\boldsymbol{x}_0}(\boldsymbol{x})$ deviates from zero only in a small three-dimensional domain $V_0$ about $\boldsymbol{x}_0$. 

\begin{figure}[htbp]
\centering
\includegraphics[scale=0.17]{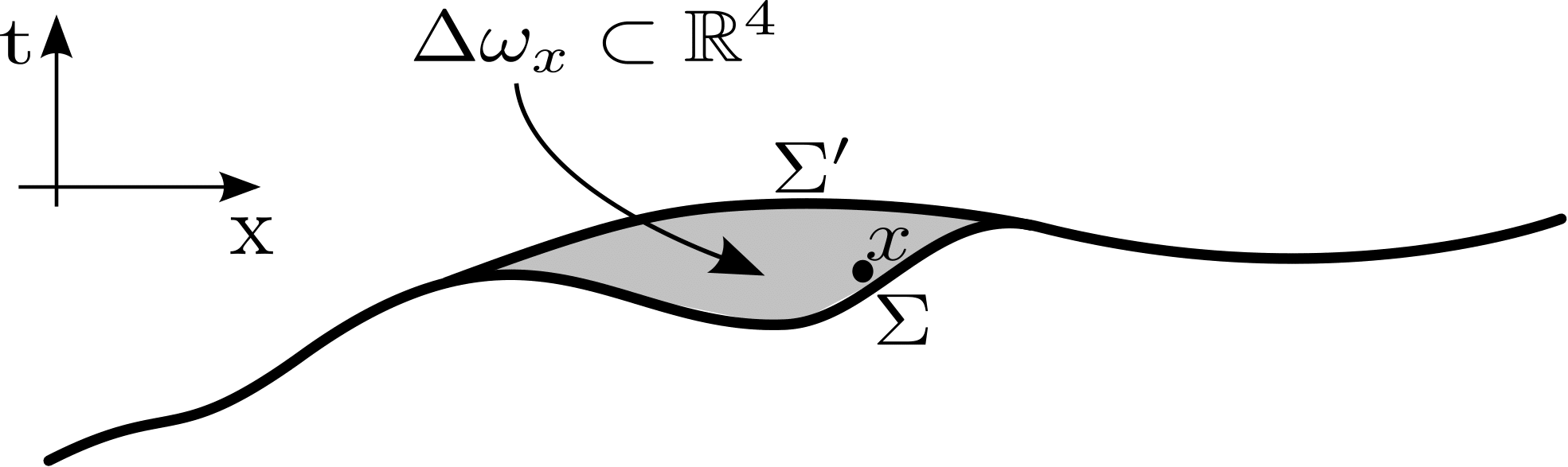}
\caption{\textbf{Small variation of space-like leaf $\Sigma$ about space-time point $x$}}
\label{tomo}
\end{figure}	 

\eqref{varderi1} can be also written more compactly: Let $\Sigma$ be the space-like leaf given by the graph of $t=\tilde{t}(\boldsymbol{x})$ and thus write $\Psi[\tilde{t}(\boldsymbol{x})]=:\Psi[\Sigma]$. Further suppose  $\Sigma'$ is a hyper-surface which deviates form $\Sigma$ only by some small space-time volume $\Delta \omega_x$ (a small world lying between $\Sigma$ and $\Sigma'$, as Tomonaga puts it) about space-time point $x$ (see figure \ref{tomo}). Then variation \eqref{varderi1} can be written as
\begin{equation}
\frac{\delta \Psi[\Sigma]}{\delta{\Sigma_x}}=\lim_{\Sigma \rightarrow \Sigma'}\frac{\Psi[\Sigma']-\Psi[\Sigma]}{\Delta \omega_x} 
\end{equation}
 and therefore our generalized Schr\"odinger equation \eqref{tomo1} as
\begin{equation}\label{tomo2}
i \frac{\delta}{\delta{\Sigma_x}}\Psi[\Sigma]=\mathscr{H}(x)\Psi[\Sigma] \mbox{ .}
\end{equation}
Now, the Tomonaga-form \eqref{tomo2} is integrable if and only if $[\mathscr{H}(x),\mathscr{H}(x')]=0$ for all pairs of space-time points $x,x'$ lying on $\Sigma$ \cite{Tomonaga} (roughly said, this ensures that the unitary evolution resulting from \eqref{tomo2} is unique, i.e.\! that we get a unique wave-function on space-like leaf $\Sigma$, although we might decompose the evolution from ''initial-leaf'' $\Sigma_0$ to $\Sigma$ into different sequences of infinitesimal variations). The commutation of the local Hamilton-densities is ensured by the requirement that the leafs $\Sigma$ are space-like hyper-surfaces.\paragraph*{}

\textbf{Tumulka:} Let us consider now the unitary wave-function dynamics for N non-interacting Dirac-particles and develop an appropriate description on arbitrary space-like leafs $\Sigma$. Let us start with the one-particle Dirac equation
\begin{equation} \label{Dirac1}
 i\gamma^{\mu}(\partial_{\mu}-ieA_{\mu})\psi = m \psi \mbox{ ,}
\end{equation}
with electromagnetic vector-potential $A^{\mu}(x)$, particle charge $e$, particle mass $m$ and the Dirac-matrices $\gamma^{\mu}$ . Now consider a space-like leaf $\Sigma$ on space-time on which we define the Hilbertspace $\mathcal{H}_{\Sigma}=L^2(\Sigma)\otimes\mathbb{C}^4$ given by the set of square-integrable $\mathbb{C}^4$-valued functions $\psi:\Sigma\rightarrow\mathbb{C}^4$ on $\Sigma$, with
\begin{equation}\label{sqint}
 \int_{\Sigma} \psi^{\dagger}(x)\gamma^0\gamma^{\mu}n_{\mu}(x)\psi(x)d^3x<\infty \mbox{ ,}
\end{equation}
 with the running variable $x \in \Sigma$, the future directed unit normal vector $n_{\mu}(x)$ on $\Sigma$ at $x$ (where ``unit vector`` means $n^{\mu}(x)n_{\mu}(x)=1$), the volume measure $d^3x$ generated by the Riemannian metric on $\Sigma$ and the adjoint spinor $\psi^{\dagger}(x)$. Observe that $\gamma^0\gamma^{\mu}n_{\mu}(x)$ is positive definite for every future oriented time-like 4-vector $n_{\mu}(x)$, that $\psi\rightarrow\bar{\psi}:=\psi^{\dagger}\gamma^0$ is a Lorentz invariant operation (in contrast to $\psi\rightarrow\psi^{\dagger}$) and that $\gamma^{\mu}n_{\mu}(x)$ heuristically picks out the ``time-component`` (with respect to $\Sigma$ at $x$) of the four-vector $\gamma^{\mu}$, i.e.\!: if we consider the Lorentz-frame in which $n_{\mu}(x)$ is the unit normal vector parallel to the time axis (i.e.\! $n_{\mu}=(1,0,0,0)$ in that frame), than $\gamma^{\mu}n_{\mu}(x)=\gamma_0$. Finally the scalar product on $\mathcal{H}_{\Sigma}$ is given by 
\begin{equation}\label{skp}
 \langle \varphi \mid \psi \rangle_{\Sigma} = \int_{\Sigma} \bar{\varphi}(x) \gamma^{\mu}n_{\mu}(x) \psi(x) d^3x \mbox{ .}
\end{equation}
Then the (one-particle) Dirac-equation \eqref{Dirac1} generates a unitary transformation $U^{\Sigma}_{\Sigma'}:\mathcal{H}_{\Sigma}\rightarrow\mathcal{H}_{\Sigma'}$ between the Hilbertspaces associated with two arbitrary distinct space-like hyper-surfaces $\Sigma$ and $\Sigma'$. The crucial unitarity is due to the validity of the continuity-equation
\begin{equation}
 \partial^{\mu}j_{\mu}(x)=:\partial^{\mu} \: \big{(}\bar{\psi}(x)\gamma_{\mu}\psi(x)\big{)}=0
\end{equation}
which follows from \eqref{Dirac1}. The rigorous proofs of existence and properties of $\mathcal{H}_{\Sigma}$ and $U^{\Sigma}_{\Sigma'}$ can be found in \cite{pointproc}. In particular the unitary operators fulfill $U^{\Sigma_2}_{\Sigma_3}U^{\Sigma_1}_{\Sigma_2}=U^{\Sigma_1}_{\Sigma_3}$ and $U^{\Sigma}_{\Sigma}=\mathds{1}_{\mathcal{H}_{\Sigma}}$

The remaining step is then to generalize this description to N-particle states. To do so consider the multi-time N-particle wave-function $\psi(x_1,...,x_N)$ on $\mathscr{M}^N$, which is solution to the N equations\footnote{One problem of multi-time equations like \eqref{dirmult} is that there is at least no obvious way to implement interaction between the particles, for an interaction potential of the form $V(\boldsymbol{x_1},...,\boldsymbol{x_N},t)$ is not consistent with multi-time formalism (and moreover such interaction would destroy the commutativity of the one-particle interaction-Hamiltonians, which is needed to ensure the integrability of the multi-time equations). But this does not harm our conceptual considerations here since wave-functions stay entangled in the absence of interaction and thus we have all we need to investigate quantum nonlocality in a relativistic context. The implementation of interaction is a different story.}
\begin{equation}\label{dirmult}
 i\gamma^{\mu}_k(\partial_{k,\mu}-ie_kA_{k,\mu})\psi = m_k \psi \quad ; \quad k=1,...,N \mbox{ .}
\end{equation}
Here $\partial_{k,\mu}$ is the four-derivative with respect to $x_k$, $e_k$ and $m_k$ the charge and mass of particle $k$, $A_{k,\mu}(x)$ an external electromagnetic potential acting on particle $k$ and the Dirac-matrices $\gamma_{k}^{\mu}$ associated with particle $k$ are defined by $\gamma_{k}^{\mu}:=\mathds{1}_{\mathcal{H}_{S_1}}\otimes...\otimes\mathds{1}_{\mathcal{H}_{S_{k-1}}}\otimes\gamma^{\mu}\otimes\mathds{1}_{\mathcal{H}_{S_{k+1}}}\otimes...\otimes\mathds{1}_{\mathcal{H}_{S_N}}$ (where $\mathcal{H}_{S_j}$ is the appropriate spin space associated with particle $j$). 

In analogy with the above construction for the one-particle case let us consider again an arbitrary space-like hyper-surface $\Sigma\subset\mathscr{M}$ and build now $\Sigma^N\subset\mathscr{M}^N$. Then the Hilbertspace $\mathcal{H}_{\Sigma^N}=L^2(\Sigma^N)\otimes\mathbb{C}^{4^N}$ of square-integrable $\mathbb{C}^{4^N}$-valued functions $\psi:\Sigma^N\rightarrow\mathbb{C}^{4^N}$ on $\Sigma^N$ (with the obvious generalizations of \eqref{sqint} and \eqref{skp}) is simply the tensor product of the one-particle Hilbertspaces.\paragraph*{} 

An important feature of both approaches is that in order to calculate local values, which we can assign to the wave-function at some space-time point, we are free to choose any desired space-like leaf containing that point to evaluate the wave-function. For example in case of a one-particle wave-function the value of the wave-function $\psi(x_0)$ at space-time point $x_0$ is independent of the space-like hyper-surface (containing $x_0$) the wave-function is associated with, i.e.\! if $x_0 \in \Sigma \cap \Sigma'$ we have\footnote{In case of more-than-one-particle wave-functions we cannot assign a value to the wave-function at one space-time point (for it is no longer a function on $\mathscr{M}$, but on $\mathscr{M}^N$) but an appropriate local analogue might then be the reduced density-matrix of one of the particles and also here the above assertion holds (in the case of purely unitary time-evolution). }
\begin{equation}\label{surfaceind}
 \psi_{\Sigma}(x_0)=\psi_{\Sigma'}(x_0)\mbox{ .}
\end{equation}
When we include collapse of the wave-function now, the crucial novelty will be, that such ''hyper-surface-independence`` no longer holds; and this fact has far reaching conceptual consequences.

\begin{center}
 \textbf{Collapse}
\end{center}
 
In order to find out how wave-function collapse is to be implemented in a Lorentz invariant and consistent way, let us consider again a picturesque example of wave-function collapse in the simple case of a one-particle wave-function (figure \ref{solution}): As above a particle is prepared and equipped with a Hamiltonian, such that the support of the wave-function is constrained to stay in distinct small spatial regions; now, say, we have two such regions around positions $\boldsymbol{x_1}$ and $\boldsymbol{x_2}$. At time $t=t_0$ a (non-demolition) position measurement is performed in the region around $\boldsymbol{x_1}$ and the particle is found there (but not absorbed from a detector). Thus we might have the following state-history:
\begin{align}
 & \mid \Psi_{(t<t_0)} \rangle =  \frac{1}{\sqrt{2}}( \mid x_1 \rangle + \mid x_2 \rangle ) \\
 & \mid \Psi_{(t>t_0)} \rangle = \mid x_1 \rangle 
\end{align}
Now we apply our conjecture that each space-like hyper-surface has to be on an equal footing, say, to the hyperplanes $\Sigma$ and $\Xi$ in figure \ref{solution}, which might be simultaneity slices in two different frames. Observe, that $\Sigma$ defines a ''simultaneous instant of time'' in some frame \textit{after} the position measurement , while $\Xi$ defines a ''simultaneous instant of time'' \textit{prior} to that measurement in another frame. Thus obviously we have to associate $\Sigma$ with the collapsed wave-function and $\Xi$ with the un-collapsed one. To put it more generally: The hyper-surface $\Sigma$ lies in the future of the (measurement-) event where the collapse is centered (i.e.\! every future-oriented time-like curve originating from that event crosses $\Sigma$) such that we have to assign the collapsed wave-function to it. In contrast $\Xi$ is prior to the ''reduction causing event'' (each past-oriented time-like curve originating from there intersects $\Xi$) such that we have to assign the initial state to it. 

Both slices intersect in space-time point $X$ and since we are concerned with a one-particle wave-function we might ask how $\Psi(X)$ looks like. But that is obviously not the right question, since $\Psi(X)$ has no meaning at all in this context; not until a hyper-surface is chosen on which $\Psi$ is evaluated. Moreover the value at $X$ strongly depends on that choice now: 

\begin{figure}[htbp]
\centering
\includegraphics[scale=0.18]{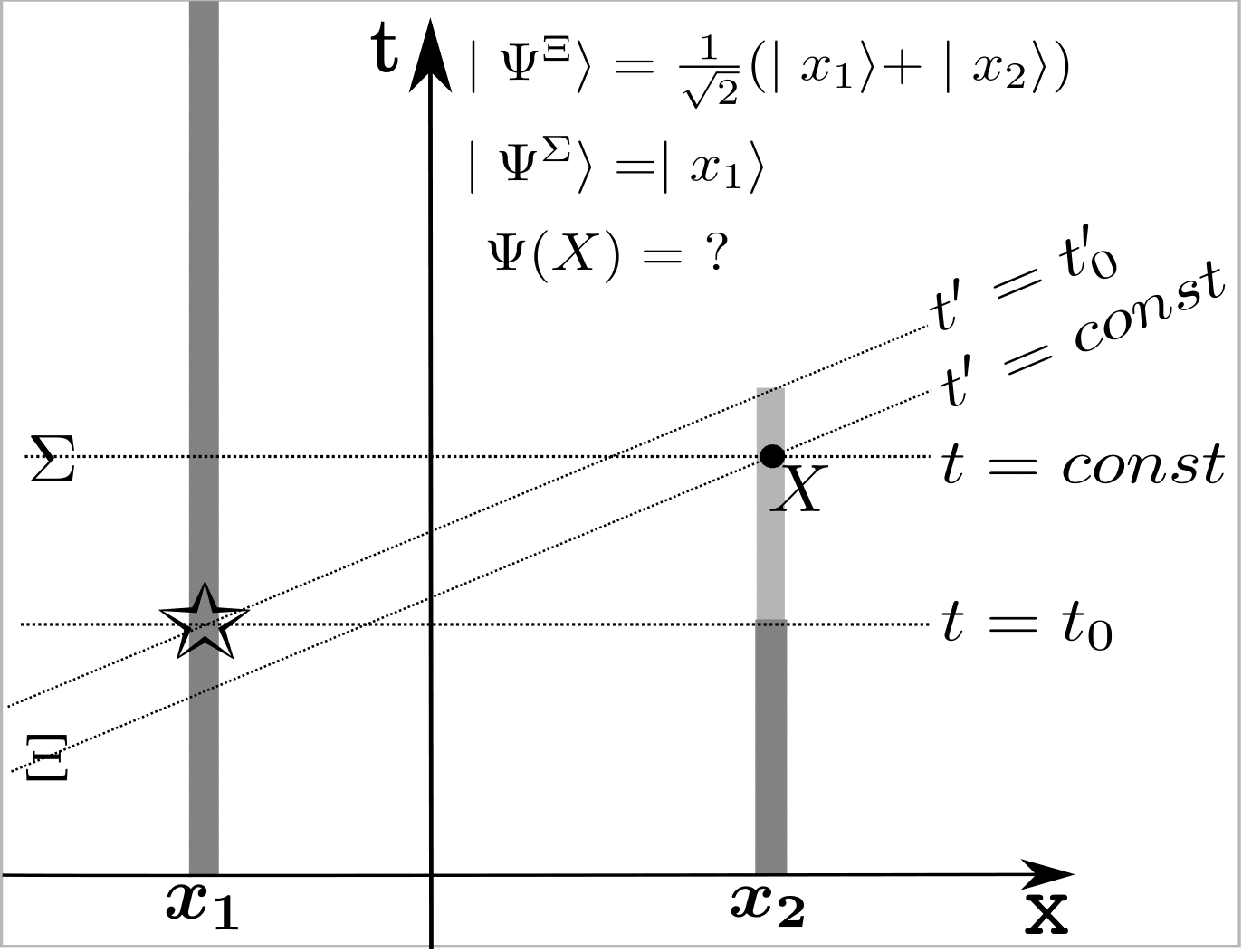}
\caption{\textbf{Towards solution:} A particle with initial state $ \frac{1}{\sqrt{2}}( \mid x_1 \rangle + \mid x_2 \rangle )$ is detected (but not absorbed) at $\v{x_1}$ at time $t=t_0$. Consequently, the wave-function associated with hyperplane $\Sigma$ vanishes at space-time point $X\in\Sigma\cap\Xi$, in contrast to the wave-function along hyperplane $\Xi$. Both might be simultaneity slices in particular frames.}
\label{solution}
\end{figure}

\begin{equation}
\int_{U^{\Xi}_{\varepsilon}(X)} \Psi^{\Xi}(x) d^3x = \frac{1}{\sqrt{2}} \neq \int_{U^{\Sigma}_{\varepsilon}(X)} \Psi^{\Sigma}(x) d^3x = 0 \mbox{ ,}
\end{equation}
where $U^{\Xi / \Sigma}_{\varepsilon}(X)$ is the neighborhood of $X$ with radius $\varepsilon$ restricted to $\Xi / \Sigma$ and $\varepsilon$ is big enough to cover the (potential) support of the wave-function around $X$ with $U^{\Xi / \Sigma}_{\varepsilon}(X)$ . And indeed we can put it a bit simpler by observing that
\begin{equation}
 (0\neq) \hspace{0.2cm} \Psi^{\Xi}(X) \neq \Psi^{\Sigma}(X) \hspace{0.2cm} (=0)
\end{equation}
 must hold for the scheme depicted in figure \ref{solution} within every theory which accounts for quantum mechanical predictions and, at the same time, does not draw on preferred space-like structures of space-time.   

Now we are in a position to straight forwardly generalize the evolution law of wave-functions to situations in which wave-function collapse occurs:

\begin{quotation}
 \textsl{The state of the system is not a function of space-time [...] but, \textit{rather}, [...] it is ineluctably a \textit{functional} on the set of space-like hyper-surfaces. [...] We shall require (so as to complete, together with \textrm{[the Tomonaga form \eqref{tomo2}]}, the description of the evolution of $\psi$ from one surface to another) some covariant prescription for the collapse within this language [...]: The state reduction occurs separately along every space-like hyper-surface which passes through the measurement event; if one hyper-surface is continuously deformed into another, the reduction occurs as the hyper-surface \textit{crosses} that event.} \cite{aa3}
\end{quotation}

This is in essence exactly the same way in which the wave-function transforms under collapse in Tumulka's relativistic rGRWf-model (see section \ref{rgrwf}): Only here, we do not need measurement-events to formulate the reduction law, since the primitive ontology of the theory is given by the collapse-events themselves (whose (stochastic) dynamics is precisely described by the theory). These events are called flashes. In the language of this theory the above law sounds like:

\begin{quotation}
\textsl{How does the wave-function transform under a change of slicing of space-time [...]? In two ways. First, some flashes may lie in the future of the new surface [...] but in the past of the old surface [...] and vice versa; consequently, the corresponding wave-functions differ by application of the collapse operators (respectively their inverses, and renormalization) belonging to these flashes. Second, on top of that the wave-function differ by the unitary Dirac propagator from one surface to the other.} \cite{rGRWf} 
\end{quotation}
Or elsewhere
\begin{quotation}
 \textsl{As we push $\Sigma$ to the future, $\Psi_{\Sigma}$ collapses whenever $\Sigma$ crosses a flash, and evolves deterministically in between.} \cite{collandrel}
\end{quotation}

Observe, that this law of wave-function dynamics is as relativistic as it could be since it does not pick out any coordinate system nor any other structures of Minkowski space-time as being physically preferred with respect to others.

 \begin{center}
  \textbf{More than One Particle}
 \end{center}

So far we have considered a one-particle wave-function where everything is unsophisticated since configuration space coincides with physical space. So let us take a brief look at an example with more than one particle to prevent possible confusion in reasoning about these things (this example is also briefly considered by Tumulka in \cite{unrompics}).

\begin{figure}[htbp]
\centering
\includegraphics[scale=0.13]{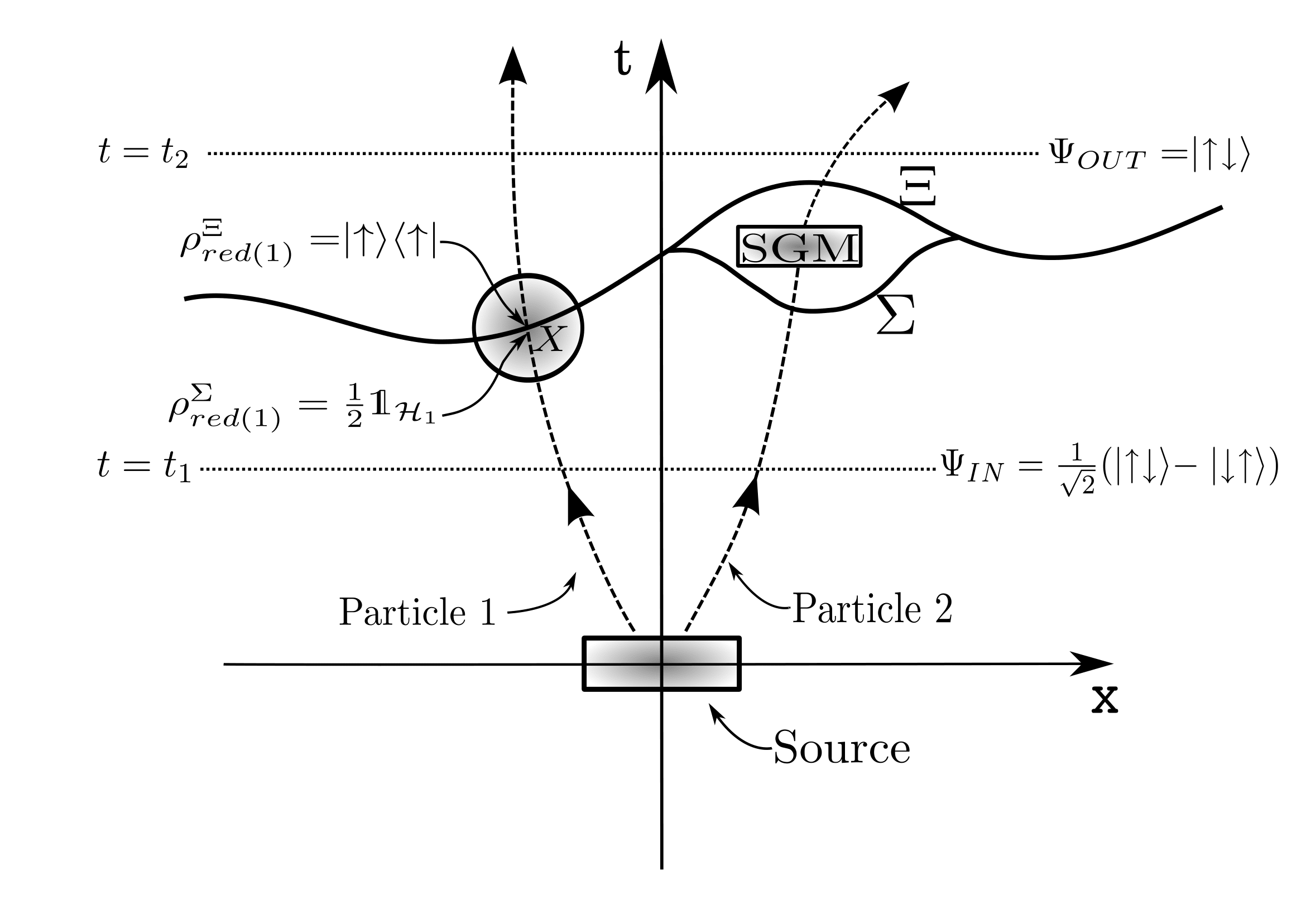}
\caption{\textbf{Hyper-surface dependence of the reduced density-matrix:} A pair of singlet particles is prepared and spatially separated. Then the z-component of the spin of particle($2$) is measured at a SGM with outcome $\sigma_z^{(2)}=-1$. In a \textit{Solution-\RM{1}}-model the reduced density matrix corresponding to particle($1$) about space-time point $X\in\Sigma\cap\Xi$ depends crucially on the choice of hypersurface, $\Sigma$ or $\Xi$, where the wave-function is taken, although these surfaces might be almost identical.}
\label{solution2}
\end{figure}

Consider again a pair of singlet-particles, where ``at some time`` the world-line of particle(2) crosses the world-line of some SGM (oriented in z-direction) and is deflected downwards (is found in the respective region directly afterwards) (see figure \ref{solution2}). Thus the wave-function undergoes a collapse
\begin{equation}
 \mid \Psi_- \rangle \longrightarrow \mid \uparrow \downarrow \rangle 
\end{equation}
and the question arises which wave-function belongs to particle(1) in some region (say, around space-time point $X$) space-like separated from the measurement of the z-spin of particle(2).     

Now, of course, we cannot evaluate the wave-function at some space-time point anymore, for the two-particle wave-function (the spatial part) is no more a function of space-time. The most appropriate analogue of the ``wave-function of particle(1)'' (i.e.\! the object which is appropriate to describe the physics of particle(1) locally as good as possible without incorporating particle(2)) is now the reduced density matrix, i.e.\! we perform the partial trace with respect to the Hilbertspace $\mathcal{H}_{S_2}$ of particle(2) onto the density matrix of the two-particle system. And with the state description developed so far the latter density matrix (and with it the former reduced one) depends strongly on the choice of a hyper-surface where the wave-function is evaluated: 

Consider the two space-like hyper-surfaces $\Sigma$ and $\Xi$ depicted in figure \ref{solution2} which are almost identical. They only differ in a small region around the event of measurement of the z-spin of particle(2). Then the density matrix belonging to $\Sigma$ is 
\begin{equation}
 \rho^{\Sigma}=\mid \Psi_- \rangle \langle \Psi_- \mid
\end{equation}
and consequently the reduced density matrix of particle(1)   
\begin{equation}
\rho^{\Sigma}_{red(1)}=\frac{1}{2} \bigg{(}(\mid \uparrow \rangle \langle \uparrow \mid)^{(1)} + (\mid \downarrow \rangle \langle \downarrow \mid)^{(1)}\bigg{)} = \frac{1}{2} \mathds{1}_{\mathcal{H}_{S_1}} \mbox{ .}
\end{equation} \vspace{0.5cm}
On the other hand the density matrix along $\Xi$ is
\begin{equation}
 \rho^{\Xi}=\mid \uparrow \downarrow \rangle \langle \uparrow \downarrow \mid
\end{equation}
 and thus
\begin{equation}
 \rho^{\Xi}_{red(1)}= (\mid \uparrow \rangle \langle \uparrow \mid)^{(1)} 
\end{equation}
which obviously describes a ``physical situation'' very different from the one described by $\rho^{\Sigma}_{red(1)}$. Nevertheless, we have to attribute the same physical significance to both descriptions, as long as we refuse to employ distinguished space-like structures of space-time which account for nonlocality.

\begin{center}
 \textbf{Consequence}
\end{center}

The most striking consequence from this reasoning is (as mentioned in the introduction) that it excludes the possibility to create the primitive ontology directly out of the wave-function in a naive way. If one follows discussions and incidental remarks in papers and text-books it seems that many physicists have in the back of their mind the idea, to identify the wave-function (in a not clearly defined way) with matter, e.g.\! as a quantity which somehow generates a density of matter. But now we encounter that such interpretation is actually impossible in relativistic quantum theory without drawing on some preferred space-time structure, since -- as we have seen -- the wave-function might vanish in some space-time regions in one frame of reference while it has finite values there in another frame. And such behavior is unacceptable for matter density. It would imply the same ontological inconsistency we were faced with in section \ref{meaningloccom} (as a consequence of non-commuting space-like separated operators, there): The distributions of matter in different Lorentz-frames would not be the respective Lorentz-transformed distributions of matter there. The world in different frames would not be the respective Lorentz-transformed world but a completely different world. \paragraph*{}

Maybe the requirement of \textbf{\textit{ontological consistency}} is a proper modest and indisputable \textbf{\textit{notion of relativity}}: \textit{Distributions of matter in one Lorentz-frame must be the respective Lorentz-transformed distributions of matter present in other Lorentz-frames.} \paragraph*{}

 Finally, to make the vague conception of the interpretation of the wave-function as something which generates a density of matter precise (i.e.\! to give a corresponding possible law for the primitive ontology), let us consider the perhaps simplest way to do so: Let us begin with non-relativistic quantum theory and think of matter as the (weighted) projection of the square of the wave-function (to do justice to Born's rule) from configuration space down to physical space, i.e.\!:
\begin{equation}\label{massont}
 \mathcal{M}(\boldsymbol{x},t)= \sum_{i=1}^N m_i \int d^3x_1...d^3x_N \mid \psi(\boldsymbol{x}_1,...,\boldsymbol{x}_N,t) \mid^2 \delta(\boldsymbol{x}-\boldsymbol{x}_i) \mbox{ ,}
\end{equation}
 where the weight $m_i$ is the mass associated with particle $i$. In other words, this model constitutes a mass distribution by evaluating the marginal mass distribution of particle $i$ and then summing up all the one-particle marginal distributions. 

Now, this is the first possible precise law for a primitive ontology we encounter within these lines and it is well suited to account for Born's probability rule (but in standard quantum theory only if we omit to think of pointers and the like as consisting of constituents guided by the laws of quantum theory). But it is not suitable to comply with ontological consistency if applied to relativistic quantum theory: to see this consider for example again the scheme of the one-particle wave-function evolution depicted in figure \ref{solution}. The mass-density \eqref{massont} in the one-particle case is simply
\begin{equation}
 \mathcal{M}(\boldsymbol{x},t)=\mid \Psi(\boldsymbol{x},t) \mid^2
\end{equation}
and lifted to Minkowski space-time this law would predict matter to be found around space-time point $X$ in the frame in which $\Xi$ is a simultaneity slice (indeed half of the matter of the considered system in the direct spatial neighborhood of that point), but no matter around $X$ in the frame in which $\Sigma$ is a simultaneity slice.\paragraph*{}       

This illustrates that the desire to treat all space-like hyper-surfaces on an equal footing, leads to the conclusion that the wave-function cannot enter into the primitive ontology in a naive way (in the sense discussed so far). 

\subsection{\label{distfol}\textit{Solution \texorpdfstring{\RM{2}}{}}: Distinguished Foliation}

In their 1981-paper Aharonov and Albert claim to infer from the possibility of the above described measurement procedure the impossibility of wave-function collapse along preferred hyper-surfaces:   

\begin{quotation}
 \textsl{The proposal that the reduction be taken to occur covariantly along the backward light-cone of \textit{[the measurement event]}, or, indeed, that it be taken to occur along \textbf{any} hyper-surface other than $t=0$ \textit{[which corresponds e.g.\! to $t=t_0$ in figure \ref{HKEPR}]}, will fail ... , since it cannot account for the results of nonlocal measurements of the kind we have described here. }
\end{quotation}

I have argued in chapter \ref{chapHK} that this kind of measurement indeed refutes backward light-cone reductions as proposed by Hellwig and Kraus. But it is a different business to take the collapse to occur along fixed (space-like) hyper-surfaces (which, unlike the Hellwig-Kraus surfaces, do not depend on measurements and might constitute a unique foliation of space-time). Hence, we should take a closer look on such a proposal.

\begin{figure}[htbp]
\centering
\includegraphics[scale=0.15]{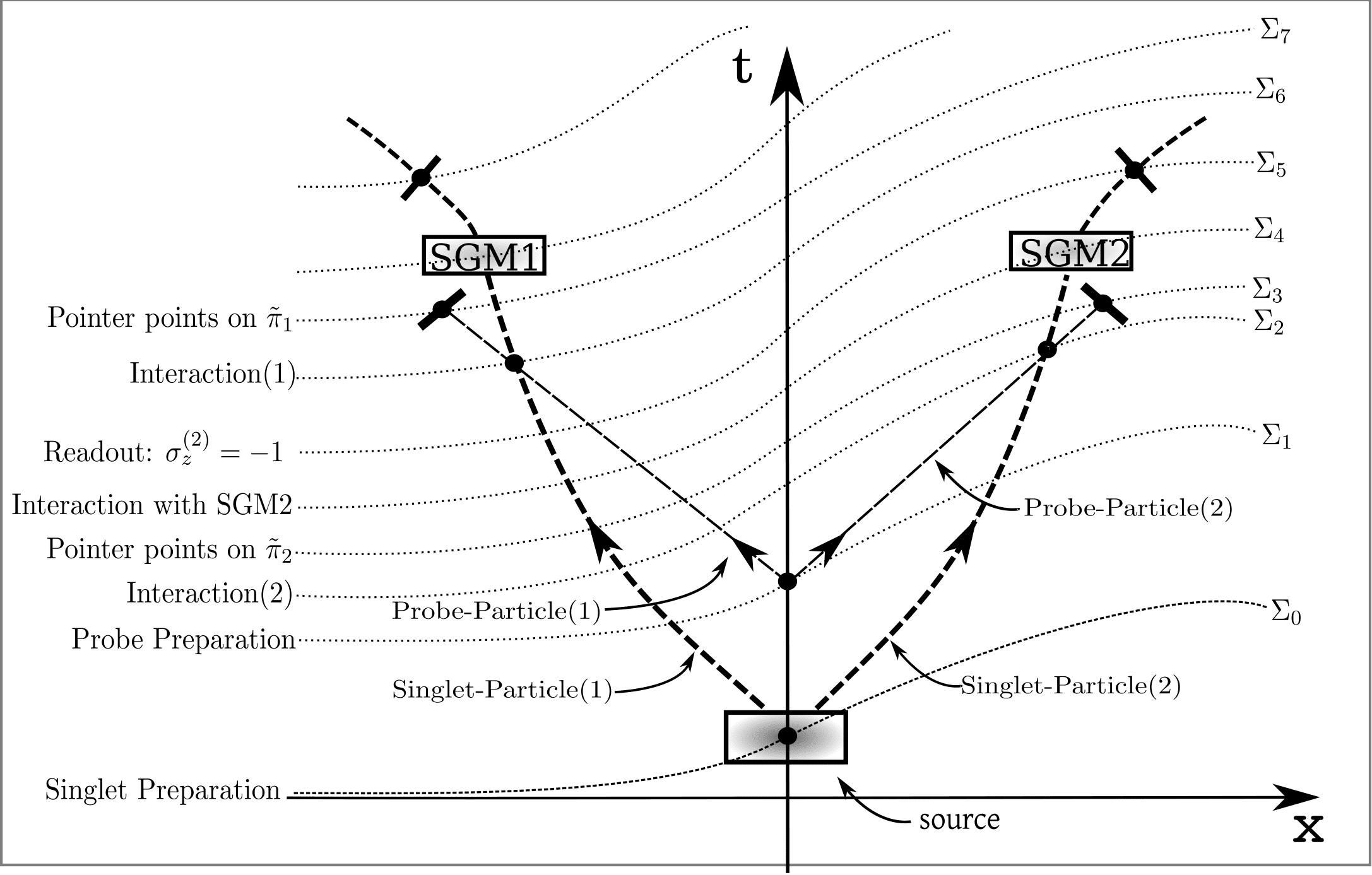}
\caption{\textbf{Distinguished Foliation:} The Aharonov-Albert experiment performed in the depicted frame where the relevant pairs of interactions are simultaneous. A possible distinguished foliation of space-time into space-like leafs is indicated and the collapse is supposed to occur along these leafs. Of particular interest is the collapse along the leaf $\Sigma_5$ according to the measurement of the $z$-component of singlet-particle($2$).}
\label{Foliation}
\end{figure}

At first glance it seems indeed to be a natural conclusion of the Aharonov-Albert procedure to exclude the existence of a preferred foliation: Consider the scheme which we already used to refute the Hellwig-Kraus proposal, depicted in figure \ref{Foliation}: The EPRB-setup is supplemented by an Aharonov-Albert experiment shortly before the interactions at the SGMs and the subsequent position-measurements which are separately resolved now. Without loss of generality assume that the outcome is $\sigma^{(1)}_z=+1$ and $\sigma^{(2)}_z=-1$. Further we consider the whole procedure, i.e.\! the readouts $\tilde{\pi}_i \mbox{, with } i=1,2$ indicated in figure \ref{Foliation} stand for three pairs of values (belonging to $\sigma^{tot}_x \mbox{, } \sigma^{tot}_y \mbox{ and } \sigma^{tot}_z$, respectively).   

The equal-time hyperplanes of the laboratory-frame (the $x-t$ frame depicted here) are not indicated, but the pairs of relevant interactions (probe/singlet interactions and the interactions with the devices) are supposed to be simultaneous in this frame. Instead, curved space-like hyper-surfaces are indicated and we shall assume now a model in which every measurement-event lying on one of the hyper-surfaces, collapses the wave-function all along that surface.

As already mentioned, it seems at first glance that this leads to a contradiction: In this picture, when probe-particle(1) interacts with singlet-particle(1), the latter is already no more a singlet-particle. According to our model, the wave-function collapses along the surface $\Sigma_5$ and the state is projected onto $(\mid \downarrow \rangle \langle \downarrow \mid )^{(2)}$ (and renormalized), such that the singlet state should be irreversibly destroyed at the time of interaction with probe-particle(1). How could then this interaction (together with the interaction of probe-particle(2) ) confirm that $\sigma^{tot}_z=\sigma^{tot}_x=\sigma^{tot}_y=0$? This seems to suggest that our model with a distinguished foliation contradicts the predictions of quantum theory (as claimed by Aharonov and Albert in the above quote) and that \textit{Solution \RM{1}} is the only way to reconcile wave-function reduction with Minkowski-space-time.

But there must be a flaw in the apparent contradiction for two reasons: First, there exists a relativistic quantum mechanical model (the so called Hypersurface-Bohm-Dirac-Model \cite{HBDM}, see chapter \ref{HBDM}) with particle trajectories, in which the non-local connection between the particles (and thereby the so called effective collapse of the wave-function which is effectively the same as the wave-function collapse in the scheme we are considering right now) is postulated to occur along the leafs of a preferred foliation. And it has been shown \cite{HBDM} that this model is empirically equivalent to quantum theory, i.e.\! it yields all the quantum mechanical predictions, independent of the actual choice of foliation. Therefore it cannot be true that a model with preferred foliation contradicts quantum mechanical predictions in principle. 

\begin{figure}[htbp]
\centering
\includegraphics[scale=0.39]{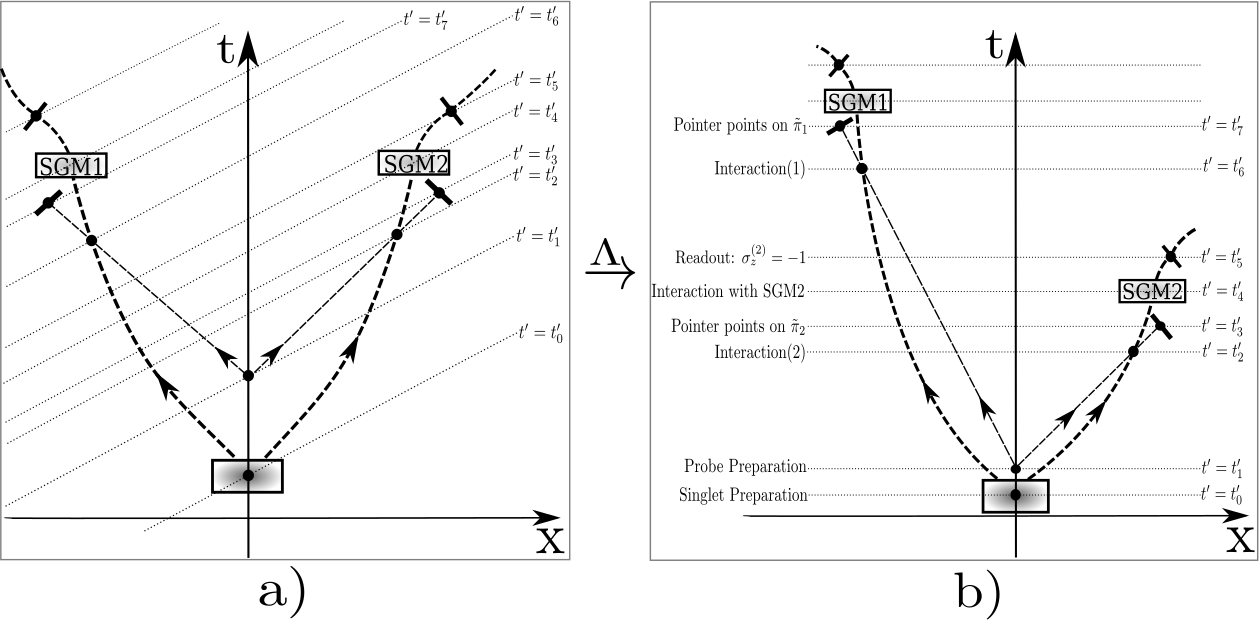}
\caption{\textbf{Flattened and transformed foliation:} Since the relevant feature of the foliation in figure \ref{Foliation} is to define the sequence of interactions and collapses, we can disregard the detailed shape of the leafs in order to calculate the state history corresponding to this sequence. This is done $a)$ by exchanging the hypersurfaces with flat hyperplanes and $b)$ by a Lorentz-transformation into the frame where these planes are constant time slices.} 
\label{FoliationCalc}
\end{figure}

Second, also a careful look onto \textit{Solution \RM{1}} shows that if the Aharonov-Albert procedure contradicts a preferred foliation in principle, it must also contradict \textit{Solution \RM{1}}: For example, if we transform the scenario illustrated in figure \ref{Foliation} (with removed foliation) into a Lorentz-frame in which the readout of $\sigma^{(2)}_z$ precedes the interaction of probe-particle(1) (like illustrated in figure \ref{FoliationCalc}) we can, according to \textit{Solution \RM{1}}, use the equal-time hyperplanes of this frame to define a state-history. But this state-history is essentially identical to the state-history as defined by the sequence of space-like leafs of the proposed distinguished foliation in figure \ref{Foliation}. If the latter would contradict quantum theory, then the former would, too.

Thus it seems worth calculating explicitly the time evolution for this scheme within our ''collapse-along-hyper-surfaces-of-figure-\ref{Foliation}``-model, to see that indeed everything comes out right and that contradictions do not show up, actually.     

Since we have no concrete mathematical expression for our foliation but only a qualitative picture, we must focus on the only relevant feature defined by it: the sequence of interactions and collapse. Apart from this sequential history the actual shape of the leafs provides no relevant contributions to our issue here, such that in order to calculate we can simply ''make them flat`` (see figure \ref{FoliationCalc}a)). The resulting hyperplanes are now equal-time hyperplanes of a distinguished frame $K'$ of instantaneous collapse and we can perform the calculation within this frame (figure \ref{FoliationCalc}b)).               

Let us start with the Aharonov-Albert procedure to measure $\sigma^{tot}_z$.

\begin{center}
 \textbf{$\sigma^{tot}_z$-measurement}
\end{center}

Now the first part of the calculation of the state evolution is the same as already calculated in chapter \ref{timedisp} (see \eqref{t3t4}): Suppose the generalized momentum of probe-particle(2) is found to have the value $\tilde{\pi}_2$ at $t'=t'_3$. Then for times $t'_3<t'<t'_4$ the state will be the entangled wave-function 
\begin{equation} \label{t34}
 \mid \Psi_{(t'_3<t'<t'_4)} \rangle = \frac{1}{\sqrt{2}}\bigg{(}  \mid \pi_1 = -\tilde{\pi}_2 + F \rangle \otimes \mid \uparrow \downarrow \rangle - \mid \pi_1 = -\tilde{\pi}_2 - F \rangle \otimes \mid \downarrow \uparrow \rangle \bigg{)} \mbox{ .}
\end{equation}
At time $t'_4$ particle(2) interacts with the magnetic field produced by SGM2  which makes the support of the spatial part of the wave-function (which is not considered in the calculation) splitting up behind SGM2. Then at time $t'_5$ a (non-demolition) position measurement of particle(2) is performed and the branch is identified in which the particle is found. Say it is found in the lower region (with respect to the z-axis, in which the SGM is oriented) behind the SGM, such that the wave-function collapses onto the first term of sum \eqref{t34}. Thus we have
\begin{equation}
 \mid \Psi_{(t'_5<t'<t'_6)} \rangle = \mid \pi_1 = -\tilde{\pi}_2 + F \rangle \otimes \mid \uparrow \downarrow \rangle
\end{equation}
At time $t'_6$ particle(1) interacts with the other probe-particle, resulting in the unitary evolution 
\begin{equation}
\begin{gathered}
 \mid \Psi_{(t'_6<t'<t'_7)} \rangle = \mathcal{U}_1 \mid \Psi_{(t'_5<t'<t'_6)} \rangle = e^{-iFq_1\sigma^{(1)}_z}\mid \pi_1 = -\tilde{\pi}_2 + F \rangle \otimes \mid \uparrow \downarrow \rangle = \\
e^{-iFq_1}\mid \pi_1 = -\tilde{\pi}_2 + F \rangle \otimes \mid \uparrow \downarrow \rangle = \mid \pi_1 = -\tilde{\pi}_2 \rangle \otimes \mid \uparrow \downarrow \rangle
\end{gathered}
\end{equation}
Thus when $\pi_1$ is measured at time $t'_7$ it is found that $\pi_1$ realizes the actual value $\tilde{\pi}_1=-\tilde{\pi}_2$, i.e.\! with \eqref{AAsigmaz}
\begin{equation}
 \sigma^{tot}_z=\sigma^{(1)}_z + \sigma^{(2)}_z=-\frac{\tilde{\pi}_1 + \tilde{\pi}_2}{F} = 0 \mbox{ .}
\end{equation}
Afterwards particle(1) is found in the upper region behind SGM1 (indicating that $\sigma^{(1)}_z=+1$), of course.

All this is not at all a miracle, since the z-component of the total spin is zero in the singlet state as well as in the state $\mid \uparrow \downarrow \rangle$, anyway. So let us check the Aharonov-Albert procedure to measure one of the more interesting other components, say $\sigma^{tot}_x$, for which that is not true anymore.

\begin{center}
 \textbf{$\sigma^{tot}_x$-measurement}
\end{center}

The initial singlet-state is symmetrical under change of basis, i.e.\! for our purpose we have
\begin{equation}
 \mid \Psi_- \rangle = \frac{1}{2} \big{(} \mid \uparrow \downarrow \rangle_x - \mid \downarrow \uparrow \rangle_x\big{)}
\end{equation}
as the reader can easily confirm with \eqref{sigmatrans} (for $\varphi=0$ and $\theta=\frac{\pi}{2}$). 

Denote again by $q^x_i$ the degree of freedom which couples to $\sigma^{(i)}_x$ and by $\pi^x_i$ its canonically conjugated generalized momentum and let us adopt otherwise the notational machinery developed in chapter \ref{timedisp} and within this chapter. Also figure \ref{FoliationCalc} (with $\tilde{\pi}_i$ replaced by $\tilde{\pi}^{x}_i$) shall be the blueprint for our calculation.

Consequently the initial state (right after the preparations) will be
\begin{equation}
\mid \Psi_{in} \rangle = \mid \Psi_{(t'_1<t'<t'_2)} \rangle = \mid \Phi_0 \rangle \otimes \mid \Psi_- \rangle 
\end{equation}
\begin{center}
 with
\end{center} 
\begin{equation}
 \mid \Phi_0 \rangle = \mid q^x_-=0 ; \pi^x_+=0 \rangle \hspace{1cm} \mbox{ and } \hspace{1cm} \mid \Psi_- \rangle = \frac{1}{\sqrt{2}}\big{(}\mid \uparrow \downarrow \rangle_x - \mid \downarrow \uparrow \rangle_x \big{)} \mbox{ ,} 
\end{equation}
where the meaning of $q^x_-$ and $\pi^x_+$ is analogous to \eqref{pi+-}.

Now we can calculate the state for times $t'_2<t'<t'_3$ in analogy with \eqref{t2t3} and \eqref{t23}:
\begin{equation}
 \begin{gathered}
 \mid \Psi_{(t'_2<t'<t'_3)} \rangle = \; \mathcal{U}^x_2 \mid \Psi_{in} \rangle=  \frac{1}{\sqrt{2}} \bigg{(} e^{-i F \hat{q}^x_2 \hat{\sigma}^{(2)}_x} \mid \Phi_{0} \rangle \otimes \mid \uparrow \downarrow \rangle_x - e^{-i F \hat{q}^x_2 \hat{\sigma}^{(2)}_x} \mid \Phi_{0} \rangle \otimes \mid \downarrow \uparrow \rangle_x \bigg{)} = \\
\mbox{. . . } = \\
  \frac{1}{\sqrt{2}}\bigg{(} \int d\pi^x_1 \int d\pi^x_2 \; \mid \pi^x_1 ; \pi^x_2 \rangle \langle \pi^x_1 ; \pi^x_2 - F \mid q^x_-=0 ; \pi^x_+=0 \rangle \otimes \mid \uparrow \downarrow \rangle_x - \\ 
\int d\pi^x_1 \int d\pi^x_2 \; \mid \pi^x_1 ; \pi^x_2 \rangle \langle \pi^x_1 ; \pi^x_2 + F \mid q^x_-=0 ; \pi^x_+=0 \rangle \otimes \mid \downarrow \uparrow \rangle_x \bigg{)}
 \end{gathered}
\end{equation}
At time $t'_3$ then $\pi^x_2$ is measured and the readout is $\tilde{\pi}^x_2$. The new state $\mid \Psi_{(t'_3<t'<t'_4)} \rangle \in \mathcal{H}_{S_1} \otimes \mathcal{H}_{S_2} \otimes \mathcal{H}_{P_1}$ is analogous to \eqref{t3t4} and we shall immediately transform it back into the $\sigma_z$-basis to be forearmed to evaluate the effect of the subsequent $\sigma^{(2)}_z$-measurement:
\begin{equation}
 \begin{gathered}\label{t'34}
  \mid \Psi_{(t'_3<t'<t'_4)} \rangle =  \frac{1}{\sqrt{2}}\bigg{(}  \mid \pi^x_1 = -\tilde{\pi}^x_2 + F \rangle \otimes \mid \uparrow \downarrow \rangle_x - \mid \pi^x_1 = -\tilde{\pi}^x_2 - F \rangle \otimes \mid \downarrow \uparrow \rangle_x \bigg{)} = \\
\frac{1}{2\sqrt{2}} \mid \pi^x_1 = -\tilde{\pi}^x_2 + F \rangle \otimes \Big{(}-\mid \uparrow \uparrow \rangle - \mid \downarrow \uparrow \rangle + \mid \uparrow \downarrow \rangle + \mid \downarrow \downarrow \rangle \Big{)} - \\
\frac{1}{2\sqrt{2}}\mid \pi^x_1 = -\tilde{\pi}^x_2 - F \rangle \otimes \Big{(}-\mid \uparrow \uparrow \rangle + \mid \downarrow \uparrow \rangle - \mid \uparrow \downarrow \rangle + \mid \downarrow \downarrow \rangle\Big{)} \mbox{ .}
 \end{gathered}
\end{equation}
Again at time $t'_4$ SGM2 splits up the support of the spatial part of the wave-function and at time $t'_5$ it is found that $\sigma^{(2)}_z=-1$. The collapsed wave-function then will be proportional to the projection of \eqref{t'34} onto $(\mid \downarrow \rangle \langle \downarrow \mid)^{(2)}$ and with normalization we have
\begin{equation}
 \begin{gathered}
 \mid \Psi_{(t'_5<t'<t'_6)} \rangle  = \\
\frac{1}{2} \bigg{(} \mid \pi^x_1 = -\tilde{\pi}^x_2 + F \rangle \otimes \Big{(} \mid \uparrow \downarrow \rangle + \mid \downarrow \downarrow \rangle \Big{)} + \mid \pi^x_1 = -\tilde{\pi}^x_2 - F \rangle \otimes \Big{(} \mid \uparrow \downarrow \rangle - \mid \downarrow \downarrow \rangle \Big{)} \bigg{)}= \\
\frac{1}{2} \bigg{(} \mid \pi^x_1 = -\tilde{\pi}^x_2 + F \rangle \otimes \Big{(} \mid \uparrow \uparrow \rangle_x + \mid \uparrow \downarrow \rangle_x \Big{)} + \mid \pi^x_1 = -\tilde{\pi}^x_2 - F \rangle \otimes \Big{(} - \mid \downarrow \uparrow \rangle_x - \mid \downarrow \downarrow \rangle_x \Big{)} \bigg{)}
 \end{gathered}
\end{equation}
which again shows entanglement of the spin-$\frac{1}{2}$ particles and the probe in the intermediate time between the interactions.
At time $t'_6$ spin-$\frac{1}{2}$-particle(1) interacts with probe-particle(1):
\begin{equation}
 \begin{gathered}
\mid \Psi_{(t'_6<t'<t_7)} \rangle = \mathcal{U}^x_1 \mid \Psi_{(t'_5<t'<t'_6)} \rangle = \\
 \frac{1}{2} e^{-iF\hat{q}^x_1 \hat{\sigma}^{(1)}_x} \mid \pi^x_1 = -\tilde{\pi}^x_2 + F \rangle \otimes \Big{(} \mid \uparrow \uparrow \rangle_x + \mid \uparrow \downarrow \rangle_x \Big{)} + \\
\frac{1}{2} e^{-iF\hat{q}^x_1 \hat{\sigma}^{(1)}_x} \mid \pi^x_1 = -\tilde{\pi}^x_2 - F \rangle \otimes \Big{(} - \mid \downarrow \uparrow \rangle_x - \mid \downarrow \downarrow \rangle_x \Big{)} = \\
\frac{1}{2} e^{-iF\hat{q}^x_1} \mid \pi^x_1 = -\tilde{\pi}^x_2 + F \rangle \otimes \Big{(} \mid \uparrow \uparrow \rangle_x + \mid \uparrow \downarrow \rangle_x \Big{)} + \\
\frac{1}{2} e^{+iF\hat{q}^x_1} \mid \pi^x_1 = -\tilde{\pi}^x_2 - F \rangle \otimes \Big{(}- \mid \downarrow \uparrow \rangle_x - \mid \downarrow \downarrow \rangle_x \Big{)} = \\
\frac{1}{2}  \mid \pi^x_1 = -\tilde{\pi}^x_2 \rangle \otimes \Big{(} \mid \uparrow \uparrow \rangle_x + \mid \uparrow \downarrow \rangle_x - \mid \downarrow \uparrow \rangle_x - \mid \downarrow \downarrow \rangle_x \Big{)} = \mid \pi^x_1 = -\tilde{\pi}^x_2 \rangle \otimes \mid \uparrow \downarrow \rangle
 \end{gathered}
\end{equation}
Thus the subsequent $\pi^x_1$-measurement will confirm the value $\tilde{\pi}^x_1=-\tilde{\pi}^x_2$ and therefore
\begin{equation}
\sigma^{tot}_x=\sigma^{(1)}_x + \sigma^{(2)}_x=-\frac{\tilde{\pi}^x_1 + \tilde{\pi}^x_2}{F} = 0 \mbox{ .} 
\end{equation}
And the measurement of $\sigma^{(1)}_z$ at SGM1 will confirm the value +1 with certainty.

The same holds for a $\sigma^{tot}_y$-measurement, of course; the calculation is in strong analogy to the just performed calculation. 

The outcomes (pointer points on value $xy$...)  have to be the same from the viewpoint of each Lorentz-frame, of course. Thus we can expect that the just calculated outcomes must be also realized in the originally considered frame of simultaneous interactions (figure \ref{Foliation}). 

Now, obviously we can verify that 
\begin{equation}
 \sigma^{tot}_z=\sigma^{tot}_x=\sigma^{tot}_y=0
\end{equation}
despite the fact that the singlet state is already collapsed in one of the relevant space-time regions. 

We see that a quantum mechanical theory which postulates wave-function collapse to occur along the leafs of a preferred foliation of $\mathscr{M}$ into space-like hyper-surfaces makes the right predictions for the outcomes in the case of an Aharonov-Albert experiment performed shortly before the single-particle spin-measurements of an EPRB-experiment. The apparent contradiction vanishes as soon as the scheme is carefully analyzed.   
  
\begin{center}
\textbf{Against State Verification Measurement}
\end{center}

In the physics literature sometimes terms or phrases are used frequently which suggest a self-evident and transparent meaning at first glance. Nevertheless it can be worth to take a second look at such expressions and to give a clear account of what the might mean and what not (as we did in the case of terms like ''causality`` or ''locality`` above). This shall be briefly done now with the expression ''state verification measurement of the singlet state`` in connection with the Aharonov-Albert procedure because this might prevent possible confusion. 

It turns out that the question whether it is justified to conclude from the ''right pointer positions`` subsequent to the Aharonov-Albert measurement, that the singlet state is verified, is more subtle to answer as it seems to be:

First, if the initial state is, say, $\mid\uparrow\downarrow\rangle$ there is non-vanishing probability that the singlet-state will be ''produced`` by the measurement and thereby the ''right pointer-positions`` (this can be straightforwardly calculated with the tools developed so far). Thus it cannot be verified that the particles has been in the singlet state prior to the measurement.

Second, it can also not be verified that the particles are in the singlet state subsequent to the readout of the two probe particles, if the interaction of the spin-$\frac{1}{2}$-particles with the probe-particles are not simultaneous (in the considered frame). To see this consider the scheme depicted in figure \ref{FoliationCalc} b) and suppose the depicted frame is the laboratory frame. As we have calculated (in connection with our reasoning about a possible foliation) this measurement yields the readout which is supposed to verify the singlet state, although the system is not in that state subsequent to the second interaction (of spin-$\frac{1}{2}$-particle(1) with probe-particle(1)) or subsequent to the second readout. The calculation makes no reference to a foliation, only the postulate of instantaneous collapse is applied in the frame depicted. Therefore also in a \textit{Solution \RM{1}}-model there is no justification to call the Aharonov-Albert procedure ''state verification``, if the interactions are not simultaneous in the considered frame. 

And finally, even if the interactions are simultaneous, a justification of the conclusion that the particles are in the singlet state subsequent to the measurement (with the right pointer positions) depends crucially on the underlaying theory: If a \textit{Solution \RM{2}}-model is taken as a basis the spin-$\frac{1}{2}$-particles depicted in figure \ref{Foliation} (where the interactions are supposed to occur simultaneously in the depicted frame) for example, will not be in the singlet state subsequent to the Aharonov-Albert measurement. Rather at the time of interaction with the probe-particles, the wave-function which describes the physics of spin-$\frac{1}{2}$-particle(1) is already the collapsed state (where the collapse is caused by the subsequent measurement of the z-component of spin-$\frac{1}{2}$-particle(2)). But nevertheless everything comes out right, i.e.\! the readout of the probe measurement will be the same as in the case of an underlying \textit{Solution \RM{1}}-model, according to which one can indeed speak of a state verification measurement of the singlet state, given the interactions are simultaneous.      

\newpage
 
\section{Appendix: Relativistic Law for the Primitive Ontology} 

\subsection{Relativistic Collapse Models}

\subsubsection{\label{rgrwf}A Point-Process on $\mathscr{M}$ \& Galaxies of Events: rGRWf}

\vspace*{.5cm}

\begin{center}
 \textbf{GRW and the flash ontology}
\end{center}

In brief and without entering mathematical description yet, the model proposed by Ghirardi, Rimini and Weber \cite{GRW} goes like this:

The wave-function dynamics is supposed to consist of two distinct dynamical processes: The unitary dynamics given by the standard Schr\"odinger-equation together with a stochastic jump process: In the spatial variable of each particle the wave-function is subjected to undergo a spontaneous and random localization process (referred to as \textsl{hitting} by some authors) at random times according to a Poisson process with mean frequency $\lambda\approx10^{-16}s^{-1}$. The localization is realized in this model by multiplication with a Gaussian localization-operator with width $\alpha\approx10^{-5}cm$ centered at random positions. The probability density $\mathbb{P}(\boldsymbol{\tilde{x}}_i)$ that a localization in the variable of particle($i$) about spatial point $\boldsymbol{\tilde{x}}_i$ occurs, is given in a way, such that there is higher probability of localization in regions in which, according to standard quantum mechanics, there is higher probability of finding the particle. 

An analysis of this model shows that its predictions coincide very well with all currently experimentally verified predictions of standard quantum theory\footnote{But nevertheless it deviates from standard quantum mechanics, such that it might be possible one day to decide empirically whether it makes the right predictions or not.} and that it has an important desired feature: The modified dynamics has little impact on microscopic objects (as it can already be seen from the very small value of the mean collapse frequency) and at the same time it destroys superpositions of different macroscopic states by some naturally arising amplification mechanism.

Bell \cite{BellJumps} realized another interesting feature of this model. He realized that the joint probability distribution
\begin{equation}\label{mtti}
\mathbb{P}\left(\left\{\boldsymbol{\tilde{x}}_{k_j}^{(j)} \in d^3x^{(j)}_{k_j} \mbox{ at } t^{(j)}_{k_j} \in dt^{(j)}_{k_j}\right\}_{j=1,...,N ; k_j=1,...,n_j} \right)
\end{equation}
of $n_j$ localizations in the variable of particle($j$) centered at positions $\boldsymbol{\tilde{x}}_{k_j}^{(j)}$ at times $t^{(j)}_{k_j}$, respectively, has a remarkable property which he calls \textsl{''relative time translation invariance``}\footnote{Bell motivated the requirement of ''relative time translation invariance`` by the observation, that, if two systems far apart in space are considered, the effect of small Lorentz transformations (i.e.\! corresponding to small velocity) to the first order results in a relative time shift between the two systems.}: This distribution does not change under relative time translations with respect to localizations associated with different particles if the particles do not interact. In other words, in this case, the prediction given for \eqref{mtti} is invariant under time-shift $t^{(j)}_{k_j}\rightarrow t^{(j)}_{k_j}+\Delta$, performed for all the times $t^{(j)}_{k_j}$ associated with localizations of a sub-system (some of the particles) of the considered system. 

This is remarkable since it opens the door to reconcile the inherent nonlocality of this model with the lack of an absolute time-order of space-like separated events in special relativity, if the space-time points $x^{(j)}_{k_j}=(t^{(j)}_{k_j},\boldsymbol{\tilde{x}}_{k_j}^{(j)})$ are taken as the physical events described by the theory --  if the events where the collapses adhere to are taken to constitute the primitive ontology or local beables of the theory. In the distribution \eqref{mtti} also space-like separated localization events are correlated in general (if the underlying wave-function is entangled), nevertheless these events are not constrained to realize some specific time-order!

 From now on, we will call the localization events $x^{(j)}_{k_j}$ \textsl{flashes} and a GRW-type-theory which postulates them to constitute matter GRWf \cite{commstruc}. With respect to the flashes Bell argues:

\begin{quotation}
 \textsl{So we can propose these events as the basis of the ''local beables`` of the theory. These are mathematical counterparts in the theory to real events at definite places and times in the real world (as distinct from the many purely mathematical constructions that occur in the working out of physical theories, as distinct from things which may be real but not localized, and as distinct from the ''observables`` of other formulations of quantum mechanics [...]. A piece of matter then is a galaxy of such events.} \cite{BellJumps}
\end{quotation}

And the ''relative time translation invariance`` of the physical description of these events suggests a good and clean potential for relativistic upgrade:
              
\begin{quotation}
 \textsl{And I am particularly struck by the fact that the model is as Lorentz invariant as it could be in the non-relativistic version. It takes away the ground of my fear that any exact formulation of quantum mechanics must conflict with fundamental Lorentz invariance.} \cite{BellJumps}
\end{quotation}
     
Motivated by the relative time translation invariance of the law of the flashes, Roderich Tumulka presented an elaborated formulation of a relativistic version of GRWf in 2006. This relativistic quantum theory is called rGRWf. It describes the (in general non-local) dynamics of N non-interacting distinguishable Dirac-particles without using spatio-temporal tools apart from the ones given by the Lorentz-metric $g^{\mu\nu}$ on $\mathbb{R}^4$. 

\begin{center}
 \textbf{The Scheme of rGRWf}
\end{center}

The simple idea is the following: The law of the theory is a probabilistic law for events occurring in space-time (i.e we are considering the set of all discrete subsets of space-time). Consider $n_j$ flashes $x^{(j)}_{k_j}\in\mathbb{R}^4$ associated with particle$(j)$, where $k_j=1,...,n_j$ and $j=1,...,N$. We will obtain then an expression for a joint probability distribution of the $\sum_{j=1}^N n_j$ flashes:
\begin{equation}\label{distr}
 \mathbb{P}\left(\left\{x^{(j)}_{k_j}\in d^4x^{(j)}_{k_j}\right\}_{j = 1,...,N  ; k_j = 1,...,n_j}\right)
\end{equation}

In order to calculate these probabilities we need initial Cauchy-data, given by the so called \textsl{seed flashes} together with the initial wave-function $\Psi_{\Sigma_0}\in\mathcal{H}_{\Sigma_0^N}$ on some arbitrary space-like hypersurface $\Sigma_0$. The wave-function is supposed to be a solution of the multi-time Dirac-equation \eqref{dirmult} and the predictions of the theory will be independent of the choice of the initial surface. The seed flashes $x_0^{(j)}$ are the last flashes before $x_1^{(j)}$ for all particles $j=1,...,N$, and they are supposed to be given.

\begin{figure}[htbp]
  \centering
    \includegraphics[scale=0.14]{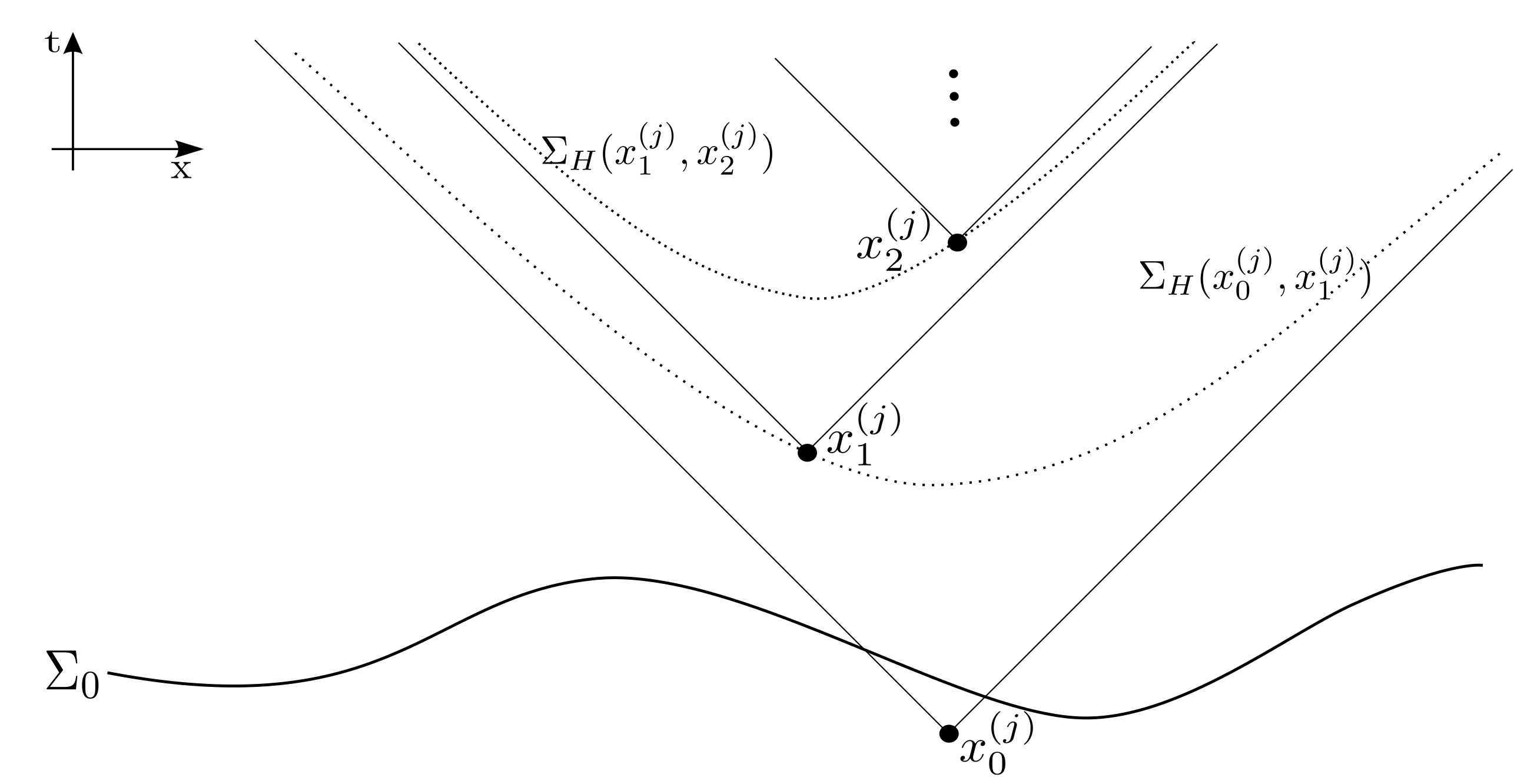}  
\caption{\textbf{Scheme for the calculation of the flash-history of particle$(j)$:} The initial conditions are given by the initial wave-function, given on some arbitrary leaf $\Sigma_0$, and the seed flash $x_0^{(j)}$. Starting from $x_0^{(j)}$ we transform the wave-function (unitarily) to the future hyperboloid $\Sigma_H(x_0^{(j)},x_1^{(j)})$ of $x_0^{(j)}$ containing the first flash $x_1^{(j)}$. There we apply an operator associated with a collapse at $x_1^{(j)}$ and transform the wave-function back to $\Sigma_0$. Then we repeat this procedure for the second flash $x_2^{(j)}$ with the future hyperboloid $\Sigma_H(x_1^{(j)},x_2^{(j)})$ of $x_1^{(j)}$ containing $x_2^{(j)}$ and so on...}
\label{rGRWfScheme}
\end{figure} 

The technical key is to consider the flash history of each particle separately (which is possible since the collapse-operators of different particles commute) and to make use of the invariant future-hyperboloids (see figure \ref{rGRWfScheme})
\begin{equation}
 \Sigma_H(x,y)=\{z\in\mathscr{F}(x)\mid (x^{\mu}-y^{\mu})(x_{\mu}-y_{\mu})=(x^{\mu}-z^{\mu})(x_{\mu}-z_{\mu})\}
\end{equation}
in the following way: Starting from the seed flash $x_0^{(j)}$ of the considered particle we evolve the initial wave-function unitarily from $\Sigma_0$ to the future-hyperboloid $\Sigma_H(x_0^{(j)},x_1^{(j)})$ of $x_0^{(j)}$ which contains $x_1^{(j)}$. There we apply an operator, associated with a collapse centered at $x_1^{(j)}$, and transform the wave-function back to $\Sigma_0$ (we can read this also in the Heisenberg picture, where the ``collapse operator`` on $\Sigma_H(x_0^{(j)},x_1^{(j)})$ is unitarily transported to $\Sigma_0$ by the described scheme). Then we repeat this procedure with the hyperboloid $\Sigma_H(x_1^{(j)},x_2^{(j)})$ and so on (see figure \ref{rGRWfScheme}). We can iterate this until we reach $x_{n_j}^{(j)}$ and do the same for the $N-1$ other particles. 

With an appropriate covariant definition of the collapse-operators acting on $\mathcal{H}_{\Sigma}$ and a reasonable covariant implementation of the exponential distribution of the flashes in time we can thereby obtain a covariant expression in order to calculate \eqref{distr}.   

\begin{center}
 \textbf{Mathematical Formulation}
\end{center}

Let us start with finding appropriate mathematical expressions corresponding to the flash history of one of the particles. Denote by $d_{\Sigma}(x,y)$ the distance of space-time points $x,y\in\Sigma$ along the surface $\Sigma$ given by the infimum of all Riemannian curve lengths along $\Sigma$ which connect $x$ and $y$. We define the collapse-operator $\left(\mathscr{K}_{\Sigma}(x)\right)^2$ about point $x\in\Sigma$ acting on $\mathcal{H}_{\Sigma}$ (or the extended operator on $\mathcal{H}_{\Sigma^N}$) by
\begin{equation}
 \big{(}\mathscr{K}_{\Sigma}(x)\big{)}^2\Psi(y):=\mathcal{N}\exp\left({-\frac{(d_{\Sigma}(x,y))^2}{2\alpha^2}}\right)\Psi(y) \mbox{ ,}
\end{equation}
with the above mentioned width of the localization $\alpha$ and normalization constant $\mathcal{N}$ such that 
\begin{equation}
 \int_{\Sigma} \left(\mathscr{K}_{\Sigma}(x)\right)^2 \: d^3x=1 \mbox{ .}
\end{equation}
Now let us define ``conditioned`` collapse-operators $\mathcal{K}(x\mid y)$ corresponding to a flash at $x\in\Sigma_H(y,x)$ if the preceding flash occurred at $y$
\begin{equation}
 \mathcal{K}(x\mid y):=\mathds{1}_{\mathscr{F}(y)}(x) \:\left(\lambda e^{-\frac{1}{2}\lambda\tau(x,y)}\right) \: \left( U^{\Sigma_H(y,x)}_{\Sigma_0}\mathscr{K}_{\Sigma_H(y,x)}(x)U^{\Sigma_0}_{\Sigma_H(y,x)}\right) \mbox{ .}
\end{equation}
The first factor is the indicator-function of the absolute future of $y$ and it ensures that two different flashes of one and the same particle are always time-like with respect to each other (otherwise the above expression and thereby the resulting probability of the corresponding flash-history will be zero). The second factor (bracket) corresponds to the Poisson distribution of the proper waiting time between two flashes $\tau(x,y)=\sqrt{\mid(x^{\mu}-y^{\mu})(x_{\mu}-y_{\mu})\mid}$ with expectation $\frac{1}{\lambda}\approx10^{16}s$ ($\approx10^7$ years). The remaining factor corresponds to the scheme illustrated above: Evolve the initial wave-function unitarily from $\Sigma_0$ to $\Sigma_H(y,x)$ by the operator $U^{\Sigma_0}_{\Sigma_H(y,x)}$ arising from the multi-time Dirac equation \eqref{dirmult}, apply the (square-root of the) collapse-operator and transform it back to $\Sigma_0$. Or we read it as the collapse-operator in the Heisenberg-picture unitarily transformed from $\Sigma_H(y,x)$ to $\Sigma_0$, where the initial wave-function is defined.  

Now we can define the flash-history operator of particle(j):
\begin{equation}
 \mathcal{K}^{(j)}\left(x_1^{(j)},...,x^{(j)}_{n_j}\right)=\mathcal{K}\left(x^{(j)}_{n_j} \mid x^{(j)}_{n_j-1}\right) \hspace{0.1cm} \mathcal{K}\left(x^{(j)}_{n_j-1} \mid x^{(j)}_{n_j-2}\right) \hspace{0.1cm} \: ... \: \hspace{0.1cm}\mathcal{K}\left(x^{(j)}_1\mid x^{(j)}_0\right)
\end{equation}
and consequently the flash-history operator of the N-particle system:
\begin{equation}
 \bigotimes_{j=1}^N\mathcal{K}^{(j)}\mbox{ .}
\end{equation}
Now, the square of this operator constitutes a positive operator valued measure (POVM) on $\mathscr{M}^{(\sum_{j=1}^Nn_j)}$ which generates, together with the initial wave-function, the desired joint probability distribution of the flashes:
\begin{equation}\label{flashlaw}
\mathbb{P}\left(\left\{x^{(j)}_{k_j}\in d^4x^{(j)}_{k_j}\right\}_{j = 1,...,N  ; k_j = 1,...,n_j}\Big{|} \Psi_{\Sigma_0},\left\{x^{(j)}_0\right\}\right)= \Bigg{\|} \bigotimes_{j=1}^N \mathcal{K}^{(j)} \Psi_{\Sigma_0} \Bigg{\|}^2 \prod_{j=1}^N\prod_{k_j=1}^{n_j} d^4x^{(j)}_{k_j}\mbox{ .}
\end{equation}

This is the law of the flashes. It defines a point process on space-time. It is invariant under change of initial-surface $\Sigma_0$, given the initial wave-function is subjected to a unitary evolution to the new surface, respectively the application of the corresponding collapse-operators (or their inverses) and normalizations for flashes laying between the old and the new surface\footnote{There are different proposals \cite{collandrel} for a definition of the collapsed wave-function (or the conditional wave-function, as Tumulka puts it) within this model, but these are rather mathematical restatements of the same physical law.}. Consequently we did not employ any intrinsic space-time structures apart from the ones given by the Lorentz-metric. The law is as Lorentz invariant as it could be. 

But nevertheless it describes in general correlations between space-like separated events (without drawing on common causes laying in the intersection of their past light-cones): The joint probability distribution for space-like separated flashes does not factorize in general, since the wave-function is not of product-form in general, but entangled. Although there is no definite time-order between space-like separated events, such events can be causally connected within this theory, if we call correlations, given by some law of nature, a \textsl{causal connection}. Obviously such kind of causation cannot being decomposed uniquely into \textsl{cause and effect}. \paragraph*{}

Now, with rGRWf we have a relativistic quantum-theory which provides a precise description of what goes on in space-time and which at the same time overcomes all the tension between quantum nonlocality and relativistic space-time structure. 

It is not able to describe particle interaction yet, this is a general problem arising from multi-time equations, but it is able to describe physical processes with underlying entangled wave-functions, as they would arise from interaction. Now it is a challenge to implement interaction (maybe not by an interaction-potential but by exchange of bosons), to formulate the model for non-distinguishable particles and to extend it to quantum field theory.

\subsubsection{rGRWm and CSL}

The GRW-theory describes the (linear and non-linear) evolution of the wave-function in a precise and coherent way. In order to relate this description to the four-dimensional physical world Bell proposed to take the \textsl{''GRW-jumps``} which \textsl{''are well localized in ordinary space''} \cite{BellJumps} as the local beables of the theory. And in the last section we encountered that this choice allows for a dynamics of these local beables which is intrinsically non-local and at the same time comprises the full spirit of special relativity, if the theory is reformulated in an appropriate way.

Another possible choice would be to take the average mass density $\mathcal{M}(\boldsymbol{x},t)$ (e.g.\! given by \eqref{massont}) arising from the wave-function as the primitive ontology. Such an ontology has been proposed by Ghirardi, Grassi and Benatti \cite{GhirardiGrassiBenatti} for continuous spontaneous localization models (\textit{CSL}) (which will be addressed in a moment):

\begin{quotation}
 \textsl{Therefore, we can guess that, within the context of the dynamical reduction program, the description of the world in terms of the mass-density function $\mathcal{M}(\boldsymbol{r})$ is a \textit{good} description.} \cite{GhirardiGrassiBenatti}
\end{quotation}

In the non-relativistic case a GRW-type-theory with a \textsl{mass-density ontology} (shorthand \textsl{GRWm}) will be empirically equivalent to GRWf, as far as predictions for outcomes of measurements are concerned \cite{commstruc} (the operator-measurement formalism can be derived in both cases the same way). But, proceeding to relativity, our previous analysis (section \ref{functional}) has shown that, in contrast to the flash ontology, it seems to be impossible to implement consistently a mass-density ontology into non-local relativistic quantum theory, if no distinguished space-like structures on Minkowski space-time are employed (even though, there might be a possibility to do so, which will be discussed at the end of this section).

But, of course, a relativistic version of GRWm (shorthand \textsl{rGRWm}) might be realized in a \textsl{Solution-\RM{2}} model: The integrations in \eqref{massont} in order to obtain, say, $\mathcal{M}(\boldsymbol{x}_0,t_0)$ is then to be performed along the leaf of a distinguished foliation containing $x=(t_0,\boldsymbol{x}_0)$. 

\begin{center}
 \textbf{CSL (continuous spontaneous localization)}
\end{center}
     
Most further developments on the issue of collapse theories overcame the jump-like character of the original GRW theory. The main motivation for the development of such CSL-models /cite{csl} has been an appropriate description of identical particles. In a CSL theory the Hamiltonian, as the generator of the time evolution of the wave-function, is supplemented by non-linear terms: Gaussian stochastic processes which couple to some operator, where the usual choice is the mass-density operator (see e.g.\! \cite{bassi}). The model provides all desired features for wave-function dynamics, in particular an (almost) linear dynamics for microscopic systems while the non-linearity destroys superpositions of differently located states of macroscopic systems. But apriori it is not clear how the wave-function relates to events in physical space-time, i.e.\! CSL has apriori no primitive ontology.

\textbf{Relativistic CSL:} Before we come to the primitive ontology I should mention one interesting feature of the wave-function in the proposals for relativistic CSL models. The first attempt to get a relativistic generalization of CSL was made by Pearle \cite{pearle1}. The model is not completely full-blown up to now, since it suffers from divergences. But this is rather a technical problem which might be solved one day. Nevertheless, the model is at least well suited to analyze some features of relativistic quantum mechanics \cite{ghirardilessons, ghirardinonlocal}.

The underlying dynamical equation for the wave-function is essentially the Tomonaga form \eqref{tomo2} with additional stochastic terms in the generator of the evolution, such that the evolution operator has non-hermitian structure. Due to this non-hermitian character the surface-independence of local values associated with the wave-function, like \eqref{surfaceind}, is violated in principle. This suggests that the model complies with a basic precondition in order to develop a \textsl{Solution \RM{1}} model, i.e.\! to develop a quantum theory which does not draw on a foliation or similar absolute physical structures on relativistic space-time.        

\textbf{The Primitive Ontology:} But, as we have argued, exactly this interesting feature of wave-function dynamics raises problems when we try to implement a mass-density ontology in a naive way into the theory -- such an ontology for CSL was suggested by Ghirardi, Grassi and Benatti \cite{GhirardiGrassiBenatti} in a non-relativistic framework (see quote above). And the flash-ontology does not work for CSL, since there are no flashes -- there is no discrete point-process on space-time in CSL. And it was crucial for the Lorentz invariant construction of rGRWf to use the invariant hyperboloids, respectively defined by two discrete flashes associated with one and the same particle, to derive the covariant law for the primitive ontology.

But, nevertheless, it might be possible to define a suitable primitive ontology in CSL without drawing on an intrinsic foliation: In \cite{ghirardilessons} Ghirardi proposed a \textsl{``criterion for events''}, which is in some sense analogous to the criterion for the \textsl{``elements of physical reality''} of EPR. Ghirardi points out that, in order to attribute ``objective`` properties to some space-time point $x$, we might consider two distinct surfaces on space-time: the backward light-cone $\Sigma(x)$ of $x$ and some space-like surface $\Sigma_0$ where the ''initial`` wave-function is specified. Then we evolve the wave-function from $\Sigma_0$ to a space-like surface arbitrarily close to $\Sigma(x)$ (the past light-cone itself is not space-like), where the nonlinear dynamics essentially strikes whenever e.g.\! measurement-like events are enclosed by $\Sigma_0$ and $\Sigma(x)$. 

If now the resulting wave-function $\psi_{\Sigma(x)}$ is an eigenstate of some operator $\hat{A}$, which is associated with a physical quantity at $x$, we can attribute the corresponding eigenvalue $\alpha$, as the value of the corresponding physical quantity, to that space-time point. If this state is not an eigenstate of the operator the local value of the corresponding physical quantity is indefinite. This concept is frame-independent since the initial surface is arbitrary and the past light-cone of $x$ is the same in all frames.

But Tumulka \cite{unrompics} pointed out the following: In order to obtain by this scheme the only events which really matter at the end of the day, i.e.\! the primitive ontology, a reasonable choice of one of these operators is enough to have a theory which describes what actually happens in space-time. If we take the mass-density operator at $x$ \textsl{''all other ``values'' are of no relevance. They are superfluous, as they do not influence how much matter is where, and thus do not influence the positions of pointers or the shape of ink on a paper. They are truly hidden variables and, indeed, can be deleted from the theory without unpleasant consequences just like the ether in relativistic mechanics -- yet unlike the particles in Bohmian mechanics.``} \cite{unrompics}

And in order to prevent possible ''indefiniteness'' of matter at some space-time points we can take the expectation value of the mass-density operator at $x$ with respect to the corresponding ``past light-cone state'' $\psi_{\Sigma(x)}$. This might indeed constitute a reasonable rGRWm theory.

It should be mentioned that this approach is not to be confused with the backward light-cone reductions approach of Hellwig and Kraus. In this model wave-functions still collapse along all space-like leafs passing through a ``collapse causing event`` in the sense of \textsl{Solution \RM{1}}.

\subsection{No Collapse but Particle Positions and Effective Wavefunctions}

\subsubsection{Bohmian Mechanics}

In quantum theory the Schr\"odinger-equation for an $N$-particles wave-function $\psi(\v{q},t)$, where $\v{q}\in\mathbb{R}^{3N}$, gives rise to a continuity equation for the probability density $\rho=\abs{\psi(\v{q},t)}^2$
\begin{equation}
 \frac{\partial}{\partial t}\rho(\v{q},t) - \div{\v{j}(\v{q},t)} =0 \mbox{ .}
\end{equation}
The current $\v{j}:\mathbb{R}^{3N}\times\mathbb{R}\rightarrow\mathbb{R}^{3N}$ is given by\footnote{For ease of notation we consider $N$ particles with the same mass $m$.} 
\begin{equation}
 \v{j}(\v{q},t)=\frac{1}{2mi}\left(\psi^{\star}\grad{\psi}-\psi\grad{\psi^{\star}}\right)=\frac{1}{m}\Im\left(\psi^{\star}\grad{\psi}\right)=\frac{\abs{\psi}^2}{m}\Im\left(\frac{\grad{\psi}}{\psi}\right)=:\rho\cdot\v{v}^{\psi} \mbox{ ,}
\end{equation}
with the velocity field $\v{v}^{\psi}:\mathbb{R}^{3N}\times\mathbb{R}\rightarrow\mathbb{R}^{3N}$
\begin{equation}\label{guiding1}
 \v{v}^{\psi}(\v{q},t)=\frac{1}{m}\Im\left(\frac{\grad{\psi}}{\psi}\right) \mbox{ .}
\end{equation}
And Bohmian mechanics \cite{BM1, BM2} means, to treat this velocity field on configuration space as the velocity field of an actual particle configuration $\v{Q}(t)$
\begin{equation}\label{guiding2}
 \d{\v{Q}}{t}=\v{v}^{\psi}(\v{Q},t) \mbox{,}
\end{equation}
i.e.\! the trajectory of particle($j$) is determined by some initial configuration and the equation
\begin{equation}\label{guiding3}
 \d{\v{Q}_j}{t}=\frac{1}{m}\Im\left(\frac{\grad_j{\psi}(\v{Q},t)}{\psi(\v{Q},t)}\right) \mbox{ .}
\end{equation}

This defines the theory; the remaining part is only analysis of the guiding equation \eqref{guiding3} (together with the Schr\"odinger equation) and the physical consequences (see e.g.\! \cite{undivuniv, quantequi, BMoperators}), some of which were mentioned in the introduction. 

\subsubsection{The Bohm-Dirac-Model}

Bohm \cite{undivuniv} proposed, in order to derive a relativistic version of the guiding law for Dirac particles, to take the preserved current arising from the $N$-particle Dirac-equation 
\begin{equation}\label{onetimeNpartDir}
 i \gamma^0_k\pd{\psi(\v{q},t)}{t} = \sum_{k=1}^N \bigg{(}-i\v{\gamma}_k \cdot \v{\partial}_{k}- e \v{\gamma}_k\cdot\v{A}(\v{q}_k,t) +e\gamma^0_k\Phi(\v{q}_k,t) + m \bigg{)} \psi(\v{q},t) 
\end{equation}
to define the Bohmian velocity field
\begin{equation}\label{BDlaw}
 \d{\v{Q}_k}{t}=\frac{\v{j}_k(\v{Q},t)}{\rho(\v{Q},t)}=\left.\frac{\bar{\psi}\left(\gamma^0_1\otimes...\otimes\gv{\gamma}_k\otimes...\otimes\gamma^0_N\right)\psi}{\bar{\psi}\gamma^0\psi}\right|_{(\v{Q},t)} \mbox{ .}
\end{equation}
Here $\gv{\gamma}_k=(\gamma^1_k,\gamma^2_k,\gamma^3_k)$, further $\gamma^i_k$ is the $i$'th Dirac matrix acting on the spin-index of particle($k$), $\gamma^0=\bigotimes_{k=1}^N\gamma^0_k$, as above we set $\bar{\psi}=\psi^{\dagger}\gamma^0$ and for ease of notation we consider $N$ particles with equal masses $m$, charges $e$ and external electromagnetic potentials $\v{A}$ and $\Phi$.

$\mathbf{N=1:}$ But this model -- as well as the one-time $N$-particle Dirac-equation -- is Lorentz invariant only if $N=1$. Then we can write the guiding law in a covariant manner
\begin{equation}
 \d{X^{\mu}}{s}\sim j^{\mu}=\bar{\psi}\gamma^{\mu}\psi \mbox{ ,}
\end{equation}
where $s\in \mathbb{R}$ is some scalar parametrization of the particle's world-line $X^{\mu}(s)$ and $j^{\mu}=(\rho,\v{j})$ is the Dirac four-current. Observe that the motion of the particle is already determined by the time-like directions tangent to its world-line, such that the length of the vectors $j^{\mu}$ has no physical significance for that motion. We can reformulate the law as $\d{X^{\mu}}{s} = \alpha j^{\mu}$ with some arbitrary scalar field $\alpha(x)$ or in a purely geometrical manner 
\begin{equation}\label{geom1}
 \d{X^{\mu}}{s} \quad \Big{\|} \quad j^{\mu}  
\end{equation}
where the symbol $\|$ means ''is parallel to``.

$\mathbf{N>1:}$ For $N>1$ equation \eqref{BDlaw} defines a law for the $N$ Dirac-particles in the distinguished frame $K$ which is defined by the constant-time slices corresponding to the time parameter in the time-derivative (observe that the derivative acts on a function of $N$ distinct position-variables $q_k$). In particular, as a result of the continuity equation, matter will be distributed according to $\rho=\abs{\psi}^2$ along all constant-time slices of $K$, but in general not according to $\abs{\psi'}^2$ in a different frame $K'$ \footnote{This observation reflects the fact that ''quantum-equilibrium cannot hold in all Lorentz-frames'', which has the status of a theorem and is analyzed in \cite{stochproc, sync}.}.

But, as Bohm argued, the distinguished frame must be ''hidden`` in this theory, as it is not amenable to experimental detection: Heuristically we can argue that predictions for the results of arbitrary measurements are inherent in distributions of matter of pointers or measurement devices subsequent to measurements. Hence these predictions can be derived from probabilities for positions according to a $\abs{\psi}^2$ distribution in the preferred frame. Since the validity of $\abs{\psi}^2$ distributions for macroscopic objects does not depend on a particular Lorentz frame in this theory, predictions for these positions of pointers and devices would be the same, if the preferred frame would be exchanged with a different one in the equations. Therefore, all kinds of measurements will validate Born's rule in each frame of reference and, in particular, it is not possible to detect the distinguished frame by means of experiments. To make this argument precise requires a bit more care (see also \cite{undivuniv, sync, HBDM}). 

The fact, that actual non-measured configurations might be distinct from the actual measured configurations, can be only understood if the non-passive character of measurements in quantum theory is appreciated.

\textbf{Product Wave-functions:} Indeed, in the case of product wave-functions the Bohm-Dirac law \eqref{BDlaw} defines a Lorentz invariant dynamics for $N>1$ particles, if we treat the wave-function as a function on $\mathscr{M}^N$ which is a solution of the multi-time Dirac-equation:   
\begin{equation}
 \psi(\v{q},t_1,...,t_n)=\prod_{k=1}^N\varphi_k(\v{q}_k,t_k)
\end{equation}
 is solution of the $N$ equations
\begin{equation}\label{multdir2}
 i\gamma^{\mu}_k(\partial_{k,\mu}-ieA_{\mu})\psi = m \psi ; \quad k=1,...,N
\end{equation}
and therefore
 \begin{equation}
  \Rightarrow i\gamma^{\mu}_k(\partial_{k,\mu}-ieA_{\mu})\varphi_k = m \varphi_k ; \quad \forall k=1,...,N \mbox{ .}
\end{equation}
In this case the guiding equations are be given by
\begin{equation} 
 \d{\v{Q}_k}{t_k}=\frac{\v{j}_k(\v{Q}_k,t_k)}{\rho(\v{Q}_k,t_k)}=\left.\frac{\bar{\varphi}_k\v{\gamma}_k\varphi_k}{\bar{\varphi}_k\gamma^0_k\varphi_k}\right|_{(\v{Q}_k,t_k)}
 \end{equation}
and this law does not depend on some particular choice of coordinates.   

\subsubsection{\label{HBDM}Hypersurface Bohm Dirac Models (HBDM's)}

But if we also want to account for entangled wave-functions the current 
\begin{equation}\label{current1}
j_k^{\mu}(\v{q}_1,t_1,...,\v{q}_N,t_N)=\bar{\psi}\left(\gamma^0_1\otimes...\otimes\gamma^{\mu}_k\otimes...\otimes\gamma^0_N\right)\psi
\end{equation}
arising from \eqref{multdir2} is no longer separable. And, in order to define velocities of the $N$ particles in the sense of \eqref{geom1}, we have somehow to connect the velocity of particle($k$) at time $t_k$ with $N-1$ four-tuples $(t_j,\v{Q}_j)$ corresponding to actual points in the world-lines of the $N-1$ other particles. This is done in the case of the Bohm-Dirac-model, equation \eqref{BDlaw}, by the constant-time slices defined by $t_1=t_2=...=t_N=t$, in which case the multi-time Dirac-equations \eqref{multdir2} reduce to the one-time Dirac-equation \eqref{onetimeNpartDir}. We can implement this restriction to a particular frame into the current \eqref{current1} by writing $\gamma^0_k=\gamma^{\mu}_k n_{\mu}$, where the constant four-vector field $n^{\mu}(x)=n^{\mu}$ picks out the zero-component of the four-vector $\gamma^{\mu}$ in the distinguished frame, i.e.\! $n^{\mu}(x)=(1,0,0,0)$ there. In other words, $n^{\mu}$ is the future oriented unit normal vector to the constant-time slices of the distinguished frame. 

Now we can rewrite the Bohm-Dirac law \eqref{BDlaw} in a way appropriate for generalization: Denote 
\begin{equation}
 j^{\mu}_k=\bar{\psi}(\gamma^{\nu}_1 n_{\nu})\otimes...\gamma^{\mu}_k...\otimes(\gamma^{\nu}_N n_{\nu})\psi \mbox{ .}
\end{equation}
Then we can write the probability density arising from the Dirac-equation as $\rho=j^{\mu}_k n_{\mu}$ (which is independent of $k$, of course) and consequently \eqref{BDlaw} as
\begin{equation}
\d{X^{\mu}_k}{t}=\frac{j^{\mu}_k}{j^{\mu}_k n_{\mu}} \mbox{ ,} 
\end{equation}
or geometrical, with some parametrization $s$ of the world-line $X^{\mu}_k(s)$ of particle($k$):
\begin{equation}
 \d{X^{\mu}_k}{s} \quad \Big{\|} \quad j^{\mu}_k \mbox{ .}
\end{equation}
It is important to note that $j^{\mu}_k$ and thereby $\d{X^{\mu}}{s}$ here depends on actual space-time positions $X^{\mu}_j$ of all $N$ particles in the following way: Denote by $\Sigma_s$ the constant time slice containing $X^{\mu}_k(s)$ in the preferred frame and by $X^{\mu}_j(\Sigma_s)$ the space-time location where the world-line of particle($j$) crosses $\Sigma_s$, for each particle $j=1,...,N$ (i.e.\! $X^{\mu}_k(s)\equiv X^{\mu}_k(\Sigma_s)$). Then the value of the current $j^{\mu}_k$ depends on these crossing-locations\footnote{This dependence enters into $j^{\mu}_k$ through the argument of the wave-function corresponding to the considered actual configuration on $\mathscr{M}^N$ and can be easily seen by comparing \eqref{BDlawfancy} with the equivalent formulation \eqref{BDlaw}}: 
\begin{equation}\label{BDlawfancy}
 \left.\d{X^{\mu}_k}{s}\right|_{s=\tilde{s}} \quad \Big{\|} \quad j^{\mu}_k\Big{(}X^{\mu}_1(\Sigma_{\tilde{s}}),...,X^{\mu}_N(\Sigma_{\tilde{s}})\Big{)} \mbox{ .}
\end{equation}

This reformulation of the Bohm-Dirac law \eqref{BDlaw} is well suited now to generalize the intrinsic dependence on some preferred frame of reference to an intrinsic dependence on some arbitrary foliation $\mathcal{F}$ of $\mathscr{M}$ into space-like hypersurfaces. \paragraph*{}

Given a foliation $\mathcal{F}$. Then there is a one to one correspondence between $\mathcal{F}$  and the time-like vector-field $n^{\mu}(x)$, defined by the unit normal vector to the leaf $\Sigma\in\mathcal{F}$ containing $x$ at $x$ for all $x\in\mathscr{M}$. Now define the current associated with particle($k$) as 
\begin{equation}
j^{\mu}_k=\bar{\psi}(\gamma^{\nu}_1 n_{\nu})\otimes...\gamma^{\mu}_k...\otimes(\gamma^{\nu}_N n_{\nu})\psi 
\end{equation}
where $n^{\mu}=n^{\mu}(x)$ is the unit normal vector-field of $\mathcal{F}$. Now consider again the world-line $X^{\mu}_k(s)$ of particle($k$), parametrized by some scalar parameter $s$ and denote by $\Sigma_s\in\mathcal{F}$ the leaf of the foliation containing $X^{\mu}_k(s)$. Also the point where the world-line of particle($j$) crosses $\Sigma_s$ is denoted by $X^{\mu}_j(\Sigma_s)$ again. Then the law of motion for particle($k$) is given by\footnote{The fact that $j^{\mu}_k$ is time-like (or light-like) and future-directed everywhere, and therefore that each particle world-line crosses each hypersurface $\Sigma\in\mathcal{F}$ exactly once, is due to the fact that $\gamma^0\gamma^{\mu}n_{\mu}$ is a positive operator on $\mathbb{C}^4$ for every time-like, future-directed unit vector $n^{\mu}\in\mathscr{M}$.}
\begin{equation}\label{hbdmlaw}
 \left.\d{X^{\mu}_k}{s}\right|_{s=\tilde{s}} \quad \Big{\|} \quad j^{\mu}_k\Big{(}X^{\mu}_1(\Sigma_{\tilde{s}}),...,X^{\mu}_N(\Sigma_{\tilde{s}})\Big{)} \mbox{ .}
\end{equation}
This is the equation of motion of the hypersurface-Bohm-Dirac-models \cite{HBDM}. They define a class of theories, since we did not fix a particular foliation yet and distinct foliations will yield distinct world-lines for identical initial conditions. 

But nevertheless an important foliation independent property of HBDM's can be proved \cite{HBDM}: There exists a distinguished probability-measure $\rho_{\psi_{\Sigma}} d^3x_1 \cdots d^3x_N$ (where $d^3x$ is the volume measure arising from the Riemann metric on $\Sigma$) for the distributions of crossings $X^{\mu}_1(\Sigma),...,X^{\mu}_N(\Sigma)$ on the leafs $\Sigma\in\mathcal{F}$ of the foliation. It is generated by the wave-function on $\Sigma^N$ and distinguished in the following sense: Given a distribution of crossings $X^{\mu}_1(\Sigma_0),...,X^{\mu}_N(\Sigma_0)$ according to $\rho_{\psi_{\Sigma_0}}$ on some leaf $\Sigma_0\in\mathcal{F}$. Then the dynamics given by \eqref{hbdmlaw} yields world-lines of the $N$ particles such that the distribution of crossings $X^{\mu}_1(\Sigma),...,X^{\mu}_N(\Sigma)$ is given by $\rho_{\psi_{\Sigma}}$, with
$\psi_\Sigma$ emerging by unitary transformation of $\psi_{\Sigma_0}$ according to \eqref{multdir2}, for all surfaces $\Sigma\in\mathcal{F}$. A probability measure with this property is called an \textsl{equivariant measure} and the distribution defined by it (and preserved by the dynamics) the \textsl{quantum equilibrium distribution}.    

This distribution is given by $\rho=j^{\mu}_k n_{\mu}$ (which is independent of $k$). In the case of a flat foliation into constant-time hyperplanes of some frame, it reduces to $\rho=\psi^{\dagger}\psi$ in the preferred frame. The crucial property in order to proof the equivariance of $\rho$ is the validity of the continuity equation $\partial^{\mu}_k j_{k,\mu}=0$ for the currents, which follows from the multi-time Dirac-equation \eqref{multdir2}. Due to the equivariance of $\rho$ the outcomes of quantum-measurements must confirm Born's probability rule in all frames of reference and it is due to this fact that the shape of the leafs of the foliation cannot be revealed by any experimental procedure (see \cite{HBDM}).

\subsubsection{The Foliation}

Now, what is the actual benefit of a fancy curved foliation possible in a HBDM? After all, in the Bohm-Dirac-models, in which only flat foliations are possible, all relevant physical properties are already present, above all an equivariant probability measure. And, to begin with, the generalization to arbitrary shapes of the space-like leafs adds no further physical contents. It only enlarges the number of possible physical theories captured by the class of theories. But by that it might give us more freedom to find a reasonable choice of a foliation.

The only element in the formulation of the theory which seems to conflict with the spirit of relativity is this absolute structure on space-time -- the foliation. Does the foliation still conflict with relativity, if it is given by a covariant law? After all, there are some decorations of space-time in our theories which are additional to the structures given by the Lorentz-metric: We describe physical fields and distributions of mass and the like in space-time and we only require to describe them by covariant laws. At the end of the day, maybe, the situation is somewhat more delicate with a foliation, since it is a decoration which introduces a kind of temporal order in space-time. But nonetheless, only in case of a complete covariant description we can speak of a relativistic theory; however, still it might be controversial how much the theory is in accord with the ''true spirit of relativity``.\paragraph*{}

To come to an end now, let us quickly present two proposals for a covariant foliation: \paragraph*{}

It is actually not true that there is no frame of reference which is physically distinguished in relativistic physics: For example, as Ward Struyve suggested \cite{Ward}, we can choose the frame in which the total energy, defined by the wave-function of the universe, is at rest: Consider the average total four-momentum $\avg{P^{\mu}}_{\Psi}$ arising from the covariant energy momentum tensor $T^{\mu\nu}$ (in the Heisenberg picture), where the average is performed with respect to the covariant wave-function $\Psi$ of the universe:
\begin{equation}
 \avg{P^\mu}_{\Psi} = \int_\Sigma d\sigma_\nu(x) \langle \Psi \mid T^{\mu\nu}(x) \mid \Psi \rangle \mbox{ .}
\end{equation}
Here $\Sigma$ is some arbitrary space-like surface and $d\sigma_\nu(x)$ is the infinitesimal normal four vector onto $\Sigma$ at $x\in\Sigma$. Because of the continuity equation $\partial_\mu T^{\mu\nu}=0$ the value of the above integral is independent of the choice of the space-like hyper-surface $\Sigma$ (see e.g.\! \cite{Schweber}). $\avg{P^\mu}_\Psi$ is a constant time-like four-vector \cite{Schweber}, i.e.\! it defines a Lorentz-frame whose constant-time slices are orthogonal to $\avg{P^{\mu}}_{\Psi}$. This frame is defined by a thoroughly covariant law and it exists in every theory with a wave-function in which an energy momentum tensor can be defined. Hence, if we choose the Bohm-Dirac model \eqref{BDlawfancy} (HBDM is not even needed here) in which $n^{\mu}=\avg{P^{\mu}}_{\Psi}\cdot\|\avg{P^{\mu}}_{\Psi}\|^{-1}$, we have a covariant theory. 

It is interesting to note, that the only physical variable which enters into the law for the foliation here, is the covariant wave-function (of the universe). Therefore, ultimately, we do not need to add additional structure to the theory. This structure is already inherent in the covariant wave-function of the universe! Although the role this structure plays in the theory is somehow special, it is not introduced as an extra element. And, in the end, the particle dynamics is completely determined by the wave-function alone. \paragraph*{}

Another attractive possibility is the following: Consider the surfaces of constant time-like distance from the big bang \cite{unrompics}. These space-like hypersurfaces actually do not constitute an extra structure of space-time: They are given purely by the big bang and the relativistic metric. If it is possible to define them as the foliation $\mathcal{F}$ which enters into the HBDM law \eqref{hbdmlaw}, we might also say that we have a relativistic Bohmian quantum theory without additional space-time structure. 

\vspace{8cm}

\newpage

\lhead{Acknowledgement}

\begin{Acknowledgement} 

\vspace{1cm}

 First I want to thank my daughter Luca for being there, for her laughing and for her talent to give me power and confidence in times of doubt.  And I want to express my deep gratitude to my teacher and friend Detlef D\"urr, who taught me much more about physics than I ever learned from lectures and textbooks and who always encouraged me to continue my way. I would like to thank Lisa Riedner for holding a very special place in my heart and in my life. I thank her also for proofreading parts of the manuscript and even for driving me  crazy sometimes, with simple and justified questions about what I'm doing there. I want to thank all of my friends. Each one of them is unique and special to me. And special thanks to Peter Niedersteiner, who has the talent to laugh at me at the right times and to take me serious at the right times; even if he calls me a desk criminal when I spend lots of time behind my desk, solving problems... I would  like to express special thanks to my friends Florian Hoffmann, Christoph Weber, Dustin Lazarovici and Niklas Boers for lots of extensive and as hard as fruitful discussions about life, physics, mathematics, philosophy, sense and nonsense and also for proofreading my work. Thanks a lot! Also special thanks to Roderich Tumulka and Shelly Goldstein for inspiring me with their work, for making good and fruitful comments on earlier versions of this manuscript and for helping me thereby to make important improvements. I want to thank Amelie, Marianne and Norbert Vanderquaden who accompanied me during an important period of my life. I would like to thank Gertrud and Hermann Emmert for supporting me and for making suggestions for improvements of  my work. Last but not least, I want to thank my parents Christa Beck and G\"unter Beck for supporting me and for always believing in me!

\end{Acknowledgement}
\newpage

\lhead{\rightmark}
\nocite{*}
\bibliographystyle{acm}
\bibliography{Bibl}

\end{document}